\documentclass[11pt,a4,twoside]{article}
\usepackage[dvips,a4paper,left=2.5cm,right=2.5cm,top=2.5cm,bottom=2.5cm,headsep=1em]{geometry}
\usepackage{fancyhdr}
\usepackage{titlesec}
\usepackage[latin1]{inputenc}
\usepackage{amsmath,amsfonts}
\usepackage{rotating}
\usepackage[font={small}]{caption}
\usepackage{natbib}
\usepackage{lmodern,slantsc}
\usepackage{multirow}
\usepackage{chngcntr}
\usepackage{placeins}
\usepackage{caption}
\usepackage[breaklinks,
colorlinks=true, linkcolor=blue, citecolor=black, urlcolor=blue,
pdfborder={0 0 0}, pdfpagelabels]{hyperref}
\usepackage{tikz}
\def\vx{\mathbf x}

\def\vr{\mathbf r}

\def\v0{\boldsymbol{0}}

\def\L{\langle}
\def\R{\rangle}

\newcommand{\vpx}[1]{\vx^{{\scriptscriptstyle (#1)}}}

\newlength{\FigureHeight}
\newlength{\FigureHeightHalf}
\newcommand{\FigureXYLabel}[5]
{\settoheight{\FigureHeight}{#1}
\setlength{\FigureHeightHalf}{0.5\FigureHeight}
\begin{center}
\raisebox{\FigureHeightHalf}{\makebox{#4\makebox[#5]{}}}
#1\\
\vspace{#3}
#2\\
\end{center}}
\numberwithin{equation}{section}
\titleformat{\section}
{\large\bfseries}{\thetitle.}{0.5em}{}
\titleformat{\subsection}
{\normalfont\itshape\filcenter}{\normalfont\thetitle.}{0.5em}{}
\begin{document}

\title{\vspace{-1.5em}Revisiting the Lie-group symmetry method for turbulent\\
channel flow with wall transpiration}
\author{George Khujadze$\,^1$\thanks{Email address for correspondence:
george.khujadze@uni-siegen.de}$\,\,$ \&$\,$ Michael Frewer$\,^2$\\ \\
\small $^1$ Chair of Fluid Mechanics, Universit\"at Siegen, 57068
Siegen, Germany\\
\small $^2$ Tr\"ubnerstra{\ss}e 42, 69121 Heidelberg, Germany}
\date{{\small\today}}
\clearpage \maketitle \thispagestyle{empty}

\vspace{-1.5em}\begin{abstract}

\noindent The Lie-group-based symmetry analysis, as first proposed
in
\href{http://journals.cambridge.org/action/displayAbstract?fromPage=online&aid=9216346&fileId=S0022112014000986}{Avsarkisov
{\it et al.}~(2014)} and then later modified in
\href{https://www.jstage.jst.go.jp/article/mer/2/2/2_15-00157/_article}{Oberlack
{\it et al.}~(2015)}, to generate invariant solutions in order to
predict the scaling behavior of a channel flow with uniform wall
transpiration, is revisited. By focusing first on the results
obtained in \cite{Oberlack14}, we failed to reproduce two key
results: (i) For different transpiration rates at a constant
Reynolds number, the mean velocity profiles (in deficit form) do
not universally collapse onto a single curve as claimed. (ii) The
universally proposed logarithmic scaling law in the center of the
channel does not match the direct numerical simulation (DNS) data
for the presented parameter values. In fact, no universal scaling
behavior in the center of the channel can be detected from their
DNS data, as it is misleadingly claimed in \cite{Oberlack14}.
Moreover, we will demonstrate that the assumption of a
Reynolds-number {\it independent} symmetry analysis is not
justified for the flow conditions considered therein. Only when
including also the viscous terms, an overall consistent symmetry
analysis can be provided. This has been attempted in their
subsequent study \cite{Oberlack15Rev}.

But, also the (viscous) Lie-group-based scaling theory proposed
therein is inconsistent, apart from the additional fact that this
study of \cite{Oberlack15Rev} is also technically flawed. The
reason for this permanent inconsistency is that their symmetry
analysis constantly involves several unphysical statistical
symmetries that are incompatible to the underlying deterministic
description of Navier-Stokes turbulence, in that they violate the
classical principle of cause and effect. In particular, as we
consequently will show, the matching to the DNS data of the scalar
dissipation, being a critical indicator to judge the prediction
quality of any theoretically derived scaling law, fails
exceedingly.

\vspace{0.5em}\noindent{\footnotesize{\bf Keywords:} {\it
Symmetries, Lie Groups, Scaling Laws, Symmetry Breaking,
Turbulence, Channel Flow, Wall Transpiration, Statistical
Mechanics, Higher-Order Moments, Closure Problem, Causality}}$\,$;
\end{abstract}

\pagenumbering{arabic}\setcounter{page}{1}

\section{Motivation and objectives\label{S1}}

The main purpose of this investigation is first to reveal in how
far the work of \cite{Oberlack14} can be reproduced. With focus on
the results obtained from Lie-group analysis, we will re-examine
all derivations and conclusions in \cite{Oberlack14}. One of the
key results obtained therein was that of a new {\it universal}
logarithmic scaling law in the center (core region) of a plane
turbulent channel flow with uniform wall-normal transpiration. The
derivation of this law, presented in \cite{Oberlack14} as
[Eq.$\,$(3.16)]
\begin{equation}
\bar{U}_1=A_1\ln\left(\frac{x_2}{h}+B_1\right)+C_1,
\label{160430:1552}
\end{equation}
where $A_1=k_{\bar{U}_1}/k_1$, $B_1=k_{x_2}/(h k_1)$ are two
group\footnote[2]{The constant $B_1$ as defined in
\cite{Oberlack14} misses a factor $1/h$ in order to be
dimensionally correct.} and $C_1$ one arbitrary integration
constant, is\pagebreak[4] based on three independent scaling
symmetries [Eqs.$\,$(3.2)-(3.4)] and two independent translation
symmetries [Eqs.$\,$(3.5)-(3.6)] of the two-point correlation
(TPC) equations [Eqs.$\,$(2.12)-(2.16)] for the purely inviscid
case $\nu=0$.\footnote[2]{Note that the large-Reynolds-number
asymptotics in the cited reference \cite{Oberlack00.1} was
performed differently than as claimed in the beginning of
Sec.$\,$3.1 on p.$\,$109 in~\cite{Oberlack14}.\linebreak Not for
$|\vr|\leq \eta$, but rather, oppositely, only for $|\vr|\geq
\eta$ it was shown that all viscous terms in the TPC equations
vanish. For a corresponding English explanation of the
``asymptotic analysis" performed in \cite{Oberlack00.1}, see e.g.
\cite{Oberlack02B,Oberlack03} or \cite{Khujadze04}. Hence,
oppositely as claimed, the symmetry analysis in \cite{Oberlack14}
was {\it not} performed on equations which have undergone a prior
singular asymptotic analysis in the sense~$\nu\rightarrow 0$, but
instead, only on equations which just result from considering the
purely inviscid (Euler) case~$\nu=0$.} The emergence of the
particular scaling law \eqref{160430:1552} from these just
mentioned symmetries is due to the externally set constant
transpiration velocity $v_0$, which acts as a symmetry breaking
parameter in the scaling of the mean wall-normal velocity
$\bar{U}_2$ through the single constraint
$k_1-k_2+k_s=0$~[Eq.$\,$(3.15)]. Central to the claim of
\cite{Oberlack14} is that when matching the new logarithmic law
\eqref{160430:1552} to direct numerical simulation (DNS) data,
then this law turns out to be a {\it universal} one when written
in its deficit form (normalized to the mean friction velocity
$u_\tau$ as defined in [Eq.$\,$(2.1)]\footnote[3]{In Appendix
\ref{A} we repeat the basic derivation of relation [Eq.$\,$(2.1)]
in \cite{Oberlack14} to acknowledge this result more carefully.})
\begin{equation}
\frac{\bar{U}_1-C_1}{u_\tau}=\frac{1}{\gamma}\ln\left(\frac{x_2}{h}+B_1\right),
\label{160519:1744}
\end{equation}
where all involved matching parameters $\gamma$, $B_1$ and $C_1$
are independent of the transpiration rate and Reynolds number. In
particular, after a fit to the given data, the following universal
values were proposed \cite[Sec.$\,$4, pp.$\,$116-119]{Oberlack14}:
\begin{equation}
\gamma=0.3, \qquad B_1=0, \qquad C_1=U_B, \label{160519:1743}
\end{equation}
where $\gamma$ is the new universal scaling coefficient to be
distinguished from the usual von Kármán constant $\kappa$ of the
near-wall logarithmic scaling law, and where $U_B$ is the mean
bulk velocity [Eq.$\,$(2.4)] which was kept universally constant
in all performed simulation runs for different transpiration rates
and Reynolds numbers (due to a fixed overall mass-flow rate
employed in the used DNS code; for more details, see also
\cite{Avsarkisov13}).

Our investigation on all these derived and proposed results
involve three independent parts. After introducing the governing
statistical equations and admitted Lie symmetries in
Section~\ref{S2} with the information only as given in
\cite{Oberlack14}, we will demonstrate the following:

(i) The DNS-data-matched value of $A_1$ in \eqref{160430:1552},
namely $A_1=u_\tau/\gamma$, is inconsistent to its theoretically
derived value $A_1=k_{\bar{U}_1}/k_1$ composed of two group
constants, which are, by construction, independent of the friction
velocity~$u_\tau$.

(ii) Fig.$\,$9 $(a)$ and $(c)$ in \cite{Oberlack14} cannot be
reproduced when using the DNS data made available by the authors
on their institutional
website~\href{http://www.fdy.tu-darmstadt.de/forschung_16/direkte_numerische_simulation/direkte_numerische_simulation.de.jsp}{[fdy]}.
Neither does the data universally collapse onto a single curve for
different blowing parameters in particular, nor does the
logarithmic scaling law \eqref{160519:1744} with the proposed
parameters \eqref{160519:1743} directly fit to this data.

(iii) For the inviscid ($\nu=0$) case, as particularly realized in
\cite{Oberlack14}, as well as for the viscous ($\nu\neq0$) case,
as subsequently modified in \cite{Oberlack15Rev}, the
Lie-group-based scaling theory shows in both cases a
methodological inconsistency in that certain higher order velocity
correlation functions cannot be matched anymore to the DNS data,
despite involving all {\it a priori} known symmetries of the
underlying statistical transport equations. The simple reason for
this inconsistency is that several participating symmetries are
unphysical in violating the classical principle of cause and
effect.

\section{Governing statistical equations and admitted
symmetries\label{S2}}

Since the aim in \cite{Oberlack14} is to investigate within the
inviscid ($\nu=0$) TPC equations [Eqs.$\,$(2.12)-(2.15)] only
large-scale quantities, such as the mean velocity or the Reynolds
stresses, we will proceed accordingly by considering these TPC
equations already in their one-point limit
($\vpx{2}\to\vpx{1}=\vx$, or in relative coordinates as
$\vr=\vpx{2}-\vpx{1}\to\boldsymbol{0}$):\footnote[2]{Similar to
the strategy as proposed, e.g., in \cite{Oberlack03}
[pp.$\,$462-466] or \cite{Khujadze04} [pp.$\,$395-399], only large
scale quantities as the mean velocity and Reynolds stresses are
investigated via the inviscid ($\nu=0$) TPC equations including
their one-point limit. For small scale quantities as the
dissipation, the viscous TPC equations are needed, which (in their
one-point limit) will be discussed later in Section~\ref{S52}.}
\begin{gather}
\frac{\partial\bar{U}_k}{\partial x_k}=0,\label{160423:2032}\\[0.25em]
\frac{\partial \bar{U}_i}{\partial t}+\bar{U}_k\frac{\partial
\bar{U}_i}{\partial x_k}+\frac{\partial\bar{P}}{\partial
x_i}+\frac{\partial\tau_{ik}}{\partial x_k}=0,\\[0.25em]
\frac{\partial \tau_{ij}}{\partial t}+\bar{U}_k\frac{\partial
\tau_{ij}}{\partial x_k}+\frac{\partial\tau_{ijk}}{\partial
x_k}+\tau_{ik}\frac{\partial\bar{U}_j}{\partial
x_k}+\tau_{jk}\frac{\partial\bar{U}_i}{\partial x_k}+
\overline{\frac{\partial p}{\partial
x_i}u_j}+\overline{u_i\frac{\partial p}{\partial x_j}}=0,
\label{160423:1915}
\end{gather}
where
\begin{equation}
\tau_{ij}=\overline{u_iu_j},\qquad
\tau_{ijk}=\overline{u_iu_ju_j},
\end{equation}
are the Reynolds stresses and the third-order (one-point) velocity
moments, respectively. Note that in this one-point limit all
higher-order continuity constraints [Eqs.$\,$(2.14)-(2.15)] either
collapsed into the single constraint \eqref{160423:2032} or turned
into trivial zero identities.

Referring to the cited study \cite{Oberlack10} in
\cite{Oberlack14}, it has been shown that that a simple and
systematic structure for all symmetries is revealed if for the
infinite hierarchy of multi-point correlation (MPC) equations the
instantaneous (full) field approach is used (instead of the
fluctuating, the so-called Reynolds-decomposed field approach as
given above). In the one-point limit the corresponding full-field
representation of the inviscid TPC equations reads: \vspace{-1em}
\begin{gather}
\frac{\partial\overline{U_k}}{\partial x_k}=0,\label{160430:1813}\\[0.25em]
\frac{\partial\overline{U_i}}{\partial t}+\frac{\partial
\overline{U_iU_k}}{\partial x_k}+\frac{\partial
\overline{P}}{\partial x_i}=0,\label{160423:1918}\\[0.25em]
\frac{\partial \overline{U_iU_j}}{\partial t}+
\underbrace{\overline{\frac{\partial U_iU_k}{\partial
x_{k\vphantom{g_{g_g}}}}U_j} + \overline{U_i\frac{\partial
U_jU_k}{\partial x_k}}}_{{\displaystyle =\frac{\partial}{\partial
x_k}\overline{U_iU_jU_k}}} +\overline{\frac{\partial P}{\partial
x_i}U_j}+\overline{U_i\frac{\partial P}{\partial
x_j}}=0,\label{160430:1814}
\end{gather}
which, of course, turns exactly into the system
\eqref{160423:2032}-\eqref{160423:1915} when decomposing the full
fields into their mean and fluctuating part, i.e., by performing a
usual Reynolds field decomposition\footnote[3]{Note that in order
to obtain the explicit form of equation \eqref{160423:1915} from
\eqref{160430:1814}, the decomposed equation \eqref{160423:1918}
\raisebox{0.1em}{has to be used as an auxiliary equation.}}
\begin{equation}
U_i=\bar{U}_i+u_i,\qquad P=\bar{P}+p.\label{160501:1208}
\end{equation}
Although both representations
\eqref{160423:2032}-\eqref{160423:1915} and
\eqref{160430:1813}-\eqref{160430:1814} are equivalent, the latter
one has the unreckoned advantage, according to \cite{Oberlack10},
of being a linear system which makes the extraction of Lie
symmetries considerably easier.

For the specific flow considered in \cite{Oberlack14}, both
systems \eqref{160423:2032}-\eqref{160423:1915} and\linebreak
\eqref{160430:1813}-\eqref{160430:1814} equivalently reduce
further. Considered is a statistically stationary plane channel
flow of width $w=2h$ with a mean constant wall-normal
transpiration $\bar{U}_2=v_0$. In the streamwise direction the
flow is driven by constant mean pressure gradient, which we will
denote as $K$, in particular $\partial\bar{P}/\partial x_1 = -K$,
where $K>0$ is some arbitrary but fixed positive value. Finally,
due\pagebreak[4] to spanwise homogeneity and a spanwise reflection
symmetry in this flow, the mean spanwise velocity as well as all
velocity moments involving an uneven number of spanwise velocity
fields vanish. Hence, for the just-stated assumptions, the
full-field system \eqref{160430:1813}-\eqref{160430:1814} reduces
to:\footnote[2]{The two assumptions that the mean pressure
$\bar{P}$ decays linearly in the streamwise direction and that the
mean wall-normal velocity $\bar{U}_2$ is constant across the
channel height will be applied at a later stage.}
\begin{gather}
\frac{\partial\overline{U_2}}{\partial x_2}=0,\label{160430:1912}\\[0.25em]
\frac{\partial \overline{U_1U_2}}{\partial x_2}+\frac{\partial
\overline{P}}{\partial x_1}=0,\qquad \frac{\partial
\overline{U_2U_2}}{\partial x_2}+\frac{\partial
\overline{P}}{\partial x_2}=0,\qquad
\overline{U_1U_3}=\overline{U_2U_3}=0,
\label{160430:1913}\\[0.25em]
\frac{\partial\overline{U_1U_2U_2}}{\partial x_2}
+\overline{\frac{\partial P}{\partial
x_1}U_2}+\overline{U_1\frac{\partial P}{\partial x_2}}=0,\qquad
\frac{\partial\overline{U_iU_jU_2}}{\partial x_2}
+\overline{\frac{\partial P}{\partial
x_i}U_j}+\overline{U_i\frac{\partial P}{\partial
x_j}}=0,\;\text{for $i=j$},\label{160430:1914}
\end{gather}
while its corresponding Reynolds decomposed system
\eqref{160423:2032}-\eqref{160423:1915} equivalently reduces to:
\begin{gather}
\frac{\partial\bar{U}_2}{\partial x_2}=0,\label{160423:2129}\\[0.25em]
\bar{U}_2\frac{\partial\bar{U}_1}{\partial
x_2}+\frac{\partial\bar{P}}{\partial x_1}+\frac{\partial
\tau_{12}}{\partial x_2}=0,\qquad
\underbrace{\bar{U}_2\frac{\partial\bar{U}_2}{\partial
x_2}}_{=0}+\frac{\partial\bar{P}}{\partial x_2}+\frac{\partial
\tau_{22}}{\partial x_2}=0,\qquad \tau_{13}=\tau_{23}=0,\label{160426:0005}\\[0.25em]
\left.
\begin{aligned}
\phantom{x}\hspace{2.5cm}\bar{U}_2\frac{\partial
\tau_{12}}{\partial x_2}+\frac{\partial\tau_{122}}{\partial
x_2}+\underbrace{\tau_{12}\frac{\partial\bar{U}_2}{\partial
x_2}}_{=0}+\tau_{22}\frac{\partial\bar{U}_1}{\partial x_2}+
\overline{\frac{\partial p}{\partial
x_1}u_2}+\overline{u_1\frac{\partial p}{\partial x_2}}=0,\hspace{0.6cm}\\[0.25em]
\bar{U}_2\frac{\partial \tau_{ij}}{\partial
x_2}+\frac{\partial\tau_{ij2}}{\partial
x_2}+\tau_{i2}\frac{\partial\bar{U}_j}{\partial
x_2}+\tau_{j2}\frac{\partial\bar{U}_i}{\partial x_2}+
\overline{\frac{\partial p}{\partial
x_i}u_j}+\overline{u_i\frac{\partial p}{\partial
x_j}}=0,\;\text{for $i=j$}.
\end{aligned}
~~~\right\}\label{160423:2130}
\end{gather}
When considering the list of TPC symmetries [Eqs.$\,$(3.2)-(3.6)]
as analyzed in \cite{Oberlack14}, then the reduced
Reynolds-decomposed system \eqref{160423:2129}-\eqref{160423:2130}
admits the symmetries\footnote[3]{Please note that since the
system \eqref{160430:1912}-\eqref{160430:1914}, or its equivalent
Reynolds decomposed system
\eqref{160423:2129}-\eqref{160423:2130}, is unclosed even if the
infinite hierarchy of equations is formally considered, all
admitted invariant transformations can only be regarded in the
weak sense as equivalence transformations, and not as true
symmetry transformations in the strong sense. For more details, we
refer to \cite{Frewer14.1,Frewer15.0,Frewer15.0x} and the
references therein. In the following, however, we will continue to
call them imprecisely as ``symmetries", like it was also done in
\cite{Oberlack14}.}
\begin{align}
\bar{T}_1: &\quad x_i^*=e^{k_1}x_i,\;\;\;
\bar{U}_i^*=e^{k_1}\bar{U}_i,\;\;\;
\bar{P}^*=e^{2k_1}\bar{P},\;\;\;
\tau_{ij}^*=e^{2k_1}\tau_{ij},\nonumber\\
&\quad \tau_{ijk}^*=e^{3k_1}\tau_{ijk},\;\;\;
\overline{u_i\frac{\partial p}{\partial x_j}}^{\, *}=e^{2k_1}\,
\overline{u_i\frac{\partial p}{\partial x_j}},\label{160430:2222}\\[1.0em]
\bar{T}_2: &\quad x_i^*=x_i,\;\;\;
\bar{U}_i^*=e^{-k_2}\bar{U}_i,\;\;\;
\bar{P}^*=e^{-2k_2}\bar{P},\;\;\;
\tau_{ij}^*=e^{-2k_2}\tau_{ij},\nonumber\\
&\quad
\tau_{ijk}^*=e^{-3k_2}\tau_{ijk},\;\;\;\overline{u_i\frac{\partial
p}{\partial x_j}}^{\, *}=e^{-3k_2}\, \overline{u_i\frac{\partial
p}{\partial x_j}},\\[1.0em]
\bar{T}^\prime_s: &\quad  x_i^*=x_i,\;\;\;
\bar{U}_i^*=e^{k_s}\bar{U}_i,\;\;\;
\bar{P}^*=e^{k_s}\bar{P},\;\;\;
\tau_{ij}^*=e^{k_s}\tau_{ij}+\big(e^{k_s}-e^{2k_s}\big)\,\bar{U}_i\bar{U}_j,
\nonumber\\[0.15em]
&\quad
\tau_{ijk}^*=e^{k_s}\tau_{ijk}+\big(e^{k_s}-e^{2k_s}\big)\big(\bar{U}_i\tau_{jk}
+\bar{U}_j\tau_{ik}+\bar{U}_k\tau_{ij}\big)\nonumber\\[0.15em]
&\hspace{2.57cm} +
\big(e^{k_s}-3e^{2k_s}+2e^{3k_s}\big)\,\bar{U}_i\bar{U}_j\bar{U}_k,
\hspace{1cm}\nonumber\\[0.15em]
&\quad \overline{u_i\frac{\partial p}{\partial x_j}}^{\,
*}=e^{k_s}\, \overline{u_i\frac{\partial p}{\partial x_j}}
+\big(e^{k_s}-e^{2k_s}\big)\,\bar{U}_i\frac{\partial\bar{P}}{\partial
x_j},\label{160430:2130}
\end{align}
\begin{align}
\bar{T}_{x_i}: &\quad x_i^*=x_i+k_{x_i},\;\;\;
\bar{U}_i^*=\bar{U}_i,\;\;\; \bar{P}^*=\bar{P},\;\;\;
\tau_{ij}^*=\tau_{ij},\nonumber\\
&\quad \tau_{ijk}^*=\tau_{ijk},\;\;\;\overline{u_i\frac{\partial
p}{\partial x_j}}^{\, *}=\overline{u_i\frac{\partial
p}{\partial x_j}},\\[1.0em]
\bar{T}_{\bar{U}_1}: &\quad x_i^*=x_i,\;\;\;
\bar{U}_1^*=\bar{U}_1+k_{\bar{U}_1},\;\;\;
\bar{U}_2^*=\bar{U}_2,\;\;\; \bar{P}^*=\bar{P},\;\;\;
\tau_{ij}^*=\tau_{ij},\hspace{1.4cm}\nonumber\\
&\quad \tau_{ijk}^*=\tau_{ijk},\;\;\;\overline{u_i\frac{\partial
p}{\partial x_j}}^{\, *}=\overline{u_i\frac{\partial p}{\partial
x_j}},\label{160430:2131}
\end{align}
which directly follows from the set of TPC symmetries
[Eqs.$\,$(3.2)-(3.6)]\footnote[2]{As it stands in
\cite{Oberlack14}, [Eq.$\,$(3.6)] is not admitted as a symmetry by
the TPC equations [Eq.$\,$(2.16)]. Only if $k_{\bar{U}_2}=0$ it
turns into a symmetry transformation. Also note that the classical
translation symmetry [Eq.$\,$(3.5)] can be extended as an
independent shift in all three coordinate directions.} in
\cite{Oberlack14} when performing the limit of zero spatial
correlation $\vr\to\boldsymbol{0}$ (one-point limit) and a
subsequent prolongation to higher-order moments. By equivalently
rewriting the moments into their full-field form, we obtain the
corresponding symmetries admitted by the reduced full-field system
\eqref{160430:1912}-\eqref{160430:1914}:
\begin{align}
\bar{T}_1: &\quad x_i^*=e^{k_1}x_i,\;\;\; \overline{U_i}^{\,
*}=e^{k_1}\overline{U_i},\;\;\; \overline{P}^{\,
*}=e^{2k_1}\overline{P},\;\;\;
\overline{U_iU_j}^{\, *}=e^{2k_1}\overline{U_iU_j},\nonumber\\
&\quad \overline{U_iU_jU_k}^{\,
*}=e^{3k_1}\overline{U_iU_jU_k},\;\;\; \overline{U_i\frac{\partial
P}{\partial x_j}}^{\, *}=e^{2k_1}\, \overline{U_i\frac{\partial
P}{\partial
x_j}},\label{160425:1017}\\[1.0em]
\bar{T}_2: &\quad x_i^*=x_i,\;\;\; \overline{U_i}^{\,
*}=e^{-k_2}\overline{U_i},\;\;\; \overline{P}^{\,
*}=e^{-2k_2}\overline{P},\;\;\;
\overline{U_iU_j}^{\, *}=e^{-2k_2}\overline{U_iU_j},\nonumber\\
&\quad \overline{U_iU_jU_k}^{\,
*}=e^{-3k_2}\overline{U_iU_jU_k},\;\;\;\overline{U_i\frac{\partial
P}{\partial x_j}}^{\, *}=e^{-3k_2}\, \overline{U_i\frac{\partial
P}{\partial x_j}},\label{160502:1402}\\[1.0em]
\bar{T}^\prime_s: &\quad  x_i^*=x_i,\;\;\; \overline{U_i}^{\,
*}=e^{k_s}\overline{U_i},\;\;\; \overline{P}^{\,
*}=e^{k_s}\overline{P},\;\;\;
\overline{U_iU_j}^{\, *}=e^{k_s}\overline{U_iU_j},\nonumber\\
&\quad \overline{U_iU_jU_k}^{\,
*}=e^{k_s}\overline{U_iU_jU_k},\;\;\; \overline{U_i\frac{\partial
P}{\partial x_j}}^{\, *}=e^{k_s}\, \overline{U_i\frac{\partial
P}{\partial x_j}},\label{160424:1418}\\[1.0em]
\bar{T}_{x_i}: &\quad  x_i^*=x_i+k_{x_i},\;\;\; \overline{U_i}^{\,
*}=\overline{U_i},\;\;\; \overline{P}^{\, *}=\overline{P},\;\;\;
\overline{U_iU_j}^{\, *}=\overline{U_iU_j},\nonumber\\
&\quad \overline{U_iU_jU_k}^{\,
*}=\overline{U_iU_jU_k},\;\;\;\overline{U_i\frac{\partial
P}{\partial x_j}}^{\, *}=\overline{U_i\frac{\partial
P}{\partial x_j}},\label{160501:1832}\\[1.0em]
\bar{T}_{\bar{U}_1}: &\quad x_i^*=x_i,\;\;\; \overline{U_1}^{\,
*}=\overline{U_1}+k_{\bar{U}_1},\;\;\; \overline{U_2}^{\,
*}=\overline{U_2},\;\;\; \overline{P}^{\,
*}=\overline{P},\nonumber\\
&\quad\overline{U_iU_j}^{\, *}=\overline{U_iU_j}+k_{\bar{U}_1}
\Big(\delta_{1i}\overline{U_j}+\delta_{1j}\overline{U_i}\Big)
+k_{\bar{U}_1}^2\delta_{1i}\delta_{1j},\nonumber\\
&\quad \overline{U_iU_jU_k}^{\,
*}=\overline{U_iU_jU_k}+2\,\overline{U_i}\;\overline{U_j}\;\overline{U_k}
-\overline{U_i}\;\overline{U_jU_k}
-\overline{U_j}\;\overline{U_iU_k}-\overline{U_k}\;\overline{U_iU_j}\nonumber\\
&\quad\hspace{3.2cm} -2\,\overline{U_i}^{\, *}\overline{U_j}^{\,
*}\overline{U_k}^{\, *}+\overline{U_i}^{\, *}\overline{U_jU_k}^{\,
*}+\overline{U_j}^{\, *}\overline{U_iU_k}^{\,
*}+\overline{U_k}^{\, *}\overline{U_iU_j}^{\, *},\nonumber\\
&\quad\overline{U_i\frac{\partial P}{\partial x_j}}^{\,
*}=\overline{U_i\frac{\partial P}{\partial
x_j}}+k_{\bar{U}_1}\delta_{1i}\frac{\partial
\overline{P}}{\partial x_j},\label{160424:1452}
\end{align}
which again, when performing the Reynolds decomposition
\eqref{160501:1208}, turn back into the symmetries
\eqref{160430:2222}-\eqref{160430:2131}. In contrast to the
translation symmetry $\bar{T}_{\bar{U}_1}$ \eqref{160424:1452},
the scaling symmetry $\bar{T}^\prime_s$~\eqref{160424:1418} gained
a very simple form in the full-field representation. This
so-called third scaling symmetry $\bar{T}^\prime_s$ in the TPC
equations was first derived and discussed in \cite{Khujadze04},
and only later generalized in \cite{Oberlack10} for the infinite
hierarchy of MPC equations.

As we will demonstrate in detail in Section \ref{S5}, since our
central aim is to coherently extend the invariance analysis in
\cite{Oberlack14} to higher-order moments in which the scaling law
for the lowest-order moment (mean velocity field) is based on a
translation symmetry,\pagebreak[4] corresponding and independent
translation symmetries are also needed for all higher-order
moments in order to generate invariant functions with arbitrary
offsets being flexible enough to match the DNS data. In other
words, to be able to robustly match higher-order invariant
functions to DNS data, higher-order translation symmetries are
needed as they were first derived in \cite{Oberlack10}.

In this regard it is worthwhile to note that the considered TPC
translation symmetry [Eq.$(3.6)$] in~\cite{Oberlack14}, does {\it
not} correspond to the symmetry ``discovered in the context of an
infinite set of statistical symmetries in \cite{Oberlack10}"
[p.$\,$110], as misleadingly claimed in \cite{Oberlack14}.
Instead, when adapted to the reduced one-point and full-field
system \eqref{160430:1912}-\eqref{160430:1914}, it is given by
[Eq.$\,$(58)] in \cite{Oberlack10}~as\footnote[2]{Due to the
particular flow configuration considered, transformation
\eqref{150501:1342} is only admitted as a symmetry by
\eqref{160430:1912}-\eqref{160430:1914} if
$c_{13}=c_{23}=c_{333}=0$, and $c_{ij3}=0$, for all $i\neq 3$ and
$j\neq 3$.}
\begin{align}
\bar{T}^\prime_{c}: &\quad x_i^*=x_i,\;\;\; \overline{U_i}^{\,
*}=\overline{U_i}+c_i,\;\;\; \overline{P}^{\,
*}=\overline{P}+d,\;\;\;
\overline{U_iU_j}^{\, *}=\overline{U_iU_j}+c_{ij},\nonumber\\
&\quad \overline{U_iU_jU_k}^{\,
*}=\overline{U_iU_jU_k}+c_{ijk},\;\;\; \overline{U_i\frac{\partial
P}{\partial x_j}}^{\, *}=\overline{U_i\frac{\partial P}{\partial
x_j}}, \label{150501:1342}
\end{align}
or, in its corresponding Reynolds decomposed form, as
\begin{align}
\!\bar{T}^\prime_{c}: &\;\,\, x_i^*=x_i,\;\;\;
\bar{U}_i^*=\bar{U}_i+c_i,\;\;\; \bar{P}^*=\bar{P}+d,\;\;\;
\tau_{ij}^*=\tau_{ij}+\bar{U}_i\bar{U}_j-\bar{U}^*_i\bar{U}^*_j+c_{ij},\nonumber\\
&\;\,\,
\tau_{ijk}^*=\tau_{ijk}+\bar{U}_i\bar{U}_j\bar{U}_k+\bar{U}_i\tau_{jk}
+\bar{U}_j\tau_{ik} +\bar{U}_k\tau_{ij}
-\bar{U}^{*}_i\bar{U}_j^{*}\bar{U}_k^*-\bar{U}^*_i\tau^*_{jk}
-\bar{U}^*_j\tau^*_{ik}
-\bar{U}^*_k\tau^*_{ij}+c_{ijk},\,\nonumber\\
&\;\,\, \overline{u_i\frac{\partial p}{\partial x_j}}^{\,
*}=\overline{u_i\frac{\partial p}{\partial
x_j}}+\bar{U}_i\frac{\partial\bar{P}}{\partial
x_j}-\bar{U}^*_i\frac{\partial\bar{P}^*}{\partial x^*_j},
\label{150424:1633}
\end{align}
which does not reduce to \eqref{160430:2131}, when specifying the
group constants correspondingly to $c_1=k_{\bar{U}_1}$ and
$c_2=d=c_{ij}=c_{ijk}=0$, and which thus is the symmetry sought
that independently translates all higher-order moments. In other
words, the single translation symmetry \eqref{160430:2131} is not
a ``first principle" symmetry as misleadingly claimed in
\cite{Oberlack14}, but was, in contrast to \eqref{150424:1633},
rather introduced in an {\it ad hoc} manner just to serve the
single purpose to generate a suitable logarithmic scaling law for
the lowest-order moment (mean velocity field) without knowing at
the same time whether this scaling is also consistent to all
higher-order moments.\footnote[3]{Regarding the justification of
the translation symmetry \eqref{160430:2131} given as [Eq.$(3.6)$]
in~\cite{Oberlack14},\linebreak it should be noted that also their
statement ``...$\,$that the first hint towards (3.6) has been
given by Kraichnan~(1965)" [p.$\,$110], is incorrect and
constitutes a misinterpretation of Kraichnan's idea to random
Galilean invariance. This misconception has been recently revealed
in \cite{Frewer15.1}.} Hence, next to the single translation
symmetry \eqref{160430:2131}, we will also apply the new
``statistical translation symmetry" \eqref{150424:1633}, first
proposed in \cite{Oberlack10}, in order to achieve a consistent
prolongation to all higher-order moments within the symmetry
analysis as particularly put forward and initialized in
\cite{Oberlack14} (see Section \ref{S51}), and then as
subsequently modified in \cite{Oberlack15Rev} (see Section
\ref{S52}).

\section{On the inconsistency between the data-matched value and
the theoretically predicted relation of
$\boldsymbol{A}_{\boldsymbol{1}}$\label{S3}}

As described in Sec.$\,$4 in \cite{Oberlack14}, the best fit to
{\it all} DNS data is obtained if the scaling coefficient in the
theoretically derived law \eqref{160430:1552} is chosen as
\begin{equation}
A_1=\frac{u_\tau}{0.3}.\label{160520:2204}
\end{equation}
Since all simulation runs in \cite{Oberlack14} were performed
under the unusual constraint of a universally fixed mean bulk
velocity $U_B=U_B^*$ for different transpiration rates
$v^+_0=v_0/u_\tau$ and Reynolds numbers $Re_\tau=u_\tau h/\nu$,
one inevitably obtains the following parametrical dependency
relationship for the mean friction velocity$\,$\footnote[2]{That
an extra parametrical relation as \eqref{160520:1004} is necessary
to follow and to understand the numerical simulation performed,
has not been directly discussed in \cite{Oberlack14}. For all
simulation runs, the value $U_B^*$ was unconventionally chosen as
$U_B^*=0.8987$. Note that this information is not given in
\cite{Oberlack14}; it~can only be found on their institutional
data repository
\href{http://www.fdy.tu-darmstadt.de/forschung_16/direkte_numerische_simulation/direkte_numerische_simulation.de.jsp}{[fdy]}.}
\begin{equation}
u_\tau=u_\tau(U_B^*,v^+_0,Re_\tau),\label{160520:1004}
\end{equation}
which for the turbulent case yet can only be determined
empirically. For the laminar case, however, a closed analytical
expression can be derived (see \eqref{160524:1523} in Appendix
\ref{B}). Relation \eqref{160520:1004} can be easily validated by
taking the non-normalized definition of the mean bulk velocity
[Eq.$\,$(2.4)] in \cite{Oberlack14} and recalling the fact that
due to the Navier-Stokes equations along with the supplemented
boundary conditions, the mean streamwise velocity profile will in
general be a function of all involved parameters of the considered
flow ($h$: channel half-hight, $K$: constant mean streamwise
pressure gradient, $v_0$: constant mean wall-normal transpiration
rate, $\nu$: kinematic viscosity):
\begin{equation}
U_B=\frac{1}{2h}\int_0^{2h}\bar{U}_1(x_2)dx_2=U_B(h,K,v_0,\nu)
=u_\tau\cdot\Pi(v_0^+,Re_\tau),\label{160520:1541}
\end{equation}
where the last relation represents its non-dimensionalized single
form (relative to $u_\tau=\sqrt{|K|h}$)\linebreak depending only
on two dimensionless variables $v_0^+=v_0/u_\tau$ and
$Re_\tau=u_\tau h/\nu$. Note that if we universally fix
$U_B=U_B^*$ in \eqref{160520:1541}, then two of the three
parameters $u_\tau$, $v_0^+$ and $Re_\tau$ can be varied
independently to satisfy this constraint. The third one is then
predetermined by solving \eqref{160520:1541} for this parameter,
e.g., if we choose $u_\tau$ as the dependent one, we obtain in
this particular normalization the relation
\begin{equation}
u_\tau=\frac{U_B^*}{\Pi(v_0^+,Re_\tau)},\label{160520:2043}
\end{equation}
which, of course, represents the unique dimensional reduction of
its generalized expression \eqref{160520:1004}.
\begin{table}
\begin{center}
\begin{tabular}{c c c | c | c}\hline\\
& & & Turbulent flow & Laminar flow\\[0.5em]
$U_B^*$ & $Re_\tau$ & $v_0^+$ & $u_\tau$ & $u^L_\tau$\\[0.5em]
0.8987  & 250       & 0.05    & 0.0577   & 0.0488\\
0.8987  & 250       & 0.10    & 0.0707   & 0.0936\\
0.8987  & 250       & 0.16    & 0.1023   & 0.1475\\
0.8987  & 250       & 0.26    & 0.1861   & 0.2373\\
0.8987  & 250       & $\infty$&          & $\infty$\\[0.5em]
0.8987  & 480       & 0.05    & 0.0551   & 0.0469\\
0.8987  & 480       & 0.10    & 0.0695   & 0.0918\\
0.8987  & 480       & 0.16    & 0.1004   & 0.1457\\
0.8987  & 480       & 0.26    & 0.1859   & 0.2355\\
0.8987  & 480       & $\infty$&          & $\infty$\\[0.5em]
0.8987  & 850       & 0.05    & 0.0501   & 0.0460\\
0.8987  & 850       & 0.16    & 0.0980   & 0.1449\\[0.5em]
0.8987  & $\infty$  & $v_0^+\neq 0$ & & $U_B^*\cdot
v_0^+$\\
0.8987  & $\infty$ & $\infty$ & & $\infty$
\end{tabular}
\caption{Calculated values for $u_\tau$ according to relation
\eqref{160520:2043} for initially given $U_B^*$, $Re_\tau$ and
$v_0^+$. The latter two values were taken from Table$\,$1
[p.$\,$106] in \cite{Oberlack14}, while the values $U_B^*$ and
$u_\tau$ for the turbulent flow case were taken from the
corresponding DNS data base disclosed by the authors on their
institutional
website~\href{http://www.fdy.tu-darmstadt.de/forschung_16/direkte_numerische_simulation/direkte_numerische_simulation.de.jsp}{[fdy]}.
The values for the corresponding laminar friction velocities
$u_\tau^L$ were calculated through the analytical formula
\eqref{160524:1524} (for given $U_B^*$, $Re_\tau$ and $v_0^+$) to
serve as a comparison to the DNS-determined mean friction
velocities $u_\tau$ in the turbulent case.} \label{tab1}
\end{center}
\vspace{-0.4em}\hrule
\end{table}
In Table \ref{tab1} we provide the set of data obtained in
\cite{Oberlack14} to illustrate the mode of action of relation
\eqref{160520:2043} for different turbulent flow conditions. The
corresponding laminar flow cases are given as a comparison, which,
in contrast to the turbulent ones, can be determined analytically,
where in particular the dimensionless function $\Pi$ in
\eqref{160520:2043} can be represented even in closed form (see
\eqref{160524:1523} in Appendix~\ref{B}). To note is the
non-intuitive result that if $v_0^+$ stays fixed, $u_\tau$
monotonically decreases as $Re_\tau$ increases; a result obviously
caused by the (universally fixed) constant mean bulk velocity
$U_B^*$ for these simulations.

Hence, according to \eqref{160520:2043}, the empirically matched
scaling coefficient $A_1$ \eqref{160520:2204} shows the following
dependency in that it can be equivalently written as
\begin{equation}
A_1=\frac{U_B^*}{0.3\cdot\Pi(v_0^+,Re_\tau)}.\label{160520:2235}
\end{equation}
However, such a dependency is inconsistent to the theoretically
derived result of $A_1$ in \eqref{160430:1552}, which is given as
\begin{equation}
A_1=\frac{k_{\bar{U}_1}}{k_1},\label{160531:1935}
\end{equation}
where $k_{\bar{U}_1}$ and $k_1$ are two group parameters which
both, due the particular symmetry analysis performed in
\cite{Oberlack14}, are independent of the Reynolds number
$Re_\tau$. The~reason is that the performed symmetry analysis in
\cite{Oberlack14} was done under the constraint of zero viscosity
($\nu=0$), thus leading to a Lie-group-based derivation of $A_1$
\eqref{160531:1935}, that, by construction, cannot depend on $\nu$
(or equivalently on $Re_\tau$).\pagebreak[4]

Although the friction velocity $u_\tau$ \eqref{160520:2043} only
shows a rather weak $Re_\tau$-dependence when compared to its
dependence on the transpiration rate $v_0^+$, as can be seen in
Table \ref{tab1}, this dependence, however, cannot be neglected:
For example, for the fixed transpiration rate $v_0^+=0.05$, we
have a change of nearly 15\% in $u_\tau$ when increasing the
Reynolds number from $Re_\tau=250$ to $850$. This change in
$u_\tau$ \eqref{160520:2043} is then directly reflected in $A_1$
according to its empirical relation \eqref{160520:2235} proposed
in \cite{Oberlack14}. And this change will continue to grow when
increasing the Reynolds number even further. But, as we can
observe from the corresponding laminar values $u_\tau^L$ in Table
\ref{tab1}, this growth is bounded, i.e., for
$Re_\tau\rightarrow\infty$ the values for $u_\tau$ will converge
to some certain finite value $u_\tau^\infty$, which is also
expected to happen in the turbulent case since the flow is
arranged under the same unusual condition of a universally fixed
bulk velocity $U_B^*$:
\begin{equation}
\lim_{Re_\tau\to\infty}
u_\tau=\lim_{Re_\tau\to\infty}\frac{U_B^*}{\Pi(v_0^+,Re_\tau)}
=\frac{U_B^*}{\Pi_\infty(v_0^+)}=u_\tau^\infty,\;\;\text{for}\;\;
0<v_0^+ <\infty.
\end{equation}
In contrast to the laminar case, where this value is analytically
accessible and particularly given as $u_\tau^\infty=U_B^* v_0^+$,
it is, of course, an unknown quantity for the turbulent case; yet
still, it will take a different (most possibly lower) value than
for any finite Reynolds number~$Re_\tau<\infty$.

Although weak, the friction velocity $u_\tau$ \eqref{160520:2043},
and thus also the empirically matched logarithmic scaling
coefficient $A_1$ \eqref{160520:2235}, nevertheless shows a
non-negligible $Re_\tau$-dependence for every initially fixed
transpiration rate $v_0^+$, a dependence which, as we will
demonstrate in Section \ref{S51}, is critical when extending the
symmetry analysis as put forward in \cite{Oberlack14} to
higher-order moments.

Hence, the central assumption in \cite{Oberlack14} to predict the
(non-normalized) mean velocity scaling behavior in the center of
the channel by an invariant log-law resulting from a non-viscous
($\nu=0$) symmetry analysis, namely by \eqref{160430:1552} where
$A_1$ and $B_1$ are independent on $Re_\tau$, is not justified.
For that, a viscous ($\nu\neq 0$) symmetry analysis has to be
performed, but then, as we will consequently show in Section
\ref{S52}, no invariant mean velocity profile for $\bar{U}_1$ can
be constructed anymore,\footnote[2]{Except only for a linear
profile \eqref{160612:1806} as derived in Section \ref{S52}, but
which, of course is not a reasonable turbulent scaling law. To
note is that in \cite{Oberlack15Rev} the authors succeeded to
derive both a logarithmic as well as an algebraic scaling law for
the mean velocity field in the viscous case. But, as we will
analytically prove in Section \ref{S52}, both results are based on
a methodological mistake. This is also expressed in the fact that
as we repeat their inconsistent analysis, certain higher order
moments cannot be matched to the DNS data.} due to the well-known
(scaling) symmetry breaking mechanism of the viscous~terms.

\section{On the problems when trying to reproduce Figure 9\label{S4}}

\begin{figure}[t]
\centering
\begin{minipage}[c]{.48\linewidth}
\FigureXYLabel{\includegraphics[width=.91\textwidth]{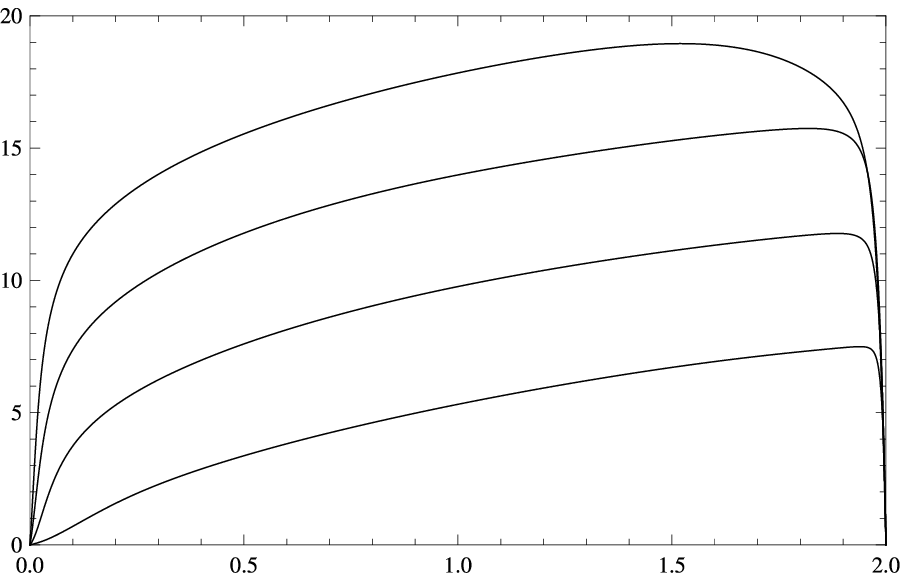}}
{${\scriptstyle\hspace{0.75cm} x_2/h}$}{-1mm}{\begin{rotate}{0}
${\scriptstyle\bar{U}_1^+}$\end{rotate}}{5mm}
\end{minipage}
\hfill
\begin{minipage}[c]{.48\linewidth}
\FigureXYLabel{\includegraphics[width=.91\textwidth]{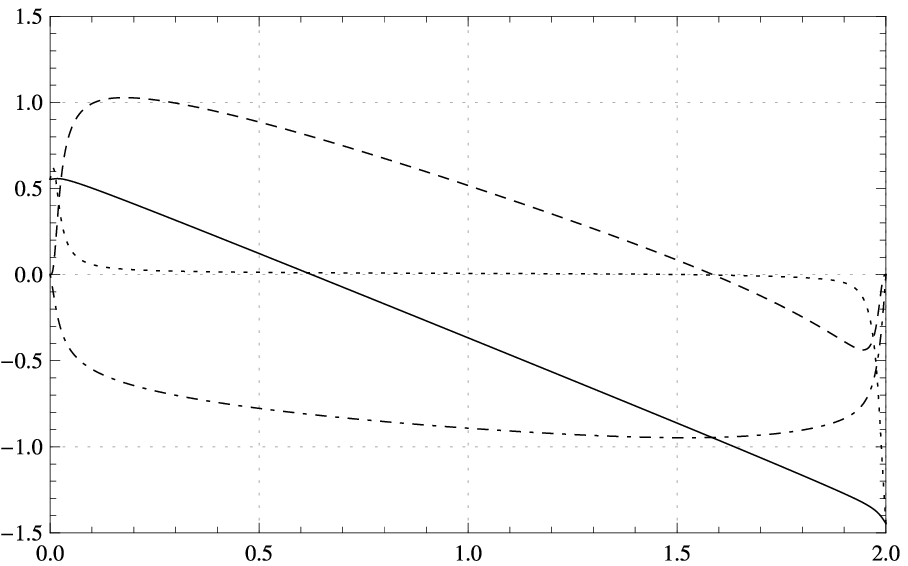}}
{${\scriptstyle\hspace{0.75cm} x_2/h}$}{-1mm}{\begin{rotate}{0}
${\scriptstyle\psi}$\end{rotate}}{2mm}
\end{minipage}
\caption{Reproduction of Fig.$\,$3 $(a)$ and $(c)$ in
\cite{Oberlack14} with the data provided by the authors on their
institutional
website~\href{http://www.fdy.tu-darmstadt.de/forschung_16/direkte_numerische_simulation/direkte_numerische_simulation.de.jsp}{[fdy]}.
{\it Left plot:} Mean streamwise velocity profile $\bar{U}_1^+$ at
$Re_\tau=480$ for $v_0^+=0.05$, $0.1$, $0.16$ and $0.26$ (from top
to bottom). $\,${\it Right plot:} The quantity $\psi$ displays
different shear stress distributions at $Re_\tau=480$ for the
fixed transpiration rate $v_0^+=0.05$: $-\overline{u_1u_2}^+$ (- -
-);$\:$ $-v_0^+\bar{U}_1^+$ (- $\cdot$ -);$\:$
$d\bar{U}_1^+/dx_2^+$ ($\cdots$);$\:$ $\tau^+ -v_0^+\bar{U}_1^+$
(---), where $\tau^+=-\overline{u_1u_2}^+ +d\bar{U}_1^+/dx_2^+$ is
the total shear stress without transpiration as defined in
[Eq.$\,$(2.11)] in \cite{Oberlack14}. The differences in this
right plot to Fig.$\,$3 $(c)$ are of minor significance: (i) The
small numerical error (misfeature) at the wall boundaries in the
profile $\tau^+ -v_0^+\bar{U}_1^+$ (solid line) has not been
displayed in Fig.$\,$3 $(c)$. Across the full channel height, a
pure straight line with the same slope of measure one has been
given instead, i.e., the true profile of this quantity has not
been plotted in \cite{Oberlack14}. (ii) A careful comparison
reveals a very small discrepancy in the vertical position of both
the dashed-dotted and the solid line. This negligible difference
may be explained by the circumstance that the data base on the
author's website may refer to a different, most possibly to a
newer simulation run with a better statistics than the one
presented in \cite{Oberlack14}: The latter version was published
in January 2014, while the released data base on their website was
created a year later in February 2015. Nevertheless, both plots
above show that Fig.$\,$3 $(a)$ and $(c)$ in \cite{Oberlack14} are
reproducible, confirming thus that we are using the correct
data~set.}\vphantom{x}\hrule \label{fig1}
\end{figure}

In this section we show the results of our effort to reproduce
Figs.$\,$9 $(a)$ and $(c)$ in \cite{Oberlack14}. The underlying
simulation data were taken from the author's institutional
website~\href{http://www.fdy.tu-darmstadt.de/forschung_16/direkte_numerische_simulation/direkte_numerische_simulation.de.jsp}{[fdy]}.
To verify and to ensure that we operate with the same data set as
presented in \cite{Oberlack14}, we first have to check if we are
able to repeat the construction of another figure. By choosing
Figs.$\,$3 $(a)$ and $(c)$ as representative test cases, our
reproduced plots in Figure \ref{fig1} undoubtedly show that we are
indeed in hold of the correct simulation data to systematically
investigate the reproducibility of all plots in \cite{Oberlack14}.

\begin{figure}[t]
\centering
\begin{minipage}[c]{.48\linewidth}
\FigureXYLabel{\includegraphics[width=.91\textwidth]{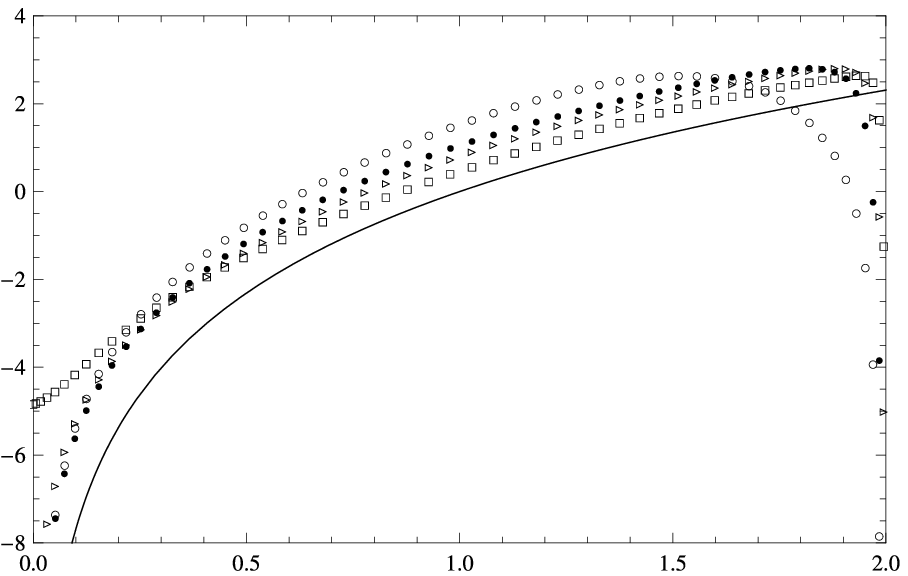}}
{${\scriptstyle\hspace{0.75cm} x_2/h}$}{-1mm}{\begin{rotate}{90}
$\hspace{-0.75cm}{\scriptstyle
(\bar{U}_1^-U_B)/u_\tau}$\end{rotate}}{1.5mm}
\end{minipage}
\hfill
\begin{minipage}[c]{.48\linewidth}
\FigureXYLabel{\includegraphics[width=.91\textwidth]{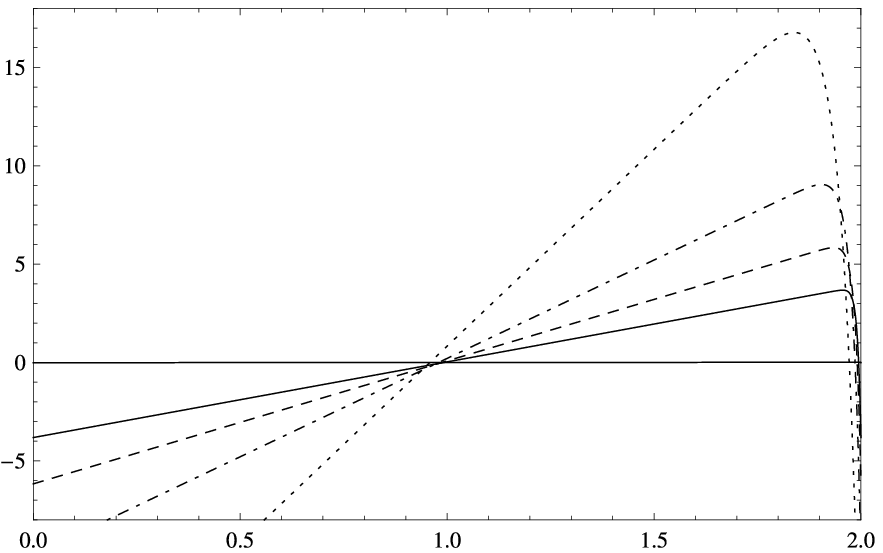}}
{${\scriptstyle\hspace{0.75cm} x_2/h}$}{-1mm}{\begin{rotate}{90}
$\hspace{-0.75cm}{\scriptstyle
(U_1^L-U_B)/u^L_\tau}$\end{rotate}}{1.5mm}
\end{minipage}
\caption{{\it Left plot:} Reproduction of Fig.$\,$9 $(a)$ in
\cite{Oberlack14} with the data provided by the authors on their
institutional
website~\href{http://www.fdy.tu-darmstadt.de/forschung_16/direkte_numerische_simulation/direkte_numerische_simulation.de.jsp}{[fdy]}.
For constant Reynolds number $Re_\tau=480$, the mean velocity
profile in deficit form is displayed for different transpiration
rates: $v_0^+=0.05$ ($\circ$);$\:$ $v_0^+=0.10$~($\bullet$);$\:$
$v_0^+=0.16$ ($\,\triangleright$);$\:$ $v_0^+=0.26$
(${\scriptstyle \square}$). For all cases the mean bulk velocity
$U_B$ \eqref{160520:1541} takes the universal value $U_B=U_B^*$,
where $U_B^*=0.8987$; in contrast to the values for $u_\tau$,
which are are not universal (see Table~\ref{tab1} for the
corresponding values). The solid line displays the new
(theoretically predicted) logarithmic scaling law
\eqref{160519:1744} for the parameters $\gamma=0.3$, $B_1=0$ and
$C_1=U_B$ as proposed in Sec.$\,$4 in \cite{Oberlack14}.
Obviously, a comparison to Fig.$\,$9 $(a)$ readily reveals that
this figure is not reproducible. It shows a strong discrepancy in
two independent aspects with the effect that an overall opposite
conclusion is obtained than as proposed in
\cite{Oberlack14}. For more details, see the main text.\\
{\it Right plot:} To have a qualitative comparison to the
turbulent case, we plotted the corresponding laminar profiles for
the same external parameters as were used in the figure on the
left-hand side. From top to bottom (relative to the positive
function values), the corresponding laminar profile structure is
displayed for increasing transpiration rates
$v_0^+=0.05,0.1,0.16,0.26$ at fixed $Re_\tau=480$ and
$U_B=U_B^*=0.8987$. The associated values for the laminar friction
velocities $u^L_\tau$ can be taken from Table \ref{tab1}, which
are based on the closed analytical expression for the laminar
velocity profile $U_1^L$ \eqref{160524:1524}. See the main text
for a comparative discussion between the turbulent case (left
plot) and its corresponding laminar case~(right~plot).}
\label{fig2}
\vspace{1.25em}
\end{figure}

\begin{figure}[h!]
\centering
\begin{minipage}[c]{.48\linewidth}
\FigureXYLabel{\includegraphics[width=.91\textwidth]{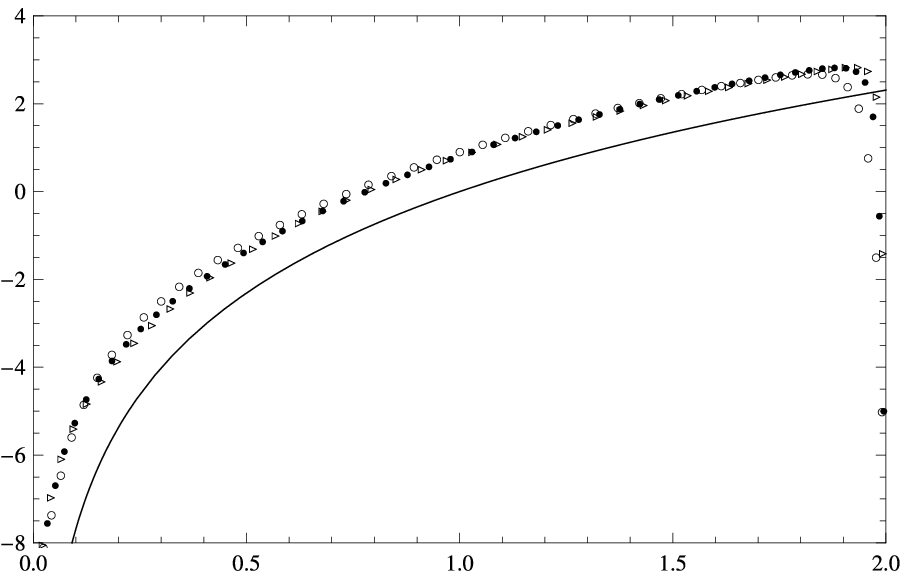}}
{${\scriptstyle\hspace{0.75cm} x_2/h}$}{-1mm}{\begin{rotate}{90}
$\hspace{-0.75cm}{\scriptstyle
(\bar{U}_1^-U_B)/u_\tau}$\end{rotate}}{1.5mm}
\end{minipage}
\hfill
\begin{minipage}[c]{.48\linewidth}
\FigureXYLabel{\includegraphics[width=.91\textwidth]{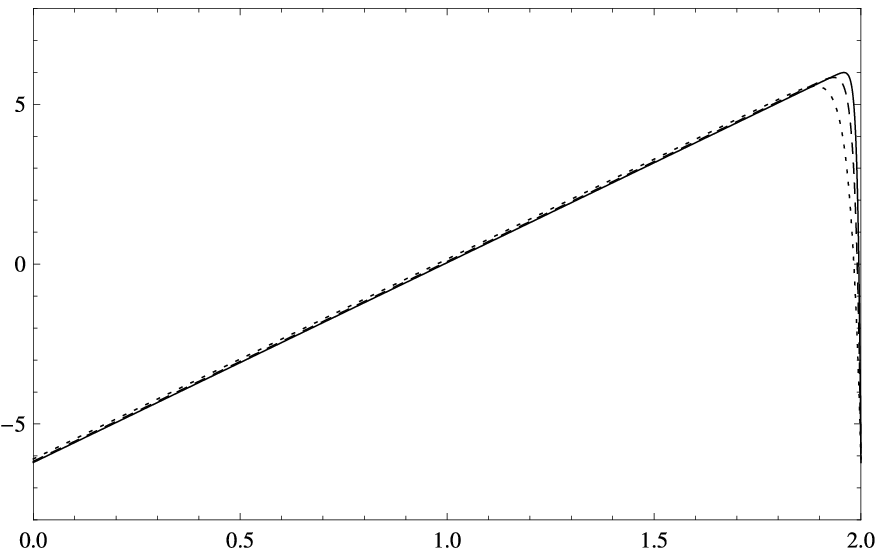}}
{${\scriptstyle\hspace{0.75cm} x_2/h}$}{-1mm}{\begin{rotate}{90}
$\hspace{-0.75cm}{\scriptstyle
(U_1^L-U_B)/u^L_\tau}$\end{rotate}}{1.5mm}
\end{minipage}
\caption{{\it Left plot:} Reproduction of Fig.$\,$9 $(c)$ in
\cite{Oberlack14} with the data provided by the authors on their
institutional
website~\href{http://www.fdy.tu-darmstadt.de/forschung_16/direkte_numerische_simulation/direkte_numerische_simulation.de.jsp}{[fdy]}.
In the same way as in Figure \ref{fig2}, the mean velocity profile
in deficit form is again displayed, but now at a constant
transpiration rate $v_0^+=0.16$ for different Reynolds numbers:
$Re_\tau=250$ ($\circ$);$\:$ $Re_\tau=480$ ($\bullet$);$\:$
$Re_\tau=850$ ($\,\triangleright$). The solid line displays again
the new (theoretically predicted) logarithmic scaling law
\eqref{160519:1744} for the corresponding parameters $\gamma=0.3$,
$B_1=0$ and
$C_1=U_B$, as proposed in Sec.$\,$4 in \cite{Oberlack14} also for this case.\\
{\it Right plot:} For the same motivation as in Figure \ref{fig2},
the corresponding laminar deficit profiles are plotted. From left
to right the Reynolds number increases $Re_\tau=250,480,850$ at
fixed transpiration rate $v_0^+=0.16$ and bulk velocity
$U_B=U_B^*=0.8987$. The laminar velocity profile $U_1^L$ and its
associated consistent friction velocity $u_\tau^L$ are given
through the analytical expressions of \eqref{160524:1524}; the
explicit values of $u^L_\tau$ for the considered parameter
combinations are given again in Table \ref{tab1}. For a
comparative discussion between the turbulent case (left plot) and
its corresponding laminar case (right plot), see again the main
text.}\label{fig3}\vphantom{x}\hrule
\end{figure}

Hence, the result of Figure \ref{fig1} allows us to make the
conclusion that both Figs.$\,$9 $(a)$ and $(c)$ in
\cite{Oberlack14} are {\it not} reproducible, when considering our
reproduction in Figure \ref{fig2} and Figure \ref{fig3},
respectively.


\pagebreak[4]
\newgeometry{left=2.5cm,right=2.5cm,top=2.5cm,bottom=2.0cm,headsep=1em}

\subsection{The nonreproducibility of Figure$\,$9 $(a)$}

Comparing our left plot in Figure \ref{fig2} with the
corresponding Fig.$\,$9 $(a)$ in \cite{Oberlack14}, readily
reveals that this figure is not reproducible. It shows a strong
discrepancy in two independent aspects with the effect that an
overall opposite conclusion is obtained than as proposed in
\cite{Oberlack14}: (i) The DNS data for the mean streamwise
velocity at a fixed Reynolds number and varying transpiration
rates, do {\it not} universally collapse onto one single curve
when formulated in its deficit form. Instead we see a monotonous
decay of the profile as the transpiration rate increases. (ii) The
theoretically predicted scaling law (solid line) does not match
the data, not even in a rough approximate sense. For that
different matching parameters need to be formulated. If $B_1$ is
continued to be chosen as zero, then both $\gamma$ and $C_1$ need
to be functions of $v_0^+$, where it should be noted that for
higher transpiration rates the matching region shifts to the
suction wall $x_2/h=2$.

The right plot in Figure \ref{fig2} serves as a comparative
reference to the left one. It allows to compare the differences
and similarities between the laminar and the turbulent flow
behavior. For the same external parameters as were used for the
turbulent case, this plot shows the corresponding laminar profiles
derived in analytically closed form in Appendix \ref{B}, with the
final result given in \eqref{160524:1524}. Interesting to see is
how the deficit profile at a constant finite Reynolds number
decays for increasing transpiration rates until it globally goes
to zero when reaching the limit~$v_0^+\rightarrow \infty$ (since
in this limit $u^L_\tau\rightarrow\infty$ and $|U_1^L|< 2U_B^*$,
in particular $u_\tau\to U_B^*v_0^+$ and thus $U_1^L\rightarrow
U_B^*\cdot x_2/h$, for $0\leq x_2/h<2$). A similar behavior,
although based on a more complex functional structure, is also to
be expected for the turbulent case:\footnote[2]{Note that global
laminar flow properties are most probably also statistically
featured by the corresponding turbulent flow condition and thus
also to be expected in a qualitative sense. The opposite
conclusion, however, is of course not true: A turbulent flow may
statistically show additional features that are not existent in
its associated laminar base flow.} Indeed, in the left plot the
onset of this global tendency in the DNS data can already be
positively observed.

\phantom{x}\vspace{-1.7em}
\subsection{The nonreproducibility of Figure$\,$9 $(c)$}

Comparing now the left plot of Figure \ref{fig3} with the
corresponding Fig.$\,$9 $(c)$ in \cite{Oberlack14}, we see that
although the DNS data in this case more or less universally
collapses onto a single curve and also coincides with the
representation and conclusion in \cite{Oberlack14}, the new
logarithmic scaling law \eqref{160519:1744} (solid line), however,
still does not match the data for the proposed parameters
$\gamma=0.3$, $B_1=0$ and $C_1=U_B$. For an unaltered $\gamma$, a
vertical upward shift of at least 0.82 units is needed, i.e., in
order to match the data, the integration constant $C_1$ needs to
be modified from $C_1=U_B$ at least to $C_1=U_B+0.82\cdot u_\tau$,
a result not obtained in \cite{Oberlack14}. Hence, we may
correctly claim that also Fig.$\,$9 $(c)$ is not reproducible.

As in Figure \ref{fig2}, the corresponding laminar deficit
profiles are presented and studied in the right plot of Figure
\ref{fig3}. Interesting to see here is that by close inspection
the deficit profiles at a constant transpiration rate do not
really collapse onto a single curve, but rather, within the range
$0\leq x_2/h<2$, slowly converge to a particular linear profile in
the limit of infinite Reynolds number $Re_\tau\to\infty$. Note
that this convergence takes place pointwise, i.e., the points
close to the blowing wall ($x_2/h=0$) converge exponentially
faster than those points close to the suction wall ($x_2/h=2$),
due to the presence of a boundary layer at this~side. In
particular, the deficit profile converges to
$(U_1^L-U_B^*)/u^L_\tau\to 1/v_0^+\cdot x_2/h-1/v_0^+$, for $0\leq
x_2/h<2$, i.e., equivalently as in the previous section for a
fixed Reynolds number, the laminar velocity profile in this range
converges again to $U_1^L\to U_B^*\cdot x_2/h$, since in this case
for a fixed transpiration rate $u^L_\tau\to U_B^*\cdot v_0^+$; see
\eqref{160524:1524}. A similar behavior, although based on a more
complex functional structure, is also to be expected for the
turbulent case: Indeed, by close inspection of the left plot of
Figure \ref{fig3}, one can observe that everywhere throughout the
channel, the DNS data does not universally lie on a single curve
as claimed in \cite{Oberlack14}, but that in fact for increasing
Reynolds number a (slow) pointwise convergence towards a
particular profile takes~place. \pagebreak[4] \restoregeometry

\section{On the inconsistency of the Lie-group-based scaling theory in turbulence\label{S5}}

In this section we will reveal the fact that when coherently
extending the Lie-group-based scaling theory as presented in
\cite{Oberlack14} for the newly proposed logarithmic law
[Eq.$\,$(3.16)] to higher orders of the one-point velocity
correlations, one unavoidably runs into a fundamental
inconsistency in that one fails to match certain theoretically
derived scaling laws to the given DNS data. We will investigate
both the inviscid (Euler, $\nu=0$) case, as particularly realized
in \cite{Oberlack14}, as well as the viscous (Navier-Stokes,
$\nu\neq 0$) case, as subsequently modified in
\cite{Oberlack15Rev}. To simplify formal expressions and
calculations, we will derive all theoretical results in the
full-field (instantaneous) representation. The corresponding
Reynolds decomposed results (later needed to directly compare to
the DNS data) are then obtained straightforwardly by just
performing the decomposition \eqref{160501:1208}. To demonstrate
our point in this section, it is fully sufficient to only consider
the turbulent transport equations up to second order, since they
already involve (unclosed) third order moments for which the
inconsistency to the DNS data is clearly pronounced.

\subsection{The inviscid case ($\nu=0$)\label{S51}}

The governing one-point equations for the flow considered in
\cite{Oberlack14} are given by the system
\eqref{160430:1912}-\eqref{160430:1914}, which, as discussed in
Section \ref{S2}, admits the continuous set of Lie-point
symmetries \eqref{160425:1017}-\eqref{160424:1452} and
\eqref{150501:1342}, where the latter symmetry is needed to
appropriately extend the construction of invariant solutions in
\cite{Oberlack14} to higher-order moments. Hence, when combining
all symmetries and following the line of reasoning in
\cite{Oberlack14}, we obtain the following invariant surface
condition
\begin{align}
\frac{dx_1}{k_1x_1+k_{x_1}}=\frac{dx_2}{k_1x_2+k_{x_2}}&=\frac{d\overline{U_i}}{\big(k_1-k_2+k_s)
\overline{U_i}+\kappa_i+c_i}=\frac{d\overline{P}}{\big(2k_1-2k_2+k_s\big)\overline{P}+\kappa^p+d}
\nonumber\\[0.5em]
&=\frac{d\big(\partial_i\overline{P}\big)}{\big(k_1-2k_2+k_s\big)\partial_i\overline{P}+\kappa^p_i}=
\frac{d\overline{U_iU_j}}{\big(2k_1-2k_2+k_s\big)\overline{U_iU_j}+\kappa_{ij}+c_{ij}}
\quad\;\nonumber\\[0.5em]
&=\frac{d\overline{U_iU_jU_k}}{\big(3k_1-3k_2+k_s\big)\overline{U_iU_jU_k}+\kappa_{ijk}+
c_{ijk}}\nonumber\\[0.5em]
&=\frac{d\overline{U_i\partial_j
P}}{\big(2k_1-3k_2+k_s\big)\overline{U_i\partial_j
P}+\kappa^p_{ij}},\; \label{160425:2207}
\end{align}
which coherently extents their corresponding condition
[Eq.$\,$(3.12)] up to third order including the pressure moments.
The functional $\kappa$-extensions result from the single
translation symmetry $\bar{T}_{\bar{U}_1}$ \eqref{160430:2131}
when written in its equivalent full-field form
\eqref{160424:1452}, and are thus given as\footnote[2]{Note that
the quadratic term $k^2_{\bar{U}_1}\delta_{1i}\delta_{1j}$ in the
transformation $\bar{T}_{\bar{U}_1}$ \eqref{160424:1452} for
$\overline{U_iU_j}$ is not contributing in its local
(infinitesimal) generator, since Lie-group symmetry theory is a
linear theory where all information of the transformations is
carried in the linear expansion terms of the group parameters.}
\begin{gather}
\left.
\begin{aligned} &\kappa_i=k_{\bar{U}_1}\delta_{1i},\qquad
\kappa_{ij}=\kappa_i\overline{U_j}+\kappa_j\overline{U_i},\\[0.5em]
&\kappa_{ijk}= \kappa_{ij}\overline{U_k}
+\kappa_{ik}\overline{U_j}+\kappa_{jk}\overline{U_i}\\
&\hspace{2.07cm}+\kappa_i\left(\overline{U_jU_k}-2\,\overline{U_j}\:\overline{U_k}\right)+
\kappa_j\left(\overline{U_iU_k}-2\,\overline{U_i}\:\overline{U_k}\right)+
\kappa_k\left(\overline{U_iU_j}-2\,\overline{U_i}\:\overline{U_j}\right),\\
&\kappa^p=0,\qquad\kappa^p_i=0,\qquad\kappa_{ij}^p=\kappa_i\frac{\partial\overline{P}}{\partial
x_j}.
\end{aligned}
~~~\right\}
\end{gather}
Progressively, we will now determine all invariant functions from
\eqref{160425:2207} and examine in how far they are compatible to
the underlying equations \eqref{160430:1912}-\eqref{160430:1914}.
The first step in this procedure is to ensure the invariance of
any existing constraints. We recall the two constraints of a mean
constant wall-normal velocity $\overline{U_2}=v_0$, or,
equivalently $d\overline{U_2}=0$, and that of a mean constant
streamwise pressure gradient $\partial\overline{P}/\partial
x_1=-K$, or, equivalently $d(\partial_1\overline{P})=0$.
Implementing the first constraint $d\overline{U_2}=0$ into
\eqref{160425:2207} will consequently result into the
corresponding combined symmetry breaking
constraint\footnote[2]{Note that both constraints in
\eqref{160425:2327} are necessary to avoid an overall zero surface
condition \eqref{160425:2207}.}
\begin{equation}
k_1-k_2+k_s=0,\;\;\text{and}\;\;\, c_2=0,\label{160425:2327}
\end{equation}
as discussed and implemented in \cite{Oberlack14}. The second
constraint $d(\partial_1\overline{P})=0$, however, will result
into an additional symmetry breaking constraint
\begin{equation}
k_1-2k_2+k_s=0,
\end{equation}
which was {\it not} discussed in \cite{Oberlack14}; an important
result indeed, since, due to \eqref{160425:2327}, it equivalently
turns into the strong constraint
\begin{equation}
k_2=0.\label{160426:0000}
\end{equation}
Note that this result could have also been obtained when directly
solving from condition \eqref{160425:2207} the invariant function
for the pressure $\overline{P}$ as a function of $x_1$ and~$x_2$
along with the first constraint of \eqref{160425:2327}. Because,
this result, when taking its gradient in the streamwise direction,
\begin{equation}
\frac{\partial\overline{P}(x_1,x_2)}{\partial
x_1}=F_1\!\left(\frac{x_1+k_{x_1}/k_1}{x_2+k_{x_2}/k_1}\right)\cdot
\big(x_2+k_{x_2}/k_1\big)^{-k_2/k_1},
\end{equation}
obviously, is only compatible to
\begin{equation}
\frac{\partial\overline{P}(x_1,x_2)}{\partial
x_1}=-K,\label{160425:2310}
\end{equation}
if the integration function $F_1$ is a global constant equal to
$-K$, and,~if $k_2=0$. Collecting now all obtained symmetry
breaking constraints
\begin{equation}
k_2=0, \qquad k_s=-k_1, \qquad c_2=0,\label{160525:0038}
\end{equation}
and applying them to the originally formulated condition
\eqref{160425:2207}, will drastically restrict the possible
structures for the considered invariant functions. For example,
for the mean streamwise velocity profile $\bar{U}_1$ only a
logarithmic function of the form
\begin{equation}
\overline{U_1}(x_2)=A_1\ln\!\left(\frac{x_2}{h}+B_1\right)+C_1,\label{160426:0933}
\end{equation}
is possible, where $C_1$ is an arbitrary integration constant and
$A_1=(k_{\bar{U}_1}+c_1)/k_1$, $B_1=k_{x_2}/(k_1h)$ two
independent parameters uniquely determined by internal group
constants. But, not only the symmetry breaking constraints, also
the underlying dynamical equations of the considered system
restrict the functions (for ODEs) or the functional possibilities
(for PDEs) even further, for example, when considering the full
derivation of the invariant correlation $\overline{U_1U_2}$: From
the defining condition \eqref{160425:2207} with inserted
constraints \eqref{160525:0038} it is initially given~by
\begin{equation}
\overline{U_1U_2}(x_2)=C_{12}\left(\frac{x_2}{h}+B_1\right)
+\tilde{A}_{12},
\end{equation}
where $C_{12}$ is again some arbitrary integration constant, and
where $\tilde{A}_{12}=-(k_{\bar{U}_1}v_0+c_{12})/k_1$, like
$B_1=k_{x_2}/(k_1h)$, is a parameter that, apart from the external
system parameter $v_0$, comprises again internal group
constants.\footnote[3]{That the parameter $\tilde{A}_{12}$
includes the external system parameter $v_0$, is denoted by the
``tilde" symbol. With the notation introduced in
\eqref{160608:1300}, this parameter can also be written as
$\tilde{A}_{12}=A_{12}+A_0v_0$, where $A_{12}=-c_{12}/k_1$ and
$A_0=-k_{\bar{U}_1}/k_1$ are then two parameters determined by
internal group parameters only. This notation will also be used
later in \eqref{160502:1202} when matching the invariant functions
to DNS data, where the ``tilde" symbol denotes essential fitting
parameters collecting all constants, independent of their nature,
into a single expression.} However, the arbitrariness of $C_{12}$
is illusive, because the underlying momentum equation
\eqref{160430:1913} restricts it to
\begin{equation}
C_{12}=K\cdot h.\label{160502:1155}
\end{equation}
All remaining invariant functions can then be generically
determined as
\begin{equation}
\left.
\begin{aligned}
&\!\!\!\overline{U_iU_j}(x_2)=C_{ij}\left(\frac{x_2}{h}+B_1\right)
+A_0\,\delta_{ijk}\overline{U_k}+A_{ij},\qquad\,
\overline{U_1U_3}(x_2)=\overline{U_2U_3}(x_2)=0,\\[0.5em]
&\!\!\!\overline{U_iU_jU_k}(x_2)=C_{ijk}\left(\frac{x_2}{h}+B_1\right)^2
+\frac{A_0}{2}\delta_{ijkmn}\left[\,
C_{mn}\left(\frac{x_2}{h}+B_1\right)+ \overline{U_m
U_n}(x_2)\,\right]+
A_{ijk},\\[0.5em]
&\!\!\!\partial_2\overline{P}(x_1,x_2)=C^p,\qquad\,\overline{U_i\partial_j
P}(x_2)=C^p_{ij}\left(\frac{x_2}{h}+B_1\right)-A_0\Big(\delta_{1i}\delta_{1j}K-
\delta_{1i}\delta_{2j}C^p\Big),\label{160608:1300}
\end{aligned}
~~\right\}
\end{equation}
where the $C$-parameters are arbitrary integration constants,
while the $A$-parameters are determined through the group
constants as
\begin{equation}
\left.
\begin{aligned}
&A_0=-\frac{k_{\bar{U}_1}}{k_1},\quad
A_1=\frac{k_{\bar{U}_1}+c_1}{k_1},\quad
A_{ij}=2A_0A_1\delta_{1i}\delta_{1j}-\frac{c_{ij}}{k_1},\\[0.5em]
&A_{ijk}=\frac{3}{2}A^2_0A_1\delta_{1i}\delta_{1j}\delta_{1k}
-\frac{c_{ijk}}{2k_1},
\end{aligned}
~~~ \right\}
\end{equation}
and, finally, the modified $\delta$-functions in
\eqref{160608:1300} are defined as
\begin{equation}
\left.
\begin{aligned}
&\delta_{ijk}=\delta_{1i}\delta_{jk}+\delta_{1j}\delta_{ik},\\[0.5em]
&\delta_{ijkmn}=
\frac{\delta_{1i}}{2}\Big(\delta_{jm}\delta_{kn}+\delta_{jn}\delta_{km}\Big)+
\frac{\delta_{1j}}{2}\Big(\delta_{im}\delta_{kn}+\delta_{in}\delta_{km}\Big)+
\frac{\delta_{1k}}{2}\Big(\delta_{im}\delta_{jn}+\delta_{in}\delta_{jm}\Big).
\end{aligned}
~~~ \right\}
\end{equation}
The arbitrary integration constants, however, are not fully
independent but show certain fixed interrelations resulting from
the underlying transport equations
\eqref{160430:1912}-\eqref{160430:1914} that the invariant
functions need to satisfy, e.g., $C^p=-C_{22}/h$, or
$2C_{122}/h+C^p_{12}+C^p_{21}=0$. When Reynolds decomposing all
derived invariant results, we straightforwardly
obtain\footnote[2]{The four ``tilde"-parameters are given as:
$\tilde{A}_{12}=v_0A_0+A_{12}$, $\tilde{A}_{22}=-v_0^2+A_{22}$,
$\tilde{A}_{222}=-v_0^3+A_{222}$, and
$\tilde{A}_{112}=A_0\tilde{A}_{12}+A_{112}$.}
\begin{gather}
\left.
\begin{aligned}
&\!\!\!\bar{U}_1=A_1\ln\!\left(\frac{x_2}{h}+B_1\right)+C_1,\qquad
\bar{U}_2=v_0,\qquad \tau_{13}=\tau_{23}=0,\\[0.5em]
&\!\!\!
\tau_{11}=C_{11}\left(\frac{x_2}{h}+B_1\right)-\bar{U}_1^2+2A_0\bar{U}_1
+A_{11},\quad\;\:\,
\tau_{12}=u_\tau^2\left(\frac{x_2}{h}+B_1\right)-v_0\bar{U}_1
+\tilde{A}_{12},\\[0.5em]
&\!\!\!\tau_{22}=C_{22}\left(\frac{x_2}{h}+B_1\right)+\tilde{A}_{22},\qquad
\tau_{33}=C_{33}\left(\frac{x_2}{h}+B_1\right)+A_{33},\\[0.5em]
&\!\!\!\tau_{112}=C_{112}\left(\frac{x_2}{h}+B_1\right)^2
+2A_0\,u_\tau^2\left(\frac{x_2}{h}+B_1\right)-2\bar{U}_1\tau_{12}
-v_0\tau_{11}-v_0\bar{U}^2_1+\tilde{A}_{112},\\[0.5em]
&\!\!\!\tau_{222}=C_{222}\left(\frac{x_2}{h}+B_1\right)^2-
3v_0\tau_{22}+\tilde{A}_{222},\quad\;\:\,
\tau_{233}=C_{233}\left(\frac{x_2}{h}+B_1\right)^2
-v_0\tau_{33}+A_{233},
\end{aligned}
~\right\}\label{160502:1202}
\end{gather}
which then can be validated against the given DNS data. Note that
we only listed those functions for which the statistical data has
been made available from the DNS in \cite{Oberlack14}. For the
scaling factor \eqref{160502:1155} we used the central definition
$u_\tau^2=K\cdot h$ (see [Eq.$\,$(2.1)]). When fitting the set of
functions \eqref{160502:1202} to the data, special attention has
to be paid to the invariant scaling laws for $\tau_{11}$,
$\tau_{12}$ and $\tau_{112}$, which all show a combination of an
algebraic and a logarithmic scaling, an awkward property, which
again only has its origin in the new statistical symmetries
$\bar{T}^\prime_s$ \eqref{160424:1418} and
$\bar{T}^\prime_c$~\eqref{150501:1342} first proposed in
\cite{Oberlack10}.

\begin{figure}[t]\hspace*{0.225cm}
\centering
\begin{minipage}[r]{.3\linewidth}
\FigureXYLabel{\includegraphics[width=.91\textwidth]{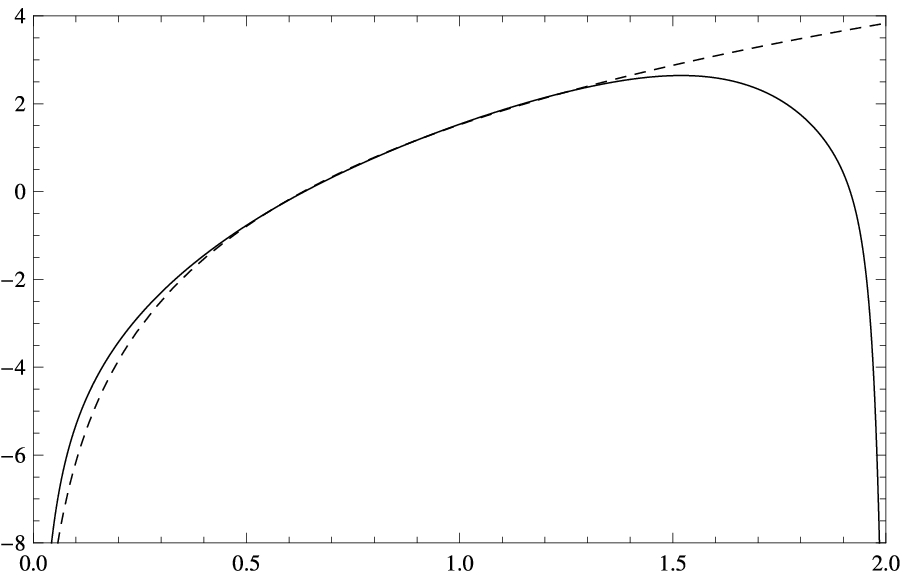}}
{${\scriptscriptstyle\hspace{0.75cm}
x_2/h}$}{-2.5mm}{\begin{rotate}{90}
$\hspace{-0.75cm}{\scriptscriptstyle (\bar{U}_1^-U_B)/u_\tau}$
\end{rotate}}{1.5mm}
\end{minipage}
\hfill
\begin{minipage}[c]{.3\linewidth}
\FigureXYLabel{\includegraphics[width=.91\textwidth]{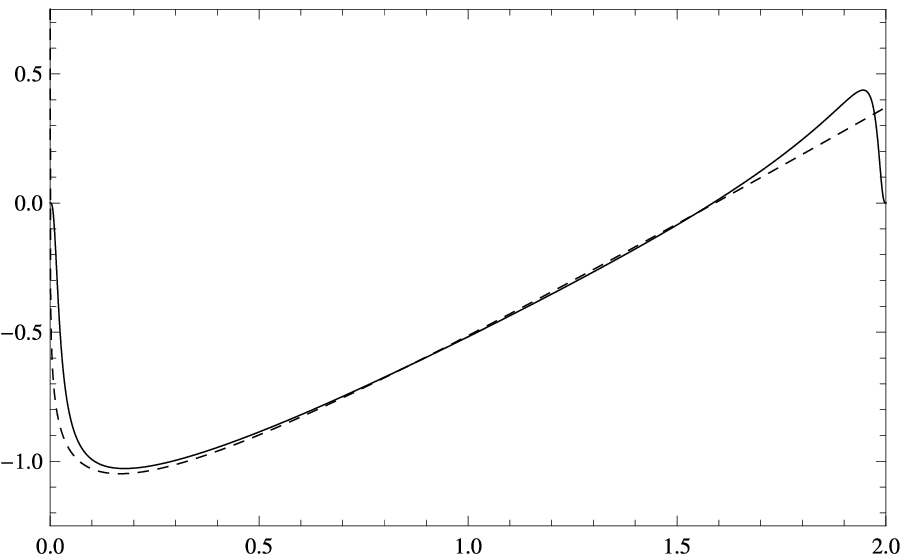}}
{${\scriptscriptstyle\hspace{0.75cm}
x_2/h}$}{-2.5mm}{\begin{rotate}{0}
$\hspace{-0.3cm}{\scriptscriptstyle
\tau_{12}^+}$\end{rotate}}{0mm}
\end{minipage}
\hfill
\begin{minipage}[c]{.3\linewidth}
\FigureXYLabel{\includegraphics[width=.91\textwidth]{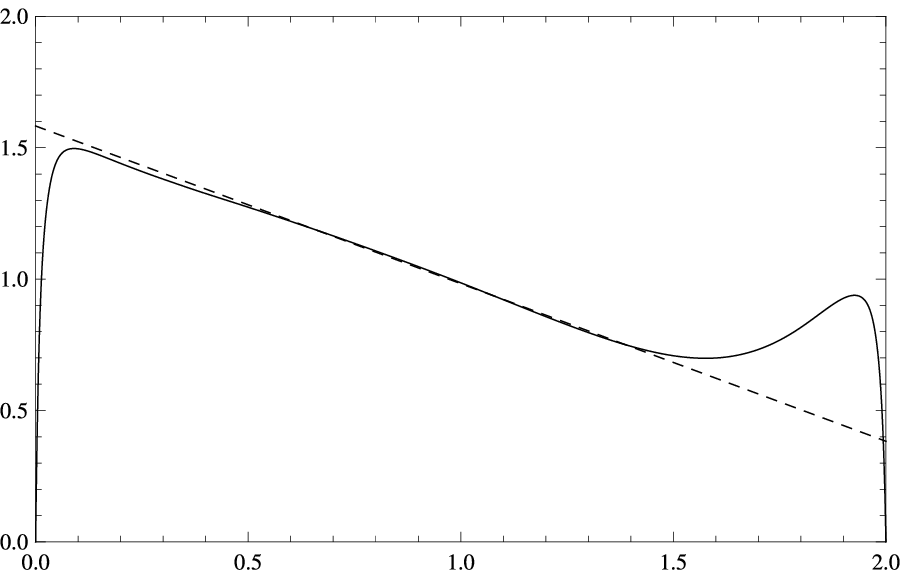}}
{${\scriptscriptstyle\hspace{0.75cm}
x_2/h}$}{-2.5mm}{\begin{rotate}{0}
$\hspace{-0.3cm}{\scriptscriptstyle\tau_{33}^+}$\end{rotate}}{0mm}
\end{minipage}
\hfill\hspace*{0.225cm}
\begin{minipage}[r]{.3\linewidth}
\FigureXYLabel{\includegraphics[width=.91\textwidth]{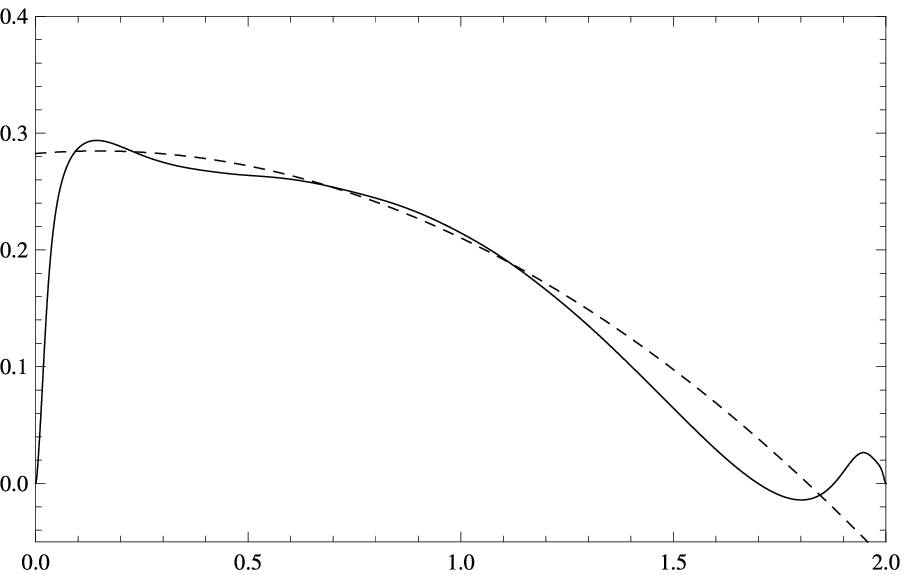}}
{$\phantom{.}$}{-2.5mm}{\begin{rotate}{0}
$\hspace{-0.5cm}{\scriptscriptstyle
\tau_{233}^+}$\end{rotate}}{0mm}
\end{minipage}
\hfill
\begin{minipage}[c]{.3\linewidth}
\FigureXYLabel{\includegraphics[width=.91\textwidth]{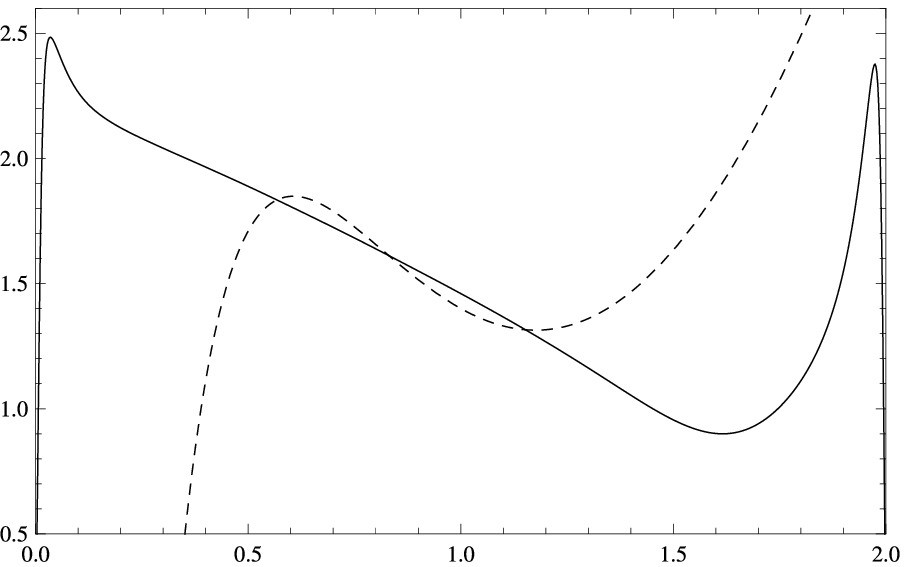}}
{$\phantom{.}$}{-2.5mm}{\begin{rotate}{0}
$\hspace{-0.3cm}{\scriptscriptstyle
\tau_{11}^+}$\end{rotate}}{0mm}
\end{minipage}
\hfill
\begin{minipage}[c]{.3\linewidth}
\FigureXYLabel{\includegraphics[width=.91\textwidth]{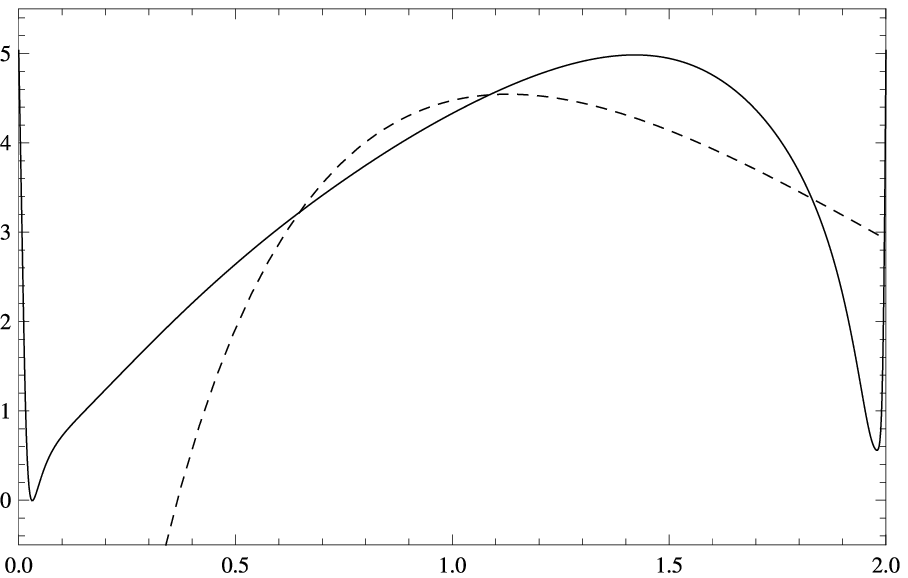}}
{$\phantom{.}$}{-2.5mm}{\begin{rotate}{0}
$\hspace{-0.5cm}{\scriptscriptstyle
\tau_{112}^+}$\end{rotate}}{0mm}
\end{minipage}
\caption{Matching of the theoretically predicted scaling laws
\eqref{160502:1202} to the DNS data for $Re_\tau=480$ and
$v_0^+=0.05$. The DNS data is displayed by solid lines, the
corresponding scaling laws by dashed lines. The associated
parameters for this best fit in each case can be taken from Table
\ref{tab2}, where the matching region was chosen in the range
$0.50\leq x_2/h\leq 1.25$. For more details and a discussion on
the fitting results obtained, see the main text.}\label{fig4}
\end{figure}
\begin{figure}[!h]\hspace*{0.225cm}
\centering
\begin{minipage}[r]{.3\linewidth}
\FigureXYLabel{\includegraphics[width=.91\textwidth]{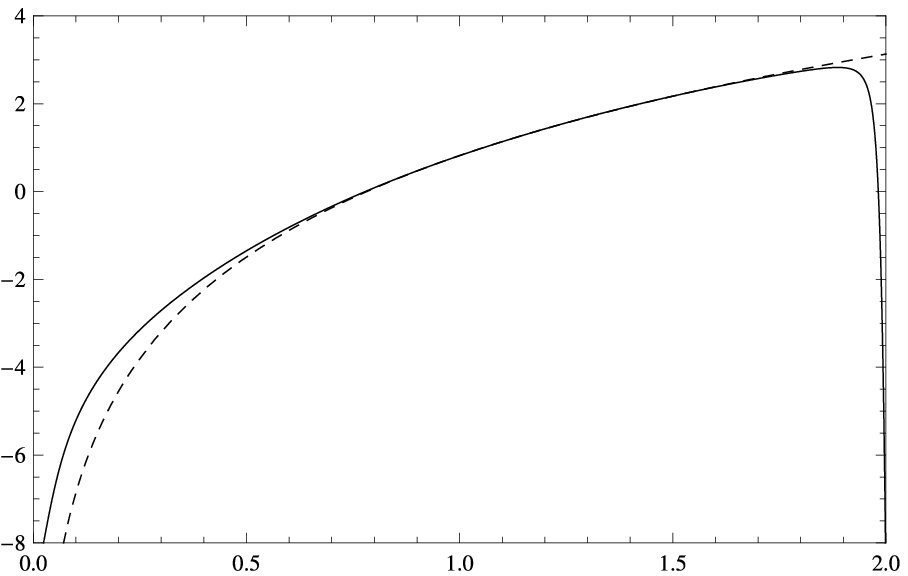}}
{${\scriptscriptstyle\hspace{0.75cm}
x_2/h}$}{-2.5mm}{\begin{rotate}{90}
$\hspace{-0.75cm}{\scriptscriptstyle (\bar{U}_1^-U_B)/u_\tau}$
\end{rotate}}{1.5mm}
\end{minipage}
\hfill
\begin{minipage}[c]{.3\linewidth}
\FigureXYLabel{\includegraphics[width=.91\textwidth]{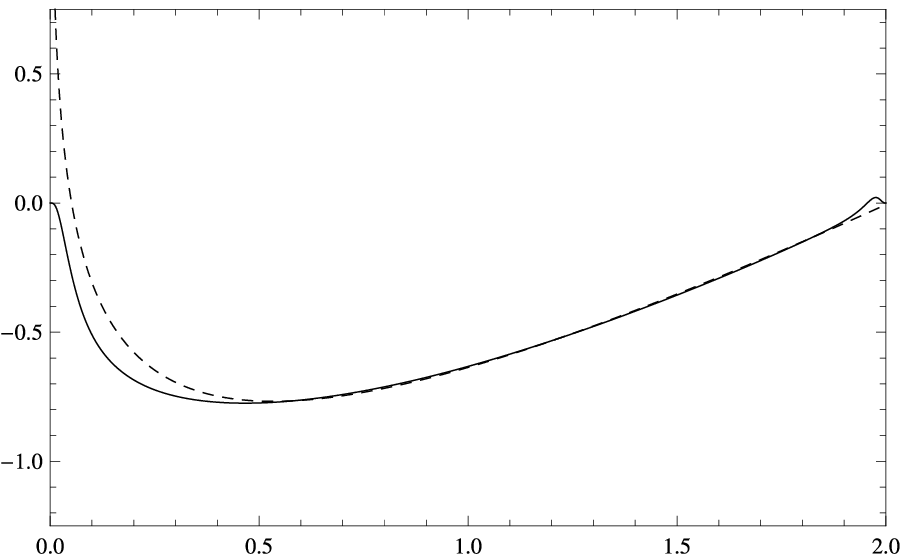}}
{${\scriptscriptstyle\hspace{0.75cm}
x_2/h}$}{-2.5mm}{\begin{rotate}{0}
$\hspace{-0.3cm}{\scriptscriptstyle
\tau_{12}^+}$\end{rotate}}{0mm}
\end{minipage}
\hfill
\begin{minipage}[c]{.3\linewidth}
\FigureXYLabel{\includegraphics[width=.91\textwidth]{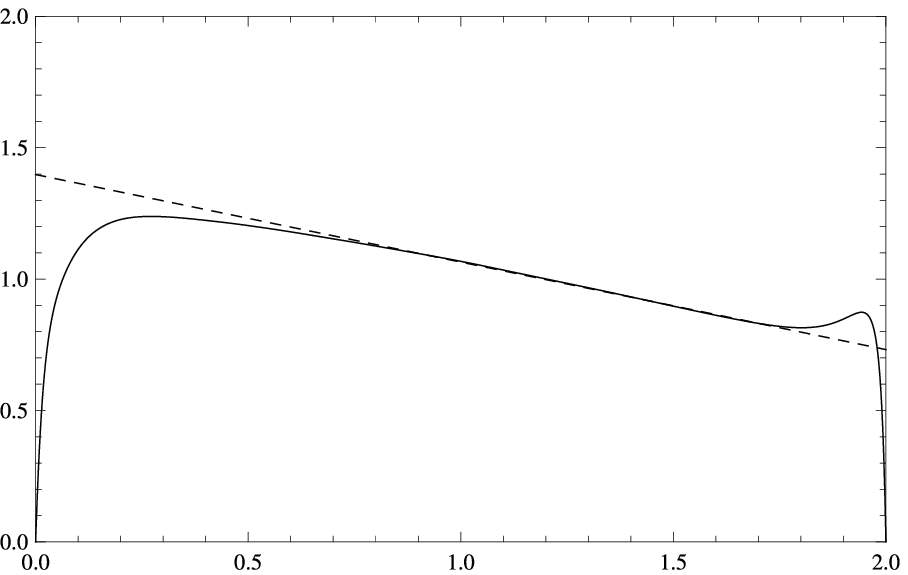}}
{${\scriptscriptstyle\hspace{0.75cm}
x_2/h}$}{-2.5mm}{\begin{rotate}{0}
$\hspace{-0.3cm}{\scriptscriptstyle\tau_{33}^+}$\end{rotate}}{0mm}
\end{minipage}
\hfill\hspace*{0.225cm}
\begin{minipage}[r]{.3\linewidth}
\FigureXYLabel{\includegraphics[width=.91\textwidth]{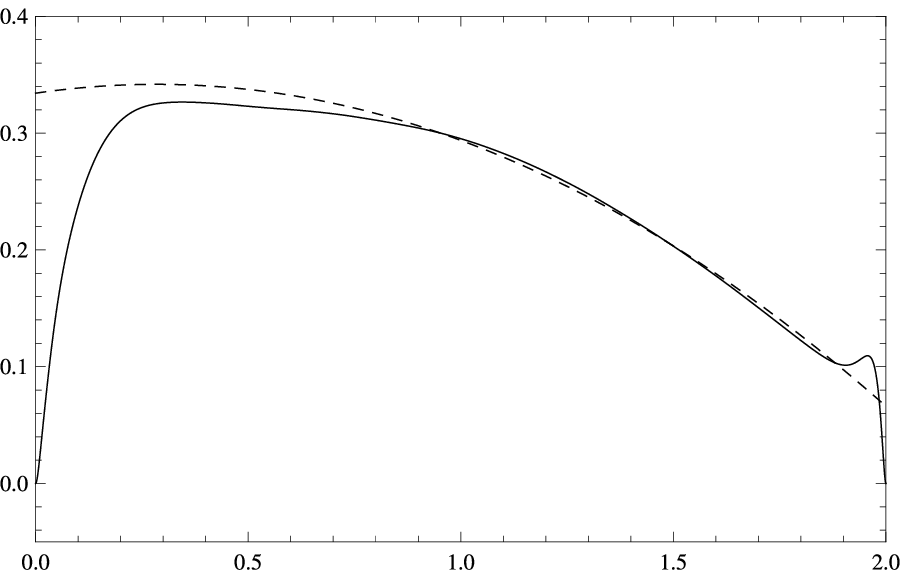}}
{$\phantom{.}$}{-2.5mm}{\begin{rotate}{0}
$\hspace{-0.5cm}{\scriptscriptstyle
\tau_{233}^+}$\end{rotate}}{0mm}
\end{minipage}
\hfill
\begin{minipage}[c]{.3\linewidth}
\FigureXYLabel{\includegraphics[width=.91\textwidth]{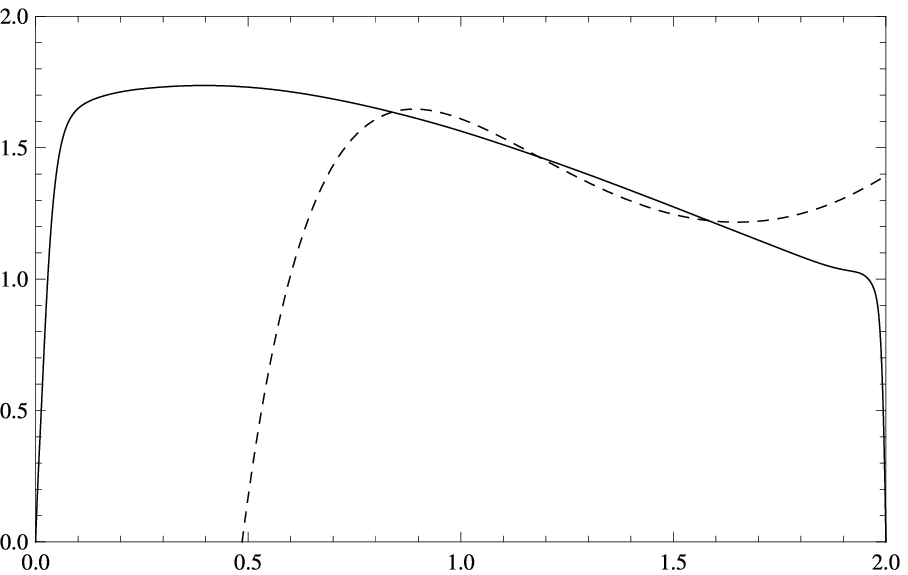}}
{$\phantom{.}$}{-2.5mm}{\begin{rotate}{0}
$\hspace{-0.3cm}{\scriptscriptstyle
\tau_{11}^+}$\end{rotate}}{0mm}
\end{minipage}
\hfill
\begin{minipage}[c]{.3\linewidth}
\FigureXYLabel{\includegraphics[width=.91\textwidth]{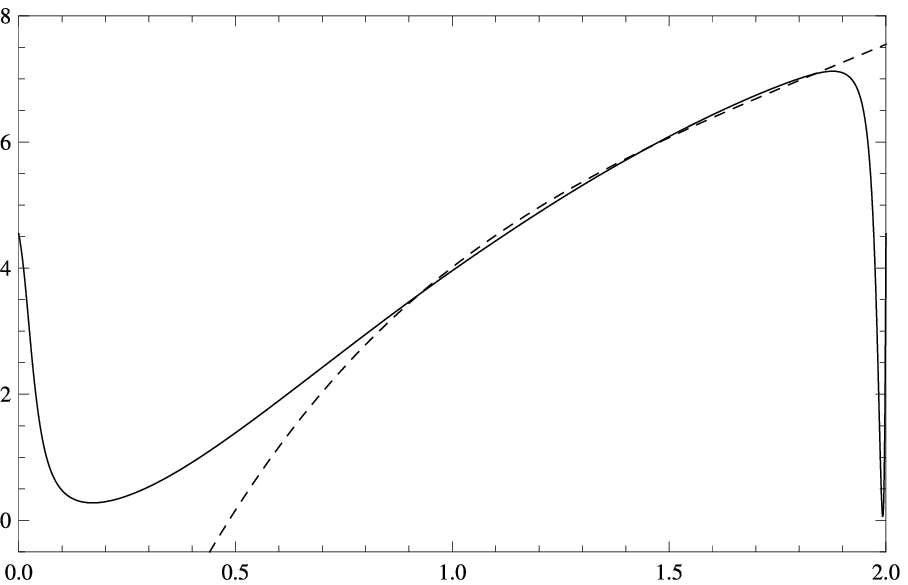}}
{$\phantom{.}$}{-2.5mm}{\begin{rotate}{0}
$\hspace{-0.5cm}{\scriptscriptstyle
\tau_{112}^+}$\end{rotate}}{0mm}
\end{minipage}
\caption{Matching of the theoretically predicted scaling laws
\eqref{160502:1202} to the DNS data for the same Reynolds number
$Re_\tau=480$ as in the figure given above, but for a higher
transpiration $v_0^+=0.16$. The DNS data is again displayed by
solid lines, the corresponding scaling laws by dashed lines. The
associated best-fitted parameters can be taken again from Table
\ref{tab2}, where the matching region was now set differently in
the range $0.75\leq x_2/h\leq 1.70$. For more details and a
comparative discussion on the fitting results obtained between the
lower (Figure \ref{fig4}) and the higher transpiration case
(Figure \ref{fig5}), see again the
main~text.}\label{fig5}\vphantom{x}\hrule
\end{figure}

\pagebreak[4]\phantom{x}\vspace{-1.6em} Figure \ref{fig4} \&
\ref{fig5} show the matching of the analytically (from ``first
principles") derived scaling laws \eqref{160502:1202} to the DNS
data at $Re_\tau=480$ for the two different transpiration rates
$v_0^+=0.05$ and $v_0^+=0.16$, respectively. The matching was
performed in the $u_\tau$-normalization, i.e., for the normalized
velocity correlations $\tau_{ij}^+=\tau_{ij}/u_\tau^2$ and
$\tau_{ijk}^+=\tau_{ijk}/u_\tau^3$, as well as for the normalized
mean velocity field in its deficit form
$(\bar{U}_1-U_B)/u_\tau=\bar{U}_1^+-U_B^+$. The particular values
for $u_\tau$ as well as for $U_B=U_B^*$ in each case can be taken
from Table \ref{tab1}.

The matching region in Figure \ref{fig4} was chosen in the range
$0.50\leq x_2/h\leq 1.25$, based on the best fit regarding the
central prediction in \cite{Oberlack14}, namely that of a new
logarithmic scaling law \eqref{160519:1744} for the mean velocity
field in the center of the channel
\begin{equation}
\frac{\bar{U}_1-U_B}{u_\tau}=\frac{1}{\gamma}\ln (x_2/h)+\lambda,
\label{160526:1806}
\end{equation}
with the fixed universal scaling coefficient $\gamma=0.3$
referring to \eqref{160519:1743}. Note that in \cite{Oberlack14}
no additional vertical shift $\lambda$ was needed, but which, as
we have demonstrated before, only leads to non-reproducible
results (see Section \ref{S4}, in particular the discussion on the
nonreproducibility of Fig.$\,$9 $(c)$ in \cite{Oberlack14}).
Depending on the transpiration rate and the Reynolds number, a
constant vertical upward shift $\lambda>0$ is necessary to match
the DNS data. Its presence, of course, re-defines the proposed
integration constant in \eqref{160519:1743} as
$C_1=U_B+\lambda\cdot u_\tau$, turning thus $C_1$ into a {\it
non}-universal constant, depending then on both the Reynolds
number $Re_\tau$ and the transpiration rate $v_0^+$, where the
latter dependency is more pronounced than the former one. A result
opposite to the one claimed in \cite{Oberlack14}, where $C_1$ was
determined as a universal constant, with the particular value
$C_1=U_B$~\eqref{160519:1743} for a universally fixed bulk
velocity $U_B=U_B^*$ as it is explicitly given in Table
\ref{tab1}.

Based on the matching region in Figure \ref{fig4}, the
corresponding region for the higher transpiration rate in Figure
\ref{fig5} was determined to lie in the range $0.75\leq x_2/h\leq
1.70$. This range was determined such that it has the same
absolute residual range $-0.015\leq
y_\text{s}-f_\text{m}(x_2/h)\leq 0.015$ as the one chosen for the
lower transpiration rate in Figure \ref{fig4} when fitting the
central scaling law \eqref{160526:1806}, where
$y_s=(\bar{U}_1-U_B)/u_\tau$ are the simulated (DNS) values and
$f_m(x_2/h)=\ln(x_2/h)/\gamma+\lambda$ the values from the
considered model function.\footnote[2]{Hence, by construction, the
quality of the fit for the mean velocity profile $\bar{U}_1$ in
Figure \ref{fig5} is thus the same as in Figure \ref{fig4}. As a
result, the mean velocity profile provided in each case the
necessary but {\it a priori} unknown matching region, which now
serves as a basis to systematically fit all remaining velocity
correlations to the DNS data. The best-fitted parameter values for
the correlations functions \eqref{160502:1202} are listed in Table
\ref{tab2}.} Such a procedure is necessary if one is interested in
how an initially chosen matching region changes when varying any
external system parameters. When comparing the matching region
$0.50\leq x_2/h\leq 1.25$ for $v_0^+=0.05$ in Figure~\ref{fig4},
with the corresponding region $0.75\leq x_2/h\leq 1.70$ for
$v_0^+=0.16$ in Figure \ref{fig5}, we clearly observe that as the
transpiration rate moderately increases at constant Reynolds
number (up to $v_0^+\leq 0.16$), the matching region not only
grows in extent, but that it also, at the same time, shifts to the
right towards the suction wall ($x_2/h\to 2$). An important result
which again has not been indicated in \cite{Oberlack14}. Instead,
only the first property of a growing validity region is reported,
which, however, cannot be true as a single statement for ever
increasing transpiration rates: In fact, since for higher rates
the validity region also shifts more and more to the fixed
right-hand boundary at the suction wall, it eventually has to
revert this growing trend at a certain transpiration rate high
enough. For example, the rate $v_0^+=0.26$ (at $Re_\tau=480$) is
already sufficient to demonstrate a {\it non}-increased validity
region when compared to all lower rates at the same Reynolds
number. Based on the same (residual) condition as for the
considered lower rates, the matching region for $v_0^+=0.26$
reduced to $1.35\leq x_2/h\leq 1.80$, where at the same time a
strong shift to the right (suction side) has occurred.

\begin{table}
\begin{center}
\begin{tabular}{c c c | c | c c | c c | c c | c c}\hline\\
\multicolumn{1}{c|}{$v_0^+$} &
\multicolumn{2}{|c|}{${\displaystyle\frac{\bar{U}_1-U_B}{u_\tau}}$}
& \multicolumn{1}{c|}{$\tau_{12}^+$} &
\multicolumn{2}{c|}{$\tau_{33}^+$} &
\multicolumn{2}{c|}{$\tau_{233}^+$} &
\multicolumn{2}{c|}{$\tau_{11}^+$} &
\multicolumn{2}{c}{$\tau_{112}^+$}\\[0.75em]\hline
\multicolumn{1}{c|}{} & & & & & & & & & & &\\[-0.75em]
\multicolumn{1}{c|}{} & $\gamma$ & $\lambda$ & $\tilde{A}_{12}^+$
& $A_{33}^+$ & $C_{33}^+$ & $A_{233}^+$ & $C_{233}^+$ & $A_{11}^+$
& $C_{11}^+$ &
$\tilde{A}_{112}^+$ & $C_{112}^+$\\[0.5em]
\multicolumn{1}{c|}{0.05} & 0.3 & 1.524 & -0.622 & 1.583 & -0.600
& 0.362 & -0.102 & -199.1 & 25.85
& -28.34 & 2.854\\[0.5em]
\multicolumn{1}{c|}{0.16} & 0.3 & 0.821 & -0.074 & 1.398 & -0.333
& 0.558 & -0.094 & -57.20 & 18.05 & -10.14 & 3.307
\end{tabular}
\caption{Best-fitted parameters of the theoretically predicted
scaling laws \eqref{160502:1202} to the DNS data as shown in
Figure \ref{fig4} \& \ref{fig5} for two different transpiration
rates $v_0^+=0.05$ and $v_0^+=0.16$, respectively, at
$Re_\tau=480$. The parameter $A_0^+$ was fitted for $\tau_{11}^+$,
and then applied in $\tau^+_{112}$: It takes the value
$A_0^+=13.82$ for $v_0^+=0.05$, and $A_0^+=6.971$ for
$v_0^+=0.16$. The matching region for the lower rate $v_0^+=0.05$
was chosen in the range $0.50\leq x_2/h\leq 1.25$, while for the
higher rate $v_0^+=0.16$ it was determined to lie in the range
$0.75\leq x_2/h\leq 1.70$. For more details, see the main text. }
\label{tab2}
\end{center}
\vspace{-0.4em}\hrule
\end{table}

In the following (comparative) discussion on the quality of the
fits for the velocity correlations
$\tau_{ij}^+$ and $\tau_{ijk}^+$ in Figure \ref{fig4} \&
\ref{fig5}, we will only focus on the systematic failure when
fitting scaling laws which show a simultaneous combination of an
algebraic and a logarithmic scaling. The fitted correlation
functions in \eqref{160502:1202} can be separated into two
classes: Those which show a strong (at least quadratic) dependency
on the mean streamwise velocity field $\bar{U}_1$, as
$\tau_{11}^+$ and $\tau^+_{112}$, and those which only show a weak
(at most linear) or no dependency at all on this field, as the
remaining ones in this list: $\tau_{12}^+$, $\tau_{33}^+$ and
$\tau_{223}^+$. While the latter correlations more or less
satisfactorily match the DNS data (where the lower order
correlations $\tau_{12}^+$ and $\tau_{33}^+$ show a better
matching than the higher order one $\tau_{223}^+$), the fitting of
the former correlations $\tau_{11}^+$ and $\tau^+_{112}$ fails to
predict the tendency of the data.

Based on our previous studies \cite{Frewer14.1,Frewer15.1}
supplemented by \cite{Frewer15.X} and \cite{Frewer16.1}, several
different mathematical proofs are given that explain this failure
and discrepancy in the matching results. The origin simply lies in
the fact that the general and explicit $\bar{U}_i$-dependency in
the scaling laws \eqref{160502:1202} for the velocity correlations
result from two ``statistical symmetries" $\bar{T}_s^\prime$
\eqref{160430:2130} and $\bar{T}^\prime_c$ \eqref{150424:1633}
that violate the classical principle of cause and effect. That is,
the $\bar{U}_1$- as well as the $\bar{U}_2$-dependency in the
velocity correlations $\tau_{ij}^+$ and~$\tau_{ijk}^+$ are simply
\emph{unphysical}. The negative results appear more strongly, of
course, in the correlations\linebreak involving the unphysical
$\bar{U}_1$-dependence.\footnote[2]{Note that we do not criticize
the functional structure of the logarithmic scaling law of
$\bar{U}_1$ itself, which can be more or less robustly matched to
the DNS data in the channel center. We rather criticize its
invariant Lie-group based derivation yielding this function with
the aid of unphysical symmetries, and its consequent unnatural
appearance in all higher order velocity correlations having a
streamwise component. This criticism is all the more significant
and pertinent as in \cite{Oberlack14} the misleading impression is
conveyed that the new logarithmic scaling law [Eq.$\,$(4.3)] for
the channel center is based on a derivation from first principles.
Yet, in this regard, it also should be clear that we do not
criticize the method of Lie-groups itself, being a very useful
mathematical tool indeed, when only applied to the right problems.
However, in \cite{Oberlack14} the method of Lie symmetry groups
has been misapplied. The reason for this mistake was, and still
is, not to recognize that every methodology in science has its
limits, in particular the fact that also the theory of Lie-groups
cannot analytically circumvent the closure problem of turbulence,
even if the infinite hierarchy of statistical equations is
formally considered. Because, instead of true symmetry
transformations only the weaker form of equivalence transformation
can be generated for such (unclosed) systems, for which, in a
strict mathematical sense, the construction of invariant solutions
is misleading and sometimes even ill-defined if no further
external information is provided: For example, as to close the
system of equations through some modelling assumptions, or, as in
the specific case of homogeneous isotropic turbulence, where one
has exclusive access to additional nonlocal invariants such as the
Birkhoff-Saffman or the Loitsyansky integral, to yield more
valuable results from such \emph{equivalence} scaling groups, in
particular the explicit values for the decay rates. For more
details, we again refer to
\cite{Frewer14.1,Frewer15.0,Frewer15.0x} and the references
therein.\\[0.75em]} The unphysical $\bar{U}_2$-dependence, however, is
less\linebreak critical for the particular flow case considered
here, since it is only a global constant, $\bar{U}_2=v_0$.
Worthwhile to note here is that the mismatch of $\tau_{11}^+$ and
$\tau^+_{112}$ in Figure \ref{fig5} is less severe for a higher
transpiration rate than in Figure \ref{fig4} for a lower one. The
reason is that the unphysical $\bar{U}_1$-dependence gets weaker
for increasing transpiration rates, simply because the mean
velocity field $\bar{U}_1^+$ itself is globally decaying for
higher rates (see Figure \ref{fig1}).

\begin{figure}[t]\hspace*{0.225cm}
\centering
\begin{minipage}[r]{.3\linewidth}
\FigureXYLabel{\includegraphics[width=.91\textwidth]{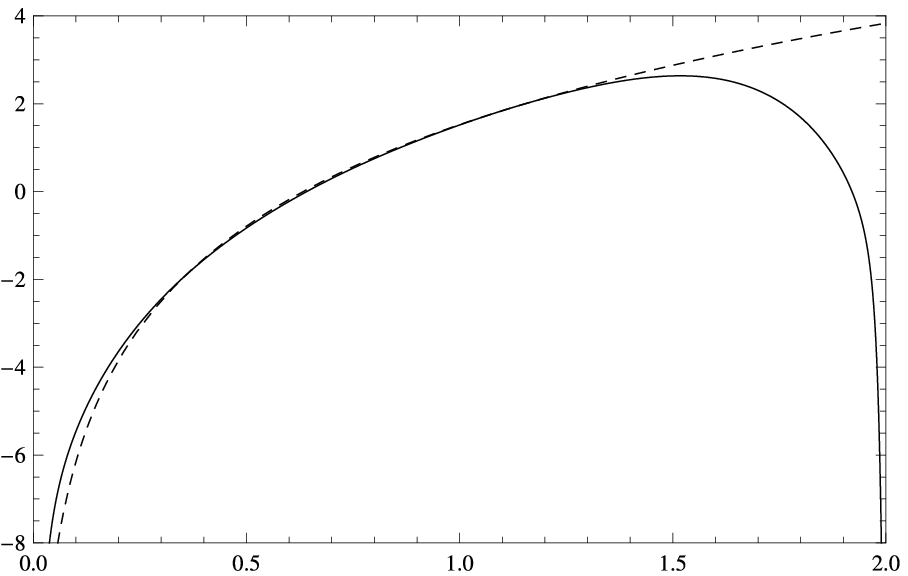}}
{${\scriptscriptstyle\hspace{0.75cm}
x_2/h}$}{-2.5mm}{\begin{rotate}{90}
$\hspace{-0.75cm}{\scriptscriptstyle (\bar{U}_1^-U_B)/u_\tau}$
\end{rotate}}{1.5mm}
\end{minipage}
\hfill
\begin{minipage}[c]{.3\linewidth}
\FigureXYLabel{\includegraphics[width=.91\textwidth]{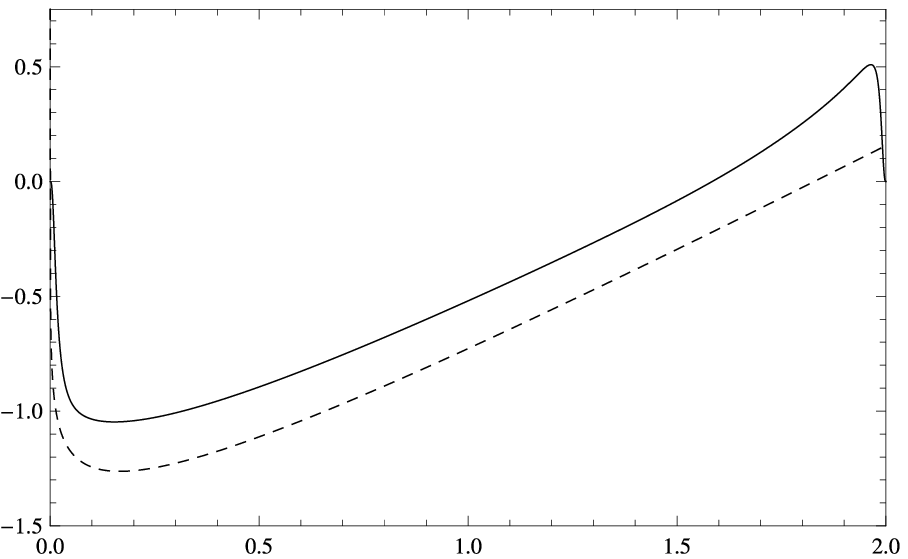}}
{${\scriptscriptstyle\hspace{0.75cm}
x_2/h}$}{-2.5mm}{\begin{rotate}{0}
$\hspace{-0.3cm}{\scriptscriptstyle
\tau_{12}^+}$\end{rotate}}{0mm}
\end{minipage}
\hfill
\begin{minipage}[c]{.3\linewidth}
\FigureXYLabel{\includegraphics[width=.91\textwidth]{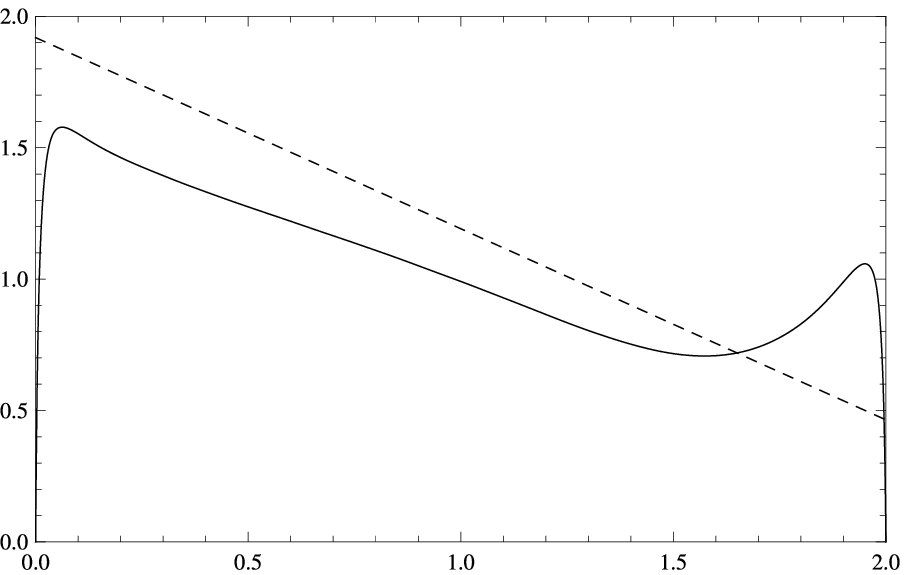}}
{${\scriptscriptstyle\hspace{0.75cm}
x_2/h}$}{-2.5mm}{\begin{rotate}{0}
$\hspace{-0.3cm}{\scriptscriptstyle\tau_{33}^+}$\end{rotate}}{0mm}
\end{minipage}
\hfill\hspace*{0.225cm}
\begin{minipage}[r]{.3\linewidth}
\FigureXYLabel{\includegraphics[width=.91\textwidth]{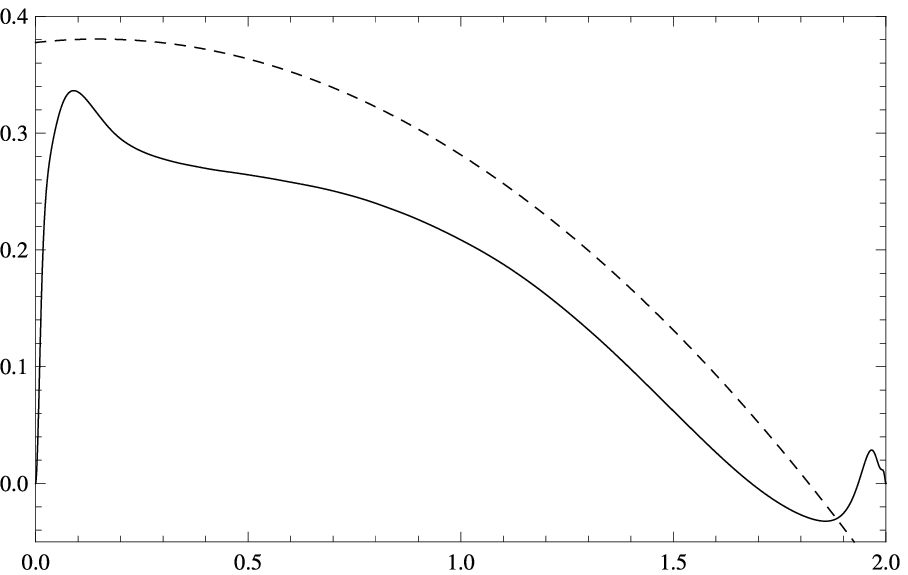}}
{$\phantom{.}$}{-2.5mm}{\begin{rotate}{0}
$\hspace{-0.5cm}{\scriptscriptstyle
\tau_{233}^+}$\end{rotate}}{0mm}
\end{minipage}
\hfill
\begin{minipage}[c]{.3\linewidth}
\FigureXYLabel{\includegraphics[width=.91\textwidth]{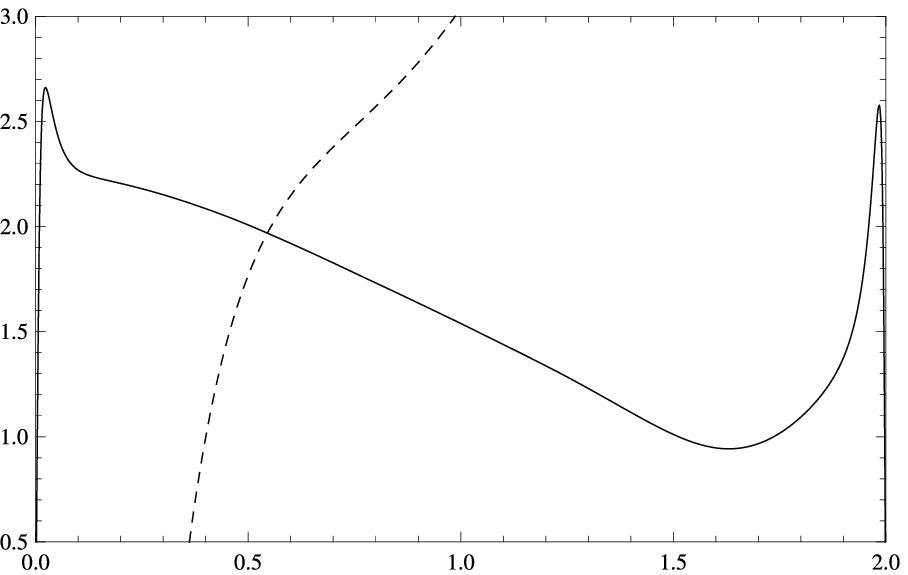}}
{$\phantom{.}$}{-2.5mm}{\begin{rotate}{0}
$\hspace{-0.3cm}{\scriptscriptstyle
\tau_{11}^+}$\end{rotate}}{0mm}
\end{minipage}
\hfill
\begin{minipage}[c]{.3\linewidth}
\FigureXYLabel{\includegraphics[width=.91\textwidth]{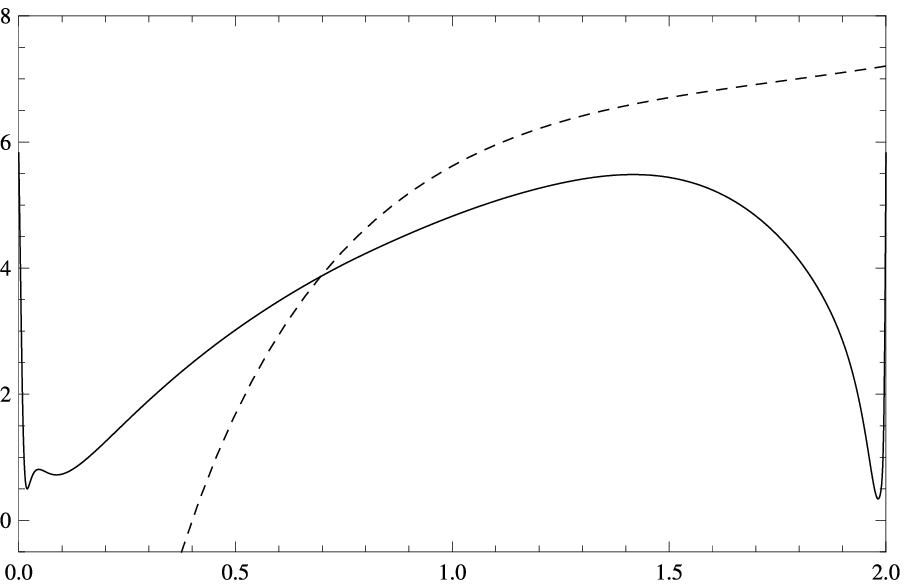}}
{$\phantom{.}$}{-2.5mm}{\begin{rotate}{0}
$\hspace{-0.5cm}{\scriptscriptstyle
\tau_{112}^+}$\end{rotate}}{0mm}
\end{minipage}
\caption{Sensitivity study on the Reynolds number $Re_\tau$ at
fixed transpiration rate $v_0^+=0.05$. The solid lines display the
DNS data for $Re_\tau=850$; the dashed lines again the
corresponding theoretically predicted ($Re_\tau$-independent)
scaling laws \eqref{160502:1202} as proposed by \cite{Oberlack14}
when coherently prolonged to higher-order moments. This figure is
to be compared with Figure \ref{fig4}, having the same
transpiration rate $v_0^+=0.05$ but at a lower Reynolds number
$Re_\tau=480$. Except for the mean velocity profile in deficit
form, the strong sensitivity on $Re_\tau$ for all higher-order
moments hence proves our conclusion in Section \ref{S3}: The
central assumption made in \cite{Oberlack14}, namely that the
scaling of a turbulent channel flow with uniform wall-normal
transpiration can be predicted by considering an {\it inviscid}
($\nu=0$) symmetry analysis, i.e., by considering invariant
solutions which, by construction, do not depend on $Re_\tau$, is
not justified at all. Important to note here is that this strong
sensitivity only lies in the matching parameters of the scaling
laws \eqref{160502:1202}, and {\it not} in the DNS data itself.
For more details on how the above figure was generated and its
connection to Figure \ref{fig4}, see the main
text.}\label{fig6}\vphantom{x}\hrule
\end{figure}

Moreover, the key assumption in \cite{Oberlack14} that an inviscid
($Re_\tau$-independent) symmetry analysis is sufficient to capture
the scaling behavior in the center of the channel, is not
justified. Instead a strong sensitivity on $Re_\tau$ in the
scaling laws \eqref{160502:1202} is observed, which, in comparison
to Figure \ref{fig4}, is shown in Figure \ref{fig6}. This figure
was generated under the assumption of \cite{Oberlack14} that the
Reynolds-number-independent scaling of the (non-normalized)
higher-order moments \eqref{160502:1202} is correct: All involved
parameters, once matched for a certain fixed Reynolds number
$Re_\tau$ and transpiration rate $v_0^+$, should then stay
invariant as $Re_\tau$ changes. Of course, this assumed invariance
should only hold for the {\it non}-normalized parameters as
formulated in~\eqref{160502:1202}, and not for the
$u_\tau$-normalized ones, simply because the friction velocity
$u_\tau$ itself changes when the Reynolds number $Re_\tau$ varies.
Although this dependence $u_\tau\sim u_\tau(Re_\tau)$ is rather
weak for a fixed transpiration rate $v_0^+$ and bulk
velocity~$U_B^*$ as can be seen in Table~\ref{tab1}, this
dependence, as was already discussed in Section \ref{S3}, cannot
be neglected, in particular not for any higher-order moments,
where the relative change
\begin{equation}
\Delta_\% u_\tau^n =
\frac{u^n_\tau(Re_\tau)-u^n_\tau(Re^*_\tau)}{u^n_\tau(Re^*_\tau)},
\label{160612:0937}
\end{equation}
in the normalization factor $u_\tau^n$ for a moment of order $n$
becomes more pronounced for increased order, as can be seen in
Figure \ref{fig6}. The Reynolds number $Re^*_\tau$ refers to some
fixed reference value, which, in the considered case for
$Re_\tau=850$ in Figure \ref{fig6}, is given by $Re^*_\tau=480$
(when compared to Figure \ref{fig4}). For example, for the given
values $u_\tau|_{Re^*_\tau=480}\sim 0.0551$ and
$u_\tau|_{Re_\tau=850}\sim 0.0501$,\linebreak taken from Table
\ref{tab1} at the fixed transpiration rate $v_0^+=0.05$, the
relative change in the normalization factor for the third order
moment is already at $\Delta_\% u_\tau^3\sim
-25\%$\hspace{0.01cm}; a change which definitely cannot be
neglected anymore.
\begin{figure}[t]\hspace*{0.225cm}
\centering
\begin{minipage}[r]{.3\linewidth}
\FigureXYLabel{\includegraphics[width=.91\textwidth]{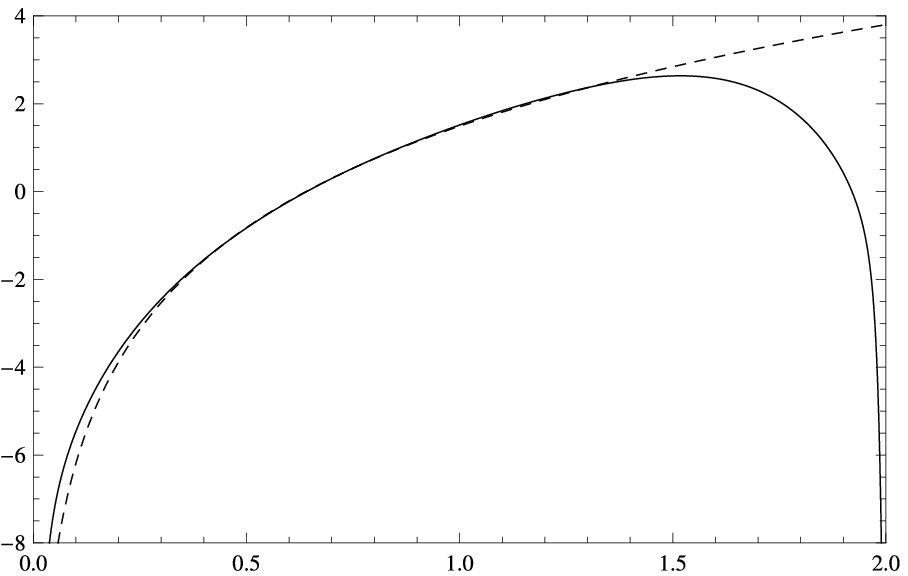}}
{${\scriptscriptstyle\hspace{0.75cm}
x_2/h}$}{-2.5mm}{\begin{rotate}{90}
$\hspace{-0.75cm}{\scriptscriptstyle (\bar{U}_1^-U_B)/u_\tau}$
\end{rotate}}{1.5mm}
\end{minipage}
\hfill
\begin{minipage}[c]{.3\linewidth}
\FigureXYLabel{\includegraphics[width=.91\textwidth]{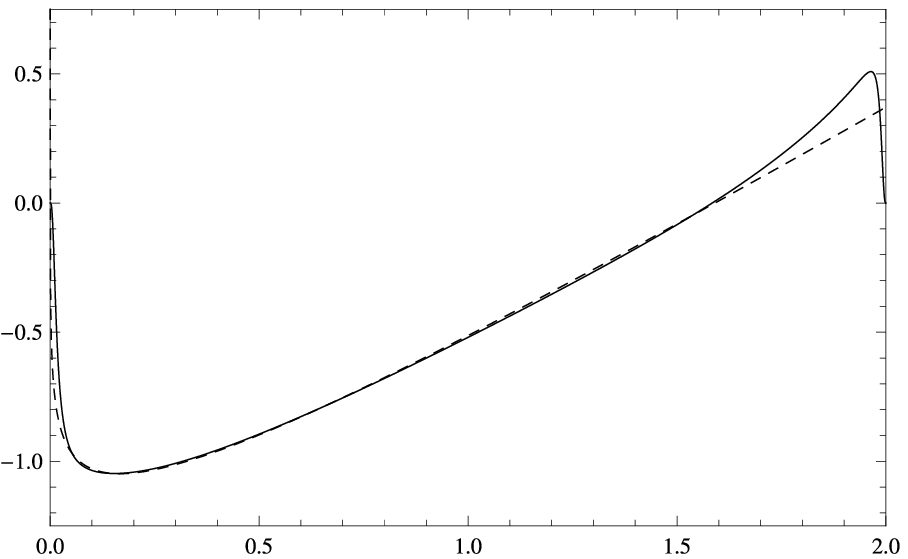}}
{${\scriptscriptstyle\hspace{0.75cm}
x_2/h}$}{-2.5mm}{\begin{rotate}{0}
$\hspace{-0.3cm}{\scriptscriptstyle
\tau_{12}^+}$\end{rotate}}{0mm}
\end{minipage}
\hfill
\begin{minipage}[c]{.3\linewidth}
\FigureXYLabel{\includegraphics[width=.91\textwidth]{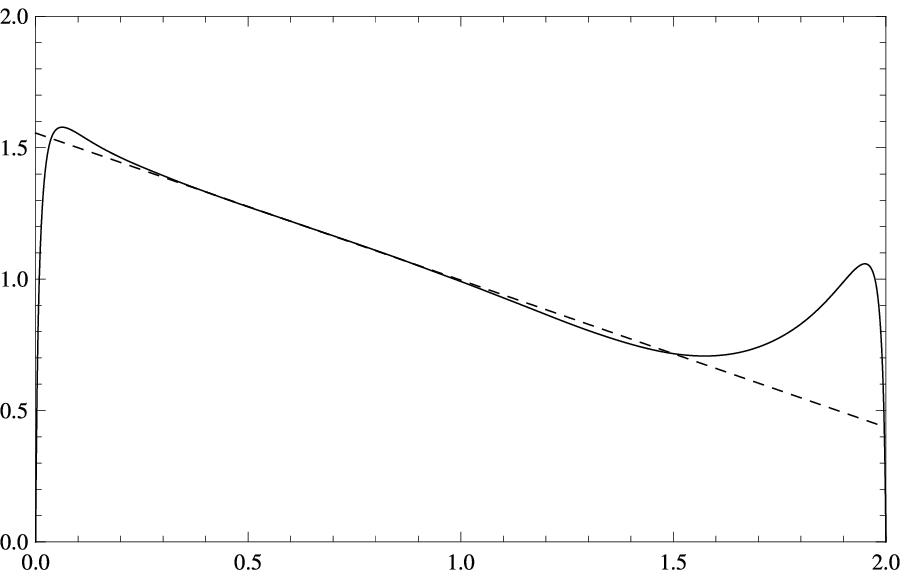}}
{${\scriptscriptstyle\hspace{0.75cm}
x_2/h}$}{-2.5mm}{\begin{rotate}{0}
$\hspace{-0.3cm}{\scriptscriptstyle\tau_{33}^+}$\end{rotate}}{0mm}
\end{minipage}
\hfill\hspace*{0.225cm}
\begin{minipage}[r]{.3\linewidth}
\FigureXYLabel{\includegraphics[width=.91\textwidth]{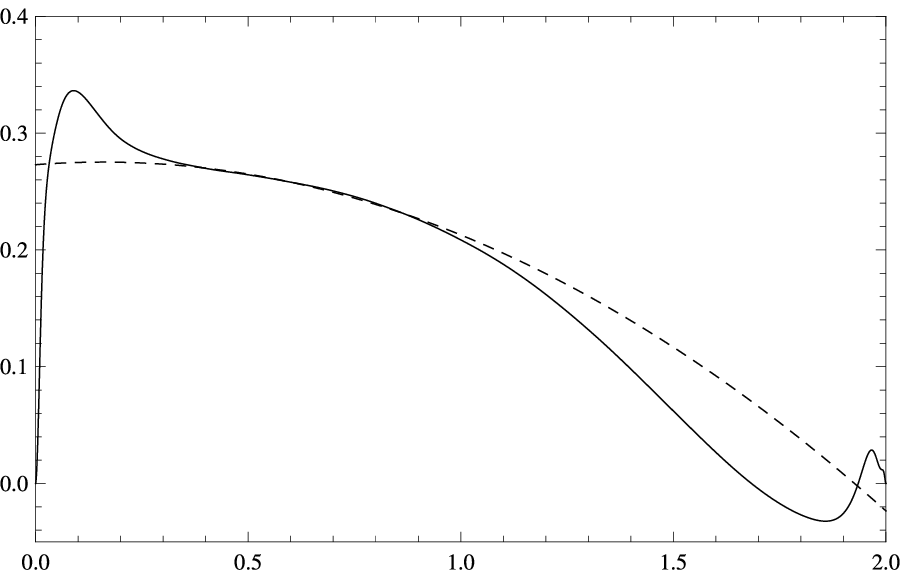}}
{$\phantom{.}$}{-2.5mm}{\begin{rotate}{0}
$\hspace{-0.5cm}{\scriptscriptstyle
\tau_{233}^+}$\end{rotate}}{0mm}
\end{minipage}
\hfill
\begin{minipage}[c]{.3\linewidth}
\FigureXYLabel{\includegraphics[width=.91\textwidth]{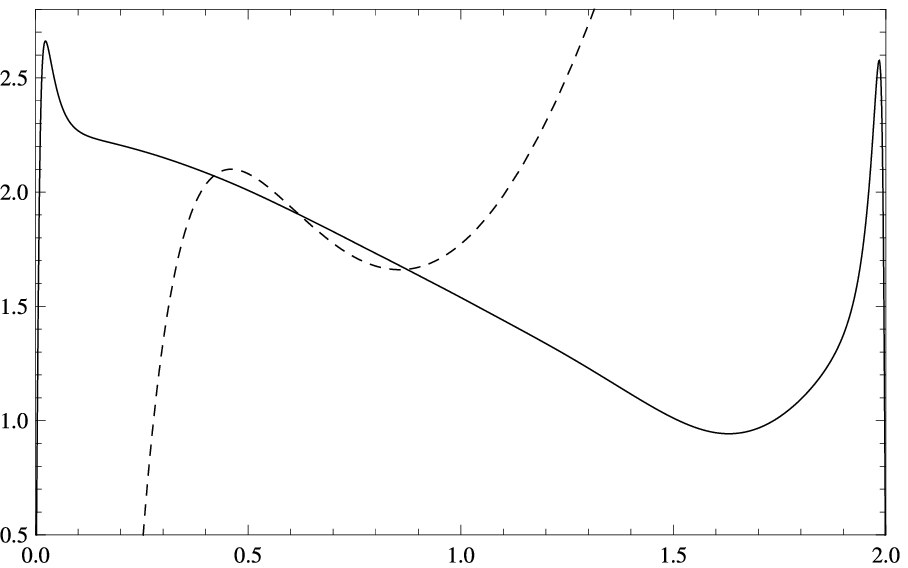}}
{$\phantom{.}$}{-2.5mm}{\begin{rotate}{0}
$\hspace{-0.3cm}{\scriptscriptstyle
\tau_{11}^+}$\end{rotate}}{0mm}
\end{minipage}
\hfill
\begin{minipage}[c]{.3\linewidth}
\FigureXYLabel{\includegraphics[width=.91\textwidth]{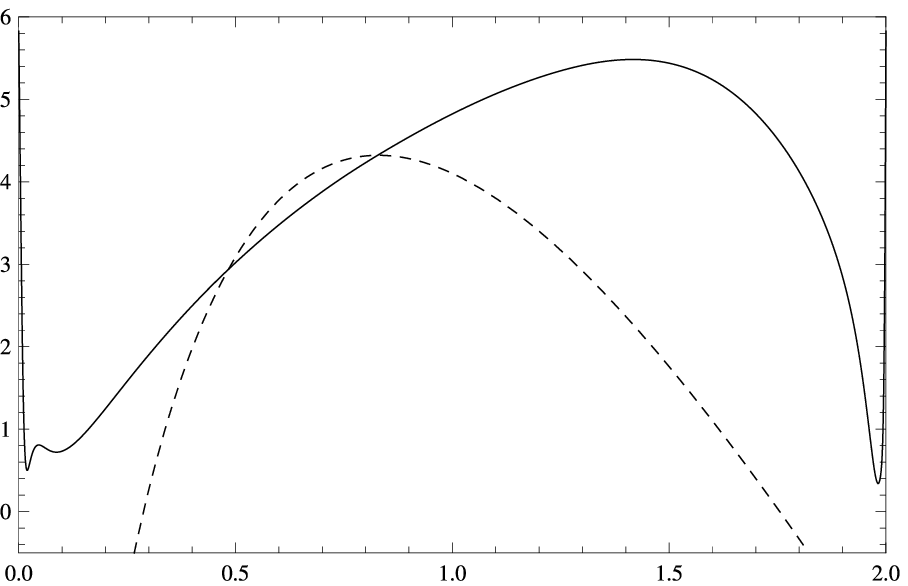}}
{$\phantom{.}$}{-2.5mm}{\begin{rotate}{0}
$\hspace{-0.5cm}{\scriptscriptstyle
\tau_{112}^+}$\end{rotate}}{0mm}
\end{minipage}
\caption{Matching of the theoretically predicted scaling laws
\eqref{160502:1202} to the DNS data for $Re_\tau=850$ and
$v_0^+=0.05$. The DNS data is displayed by solid lines, the
corresponding scaling laws by dashed lines. The associated
(non-normalized) parameters for this best fit in each case can be
taken from Table \ref{tab3}, where the matching region was chosen
in the same range $0.50\leq x_2/h\leq 1.25$ as for  $Re_\tau=480$
in Figure \ref{fig4}. For~more details and a discussion on the
fitting results obtained, see the main
text.}\label{fig7}\vphantom{x}\hrule
\end{figure}

Now, while the matching to the DNS data in Figure \ref{fig4} was
performed in the $u_\tau$-normalization for $Re^*_\tau=480$, and
since for its comparison to a higher Reynolds number in Figure
\ref{fig6} a corresponding $u_\tau$-normalization for
$Re_\tau=850$ is needed, all ``+"-parameters given in Table
\ref{tab2} have to be re-scaled by the ratio factor
$\vartheta=u_\tau|_{Re^*_\tau=480}/u_\tau|_{Re_\tau=850}\sim
1.1011$ in order to shift the (assumed invariant) non-normalized
parameters as formulated in \eqref{160502:1202} from
$Re^*_\tau=480$ to $Re_\tau=850$. As mentioned before, although
this factor $\vartheta$ is more or less close to one, it is not so
anymore for the parametric values of any higher-order moments,
where the change is significant. For example, for the normalized
value $\tilde{A}^+_{12}$ of the second moment $\tau_{12}^+$,
fitted in Figure \ref{fig4} and listed in Table \ref{tab2}, the
relative change is already about $20\%$:
\begin{equation}
\!\!\!\tilde{A}_{12}^+\Big|_{{Re^*_\tau=480}}\sim-0.622
\;\;\xrightarrow{u^2_\tau|_{Re^*_\tau=480}}\;\; \tilde{A}_{12}\sim
-0.002\;\;\xrightarrow{1/u^2_\tau|_{Re^*_\tau=850}}\;\;
\tilde{A}_{12}^+\Big|_{{Re_\tau=850}}\sim
-0.754.\label{160612:0133}
\end{equation}
It should be clear, that the scaling laws \eqref{160502:1202} in
Figure \ref{fig6} were not fitted to the DNS data, but the fact
that they were obtained from the fitted results in Figure
\ref{fig4} by up-scaling the determined parameters of Table
\ref{tab2} from $Re^*_\tau=480$ to $Re_\tau=850$ in using the
procedure outlined in \eqref{160612:0133}. Important to note here
is that the up-scaling for the $u_\tau$-normalized mean velocity
field $\bar{U}_1^+$, to be needed in the higher-order moments
$\tau_{11}^+$, $\tau_{12}^+$ and $\tau_{112}^+$, has been
performed according to the (more or less correct) assumption of
\cite{Oberlack14} that the scaling law for the velocity field
$\bar{U}_1$ in its deficit form \eqref{160526:1806} stays
invariant for different Reynolds numbers at a fixed transpiration
rate:
\begin{equation}
\frac{\bar{U}_1-U_B^*}{u_\tau|_{Re^*_\tau=480}}=
\frac{1}{\gamma}\ln\left(x_2/h\right)+\lambda
=\frac{\bar{U}_1-U_B^*}{u_\tau|_{Re_\tau=850}},
\end{equation}
which then can be solved to give the up-scaling relation for
$\bar{U}_1^+$
\begin{align}
\bar{U}^+_1\Big|_{Re_\tau=850}&=\bar{U}^+_1\Big|_{Re^*_\tau=480}
-\frac{U_B^*}{u_\tau|_{Re^*_\tau=480}}+\frac{U_B^*}{u_\tau|_{Re_\tau=850}}
\nonumber\\[0.5em]
&=\left(\frac{1}{\gamma}\ln\left(x_2/h\right)+\lambda
+\frac{U_B^*}{u_\tau|_{Re^*_\tau=480}}\right)
-\frac{U_B^*}{u_\tau|_{Re^*_\tau=480}}+\frac{U_B^*}{u_\tau|_{Re_\tau=850}}
\qquad\nonumber\\[0.75em]
& = \frac{1}{\gamma}\ln\left(x_2/h\right)+\lambda +
\frac{U_B^*}{u_\tau|_{Re_\tau=850}}.
\end{align}
All these steps finally reveal the sensitivity of the invariant
functions \eqref{160502:1202} on the Reynolds number $Re_\tau$, as
can be explicitly seen in Figure \ref{fig6}. Another option to
study this sensitivity, is to re-fit again the scaling laws
\eqref{160502:1202} to the DNS data for $Re_\tau=850$, and to see
how far the best-fitted values are off from the ones listed in
Table~\ref{tab2} relative to the reference Reynolds-number
$Re^*_\tau=480$. As to be expected, the quality of the fit is
similar to that of Figure \ref{fig4}, as can be seen in Figure
\ref{fig7}, but it was achieved for different (non-normalized)
parametric values which, as can be compared in Table \ref{tab3},
changed significantly, in particular the values for the two
highest order moments $\tau_{112}^+$ and $\tau_{233}^+$.

\begin{table}[t]
\begin{center}
\begin{tabular}{c c c | c | c c | c c | c c | c c}\hline\\
\multicolumn{1}{c|}{$Re_\tau$} & $\gamma$ & $\lambda$ &
$\tilde{A}_{12}$ & $A_{33}$ & $C_{33}$ & $A_{233}$ & $C_{233}$ &
$A_{11}$ & $C_{11}$ &
$\tilde{A}_{112}$ & $C_{112}$\\[0.0em]
\multicolumn{1}{c|}{} & & & ${\scriptstyle [10^{-3}]}$ &
${\scriptstyle [10^{-3}]}$ & ${\scriptstyle [10^{-3}]}$ &
${\scriptstyle [10^{-5}]}$ & ${\scriptstyle [10^{-5}]}$ &
${\scriptstyle [10^{-1}]}$ & ${\scriptstyle [10^{-1}]}$ &
${\scriptstyle [10^{-3}]}$ &
${\scriptstyle [10^{-3}]}$\\[0.5em]\hline
\multicolumn{1}{c|}{} & & & & & & & & & & &\\[-0.75em]
\multicolumn{1}{c|}{480} & 0.3 & 1.524 & -1.888 & 4.807 & -1.822 &
6.053 & -1.710 & -6.046 & 0.785
& -4.744 & 0.478\\[0.5em]
\multicolumn{1}{c|}{850} & 0.3 & 1.491 & -1.355 & 3.897 & -1.402 &
4.398 & -1.105 & -5.437 & 0.873 & -3.495 & 0.264
\end{tabular}
\caption{Best-fitted (non-normalized) parameters of the
theoretically predicted scaling laws \eqref{160502:1202} to the
DNS data as shown in Figure \ref{fig4} \& \ref{fig7} for two
different Reynolds numbers $Re_\tau=480$ and $Re_\tau=850$,
respectively, at $v_0^+=0.05$. In both cases, the parameter
$A^+_0$ was fitted for $\tau_{11}^+$, and then applied in
$\tau^+_{112}$:\\ It takes the (non-normalized) value $A_0=0.761$
for $Re_\tau=480$, and $A_0=0.723$ for $Re_\tau=850$. Also, in
both cases, the overall matching region was chosen in the range
$0.50\leq x_2/h\leq 1.25$. Except for $A_0$ and the parameters of
the velocity profile ($\gamma$ and $\lambda$), a strong Reynolds
number dependence is observed throughout all scaling law
parameters. Although the inviscid $(\nu=0)$ assumption in
\cite{Oberlack14} is more or less valid for the lowest order
moment (the mean velocity), it is incorrect for all higher-order
moments, in particular as the order of the moments increases, the
$Re_\tau$-dependence becomes more and more pronounced, e.g., for
the third order parameter $A_{233}$ we observe a relative change
of nearly~$30\%$.} \label{tab3}
\end{center}
\vspace{-0.4em}\hrule\vspace{-0.1em}
\end{table}

Anyhow, except for the mean velocity in deficit form, all
higher-order moments show a strong sensitivity on the Reynolds
number at fixed transpiration rate, thus clearly invalidating the
inviscid assumption in \cite{Oberlack14}. It should be clear that
this strong sensitivity\pagebreak[4] only lies in the matching
parameters of the Lie-group generated scaling laws
\eqref{160502:1202}, and {\it not} in the DNS data itself. In
other words, these scaling laws cannot be robustly matched to the
DNS data when assuming independence in one of its external system
parameters.

Returning to the inconsistent symmetry analysis performed in
\cite{Oberlack14}, in particular to the application of the
unphysical scaling symmetry $\bar{T}_s^\prime$
\eqref{160424:1418}, the mathematical proof in \cite{Frewer14.1},
Appendix~D, clearly shows that independent of the particular flow
configuration, the Lie-group based turbulent scaling laws for all
higher order velocity correlations as derived in
\eqref{160502:1202} are not consistent to the scaling of the mean
velocity field itself. In other words, the proof in
\cite{Frewer14.1} shows that for the lowest correlation order
$n=1$ (defined as the mean velocity field) no contradiction
exists, only as from $n=2$ onwards the contradiction starts, i.e.,
while the mean velocity field can be robustly matched to the DNS
data, it consistently fails for all higher order correlation
functions and gets more pronounced the higher the correlation
order~$n$~is.

Hence, to justify their new scaling law [Eq.$\,$(4.3)] in
\cite{Oberlack14} by saying that it ``was successfully validated
with DNS data for moderate transpiration rates" [p.$\,$119] is
based on a fallacy. The problem is that this ``validation" in
\cite{Oberlack14} was only performed for the lowest order moment,
which, of course, can always be matched to the DNS data since
there are enough free parameters available to be fitted. But, as
soon as any higher order correlations functions get fitted, not
enough free parameters are available anymore and the curve-fitting
procedure consistently fails in \cite{Oberlack14}, as shown
in~Figure~\ref{fig4}~\&~\ref{fig5}.

Therefore, no true validation of the Lie-group-based scaling
theory has been performed in \cite{Oberlack14}. For that also the
theoretically predicted pressure-velocity correlations need to be
validated against the DNS data to check in how far the best-fitted
parameters are consistent with the parametric relations resulting
from the underlying statistical transport equations
\eqref{160430:1912}-\eqref{160430:1914} including {\it all}
correlations, velocity as well as pressure. In particular, as the
study of \cite{Oberlack14} is based on the findings of
\cite{Oberlack10} which specifically considers the infinite
(unclosed) system of {\it all} multi-point correlation equations
and which thus is designed and laid-out to be a ``first principle"
scaling theory for all higher order correlations (including
velocity and pressure), special attention has to be devoted to the
prediction value of all those correlation functions which go
beyond the lowest or next to the lowest order. And exactly this
has been investigated by us in the present study, however, yet
only for the velocity correlation functions up to third order, but
which already gives a different picture than the ``validation"
procedure in \cite{Oberlack14} is trying to suggest. The same
issue we also face in their subsequent publication
\cite{Oberlack15Rev}, which we will~discuss~next.\pagebreak[4]

\subsection{The viscous case ($\nu\neq 0$)\label{S52}}

The viscous scaling theory to turbulent channel flow with constant
wall-normal transpiration has not been studied in
\cite{Oberlack14}. It only can be found in their subsequent
publication \cite{Oberlack15Rev}, where it is discussed in
Sec.$\,$6.2. For the same flow conditions as described in Section
\ref{S2}, the viscous transport equations corresponding to the
inviscid ones~\eqref{160430:1912}-\eqref{160430:1914} will have
the extended form
\begin{gather}
\frac{\partial\overline{U_2}}{\partial x_2}=0,\label{160502:1240}\\[0.5em]
\frac{\partial \overline{U_1U_2}}{\partial x_2}+\frac{\partial
\overline{P}}{\partial
x_1}-\nu\frac{\partial^2\overline{U_1}}{\partial x_2^2}=0,\qquad
\frac{\partial \overline{U_2U_2}}{\partial x_2}+\frac{\partial
\overline{P}}{\partial x_2}=0,\qquad
\overline{U_1U_3}=\overline{U_2U_3}=0,
\label{160502:1241}\\[0.5em]
\left.
\begin{aligned}
\frac{\partial\overline{U_1U_2U_2}}{\partial x_2}
+\overline{\frac{\partial P}{\partial
x_1}U_2}+\overline{U_1\frac{\partial P}{\partial x_2}}
-\nu\overline{U_1 \Delta U_2}-\nu\overline{U_2 \Delta U_1}
=0,\hspace{0.7cm}\\[0.5em]
\frac{\partial\overline{U_iU_jU_2}}{\partial x_2}
+\overline{\frac{\partial P}{\partial
x_i}U_j}+\overline{U_i\frac{\partial P}{\partial
x_j}}-\nu\overline{U_i \Delta U_j}-\nu\overline{U_j \Delta
U_i}=0,\;\text{for $i=j$},
\end{aligned}
~~~\right\} \label{160502:1242}
\end{gather}
where the second order viscous terms can also be equivalently
written within their full-field form~as
\begin{equation}
\nu\overline{U_i \Delta U_j}+\nu\overline{U_j \Delta
U_i}=\nu\frac{\partial^2\overline{U_iU_j}}{\partial x_2^2}-2\nu
\overline{\frac{\partial U_i}{\partial x_k}\frac{\partial
U_j}{\partial x_k}}.\label{150503:1037}
\end{equation}
Decomposing this system into mean and fluctuating fields according
to \eqref{160501:1208}, we obtain the corresponding viscous
Reynolds transport equations
\begin{gather}
\frac{\partial\bar{U}_2}{\partial x_2}=0,\label{160502:1320}\\[0.5em]
\bar{U}_2\frac{\partial\bar{U}_1}{\partial
x_2}+\frac{\partial\bar{P}}{\partial x_1}+\frac{\partial
\tau_{12}}{\partial x_2}-\nu\frac{\partial^2\bar{U}_1}{\partial
x_2^2}=0,\qquad \frac{\partial\bar{P}}{\partial
x_2}+\frac{\partial
\tau_{22}}{\partial x_2}=0,\qquad \tau_{13}=\tau_{23}=0,\label{160502:1321}\\[0.5em]
\left.
\begin{aligned}
\bar{U}_2\frac{\partial \tau_{12}}{\partial
x_2}+\frac{\partial\tau_{122}}{\partial
x_2}+\tau_{22}\frac{\partial\bar{U}_1}{\partial x_2}+
\overline{\frac{\partial p}{\partial
x_1}u_2}+\overline{u_1\frac{\partial p}{\partial x_2}}
-\nu\frac{\partial^2\tau_{12}}{\partial x_2^2}
+\varepsilon_{12}=0,\hspace{1.15cm}\\[0.5em]
\bar{U}_2\frac{\partial \tau_{ij}}{\partial
x_2}+\frac{\partial\tau_{ij2}}{\partial
x_2}+\tau_{i2}\frac{\partial\bar{U}_j}{\partial
x_2}+\tau_{j2}\frac{\partial\bar{U}_i}{\partial x_2}+
\overline{\frac{\partial p}{\partial
x_i}u_j}+\overline{u_i\frac{\partial p}{\partial
x_j}}-\nu\frac{\partial^2\tau_{ij}}{\partial x_2^2}
+\varepsilon_{ij}=0,\;\text{for $i=j$},
\end{aligned}
~~~\right\}\label{160502:1322}
\end{gather}
where $\varepsilon_{ij}$ is the well-known dissipation tensor
\begin{equation}
\varepsilon_{ij}=2\nu\overline{\frac{\partial u_i}{\partial
x_k}\frac{\partial u_j}{\partial x_k}}.\label{160604:1332}
\end{equation}
The set of symmetries admitted by
\eqref{160502:1240}-\eqref{150503:1037} stays unchanged to the
ones for the inviscid case used in the previous subsection, except
for the two Euler scaling symmetries $\bar{T}_1$
\eqref{160425:1017} and $\bar{T}_2$~\eqref{160502:1402} which both
break due the appearance of the viscous terms. Nevertheless, they
recombine to give the classical Navier-Stokes scaling symmetry
\begin{align}
\bar{T}_{\text{NS}}: &\quad x_i^*=e^{k_{\text{NS}}}x_i,\;\;\;
\overline{U_i}^{\, *}=e^{-k_{\text{NS}}}\overline{U_i},\;\;\;
\overline{P}^{\, *}=e^{-2k_{\text{NS}}}\overline{P},\;\;\;
\overline{U_iU_j}^{\,
*}=e^{-2k_{\text{NS}}}\overline{U_iU_j},\nonumber\\[0.5em]
&\quad \overline{U_iU_jU_k}^{\,
*}=e^{-3k_{\text{NS}}}\overline{U_iU_jU_k},\;\;\;
\overline{U_i\frac{\partial P}{\partial x_j}}^{\,
*}=e^{-4k_\text{NS}}\, \overline{U_i\frac{\partial P}{\partial
x_j}},\nonumber\\[0.5em]
&\quad \overline{\frac{\partial U_i}{\partial x_k}\frac{\partial
U_j}{\partial x_k}}^{\, *}
=e^{-4k_\text{NS}}\,\overline{\frac{\partial U_i}{\partial
x_k}\frac{\partial U_j}{\partial x_k}}.\label{160502:1424}
\end{align}
In \cite{Oberlack15Rev} another yet unmentioned symmetry is used
which provokes to be analyzed in more detail. Due to the linearity
of the MPC equations in their full-field representation, the
governing system \eqref{160502:1240}-\eqref{150503:1037} admits
the additional rather generic symmetry
\begin{align}
\bar{T}^\prime_{+}: &\quad  x_i^*=x_i,\;\;\; \overline{U_i}^{\,
*}=\overline{U_i}+F_i,\;\;\; \overline{P}^{\,
*}=\overline{P}+G,\;\;\;
\overline{U_iU_j}^{\, *}=\overline{U_iU_j}+F_{ij},\nonumber\\
&\quad \overline{U_iU_jU_k}^{\,
*}=\overline{U_iU_jU_k}+F_{ijk},\;\;\;\overline{U_i\frac{\partial
P}{\partial x_j}}^{\, *}=\overline{U_i\frac{\partial P}{\partial
x_j}}+G_{ij},\nonumber\\
&\quad\overline{\frac{\partial U_i}{\partial x_k}\frac{\partial
U_j}{\partial x_k}}^{\, *}=\overline{\frac{\partial U_i}{\partial
x_k}\frac{\partial U_j}{\partial x_k}}+L_{ij},\label{160501:1952}
\end{align}
where the functions $F_i$, $F_{ij}$, $F_{ijk}$, $G$, $G_{ij}\neq
G_{ji}$ and $L_{ij}$ are any particular solutions of the governing
system of equations \eqref{160502:1240}-\eqref{150503:1037}, i.e.,
where these functions satisfy again the equations
\begin{gather}
\frac{\partial F_2}{\partial x_2}=0,\label{160501:1918}\\[0.25em]
\frac{\partial F_{12}}{\partial x_2}+\frac{\partial G}{\partial
x_1}-\nu\frac{\partial^2 F_1}{\partial x_2^2}=0,\qquad
\frac{\partial F_{22}}{\partial x_2}+\frac{\partial G}{\partial
x_2}=0,\qquad F_{13}=F_{23}=0,
\label{160501:1919}\\[0.25em]
\left. \begin{aligned} \frac{\partial F_{122}}{\partial x_2} +
G_{12}+G_{21}-\nu\frac{\partial^2 F_{12}}{\partial x_2^2}+2\nu
L_{12}=0,\hspace{0.75cm}\\[0.5em]
\hspace{0.75cm}\frac{\partial F_{ij2}}{\partial x_2}+
G_{ij}+G_{ji}-\nu\frac{\partial^2 F_{ij}}{\partial x_2^2}+2\nu
L_{ij}=0,\;\text{for $i=j$}.
\end{aligned}
~~~\right\}\label{160501:1920}
\end{gather}
This symmetry just reflects the superposition property which is
featured by all (homogeneous) linear differential equations.
However, this symmetry is of no value, because, as correctly
already~noted in \cite{Oberlack10}, it ``cannot directly be
adopted for the practical derivation of group invariant solutions"
[p.$\,$463]. The simple reason it that the system of equations
\eqref{160501:1918}-\eqref{160501:1920} is unclosed and no
analytical solution is known yet which is consistent up to all
higher orders in its infinite hierarchy, otherwise one would have
found a solution to the still unsolved closure problem of
turbulence. Any guessed solution, which only satisfies the system
\eqref{160501:1918}-\eqref{160501:1920} up to a fixed order
$n=n_0$ in its infinite hierarchy, is of no value if we cannot
guarantee that (i) this solution is also consistent for all higher
orders $n>n_0$, and (ii) that it also represents a physical
solution which is consistent to the DNS data; because, for such
unclosed systems, infinitely many different and independent
mathematical solutions can be generated which all in the end are
{\it not} reflected in the DNS data
\citep{Frewer14.1,Frewer15.0,Frewer15.0x}.

Despite their concern in \cite{Oberlack10} that the superposition
symmetry due to the closure problem cannot be exploited, it
nevertheless was used in \cite{Oberlack15Rev} to generate
invariant solutions. Therein the superposition symmetry
$\bar{T}^\prime_{+}$ \eqref{160501:1952} shows its existence
through the arbitrarily chosen symmetries
$Z_{\!{\scriptscriptstyle z}ij}$ in [Eq.$\,$(353)], where some of
the functional translations in \eqref{160501:1952} were
arbitrarily fixed as linear functions (in the one-point limit
$\vr\rightarrow\boldsymbol{0}$):
\begin{gather}
\left.
\begin{aligned}
F_i=G_{ij}=L_{ij}=F_{ijk}=0,\qquad G=-k_{{\scriptscriptstyle
z}12}x_1-2k_{{\scriptscriptstyle
z}22}x_2,\hspace{1.25cm}\\[0.5em]
F_{11}=2k_{{\scriptscriptstyle z}11}x_2,\qquad
F_{12}=k_{{\scriptscriptstyle z}12}x_2,\qquad
F_{22}=2k_{{\scriptscriptstyle z}22}x_2,\qquad
F_{33}=2k_{{\scriptscriptstyle z}33}x_2.\label{160601:1953}
\end{aligned}
~~~\right\}
\end{gather}
The motivation to choose this particular set of (linear) functions
\eqref{160601:1953} and not any other set of functions that may
also solve the system \eqref{160501:1918}-\eqref{160501:1920}, is
not clear. However, if the motivation was such as to only gain a
better matching of the invariant functions to the DNS data, then
the procedure proposed in \cite{Oberlack15Rev} has nothing to do
with a theoretical prediction or forecasting of turbulent scaling
laws as claimed therein. Because, such an approach would then just
be based on a trial and error procedure which incrementally
improves the prediction of turbulent scaling only {\it a
posteriori}, and not {\it a priori}, as required for a true
theoretical and ``first principle" ansatz. In other words, since
we don't see in \cite{Oberlack15Rev} any clear motivation {\it
a~priori} for a linear solution ansatz of the functional
translational symmetries $Z_{\!{\scriptscriptstyle
z}ij}$~[Eq.$\,$(353)], it seems that they were chosen {\it a
posteriori} to only enhance the matching to the\pagebreak[4] DNS
data. But, as already said, such an approach is not in the sense
of the inventor to forecast turbulent scaling laws from a ``first
principle" theory which is ``fully algorithmic" and where ``no
intuition is needed" \cite[p.$\,$321]{Oberlack01}. Hence, through
the use of the superposition principle $\bar{T}^\prime_{+}$
\eqref{160501:1952} in an unclosed system
\eqref{160502:1240}-\eqref{150503:1037}, the Lie-group symmetry
approach in \cite{Oberlack15Rev} degenerates down to a {\it
non}-predictive incremental trial-and-error method.

But, as already discussed in the conclusion of the previous
subsection, no matter how great the effort to incrementally
improve the predictive ability of Lie-group generated invariant
solutions for turbulent scaling, the methodological approach
itself, as initially proposed in \cite{Oberlack10} and last
applied in \cite{Oberlack15Rev}, will always be inconsistent in
that a comparison to DNS data will always fail when considering
correlation orders higher than the well-matched threshold level of
a lower correlation order. The simple reason is that the analysis
is permanently set up by two unphysical ``statistical symmetries"
$\bar{T}_s^\prime$ \eqref{160424:1418} and $\bar{T}^\prime_c$
\eqref{150501:1342} that perpetually violate the classical
principle of cause and effect. The only way thus to obtain an
overall {\it consistent} symmetry analysis, is to discard all
unphysical symmetries
\citep{Frewer14.1,Frewer15.1,Frewer15.X,Frewer16.1}.

When combining all symmetry groups
\eqref{160424:1418}-\eqref{150501:1342}, \eqref{160502:1424} and
\eqref{160501:1952} with \eqref{160601:1953}, we now obtain,
instead of the inviscid invariant surface condition
\eqref{160425:2207}, the viscous condition
\begin{align}
\frac{dx_1}{k_\text{NS}x_1+k_{x_1}}=\frac{dx_2}{k_\text{NS}x_2+k_{x_2}}
&=\frac{d\overline{U_i}}{\big(\! -k_\text{NS}+k_s)
\overline{U_i}+\omega_i+\kappa_i+c_i}=\frac{d\overline{P}}{\big(\!
-2k_\text{NS}+k_s\big)\overline{P}+\omega^p+\kappa^p+d}
\nonumber\\[0.5em]
&=\frac{d\big(\partial_i\overline{P}\big)}{\big(\!
-3k_\text{NS}+k_s\big)
\partial_i\overline{P}+\omega^p_i+\kappa_i^p}\nonumber\\[0.5em]
&= \frac{d\overline{U_iU_j}}{\big(\!
-2k_\text{NS}+k_s\big)\overline{U_iU_j}+\omega_{ij}+\kappa_{ij}+c_{ij}}
\quad\;\nonumber\\[0.5em]
&=\frac{d\overline{U_iU_jU_k}}{\big(\!
-3k_\text{NS}+k_s\big)\overline{U_iU_jU_k}+\omega_{ijk}+\kappa_{ijk}+
c_{ijk}}\nonumber\\[0.5em]
&=\frac{d\overline{U_i\partial_j P}}{\big(\!
-4k_\text{NS}+k_s\big)\overline{U_i\partial_j P}+\omega_{ij}^p+\kappa_{ij}^p}\nonumber\\[0.5em]
&=\frac{d\overline{\partial_k U_i\cdot\partial_k U_j}}{\big(\!
-4k_\text{NS}+k_s\big)\overline{\partial_k U_i\cdot\partial_k
U_j}+\omega_{ij}^\nu+\kappa_{ij}^\nu}, \label{160502:1454}
\end{align}
where the functional $\omega$-extensions, resulting from the
linear superposition symmetry $\bar{T}^\prime_{+}$
\eqref{160501:1952}, are given as
\begin{gather}
\left.
\begin{aligned}
&\omega_i=0,\\[0.5em]
&\omega_{ij}=2k_{{\scriptscriptstyle
z}11}x_2\delta_{1i}\delta_{1j}+2k_{{\scriptscriptstyle
z}22}x_2\delta_{2i}\delta_{2j}+2k_{{\scriptscriptstyle
z}33}x_2\delta_{3i}\delta_{3j}+k_{{\scriptscriptstyle
z}12}x_2\big(\delta_{1i}\delta_{2j}+\delta_{1j}\delta_{2i}\big),\\[0.5em]
&\omega_{ijk}=0,\qquad\omega^p=-k_{{\scriptscriptstyle z}12}x_1
-2k_{{\scriptscriptstyle
z}22}x_2,\qquad\omega^p_i=-k_{{\scriptscriptstyle z}12}\delta_{1i}
-2k_{{\scriptscriptstyle
z}22}\delta_{2i},\\[0.5em]
&\omega_{ij}^p=0,\qquad \omega_{ij}^\nu=0,
\end{aligned}
~~~\right\}
\end{gather}
and the functional $\kappa$-extensions, resulting again from the
translation symmetry $\bar{T}_{\bar{U}_1}$ \eqref{160424:1452}, as
\begin{gather}
\left.
\begin{aligned} &\kappa_i=k_{\bar{U}_1}\delta_{1i},\qquad
\kappa_{ij}=\kappa_i\overline{U_j}+\kappa_j\overline{U_i},\\[0.5em]
&\kappa_{ijk}= \kappa_{ij}\overline{U_k}
+\kappa_{ik}\overline{U_j}+\kappa_{jk}\overline{U_i}\\
&\hspace{2.07cm}+\kappa_i\left(\overline{U_jU_k}-2\,\overline{U_j}\:\overline{U_k}\right)+
\kappa_j\left(\overline{U_iU_k}-2\,\overline{U_i}\:\overline{U_k}\right)+
\kappa_k\left(\overline{U_iU_j}-2\,\overline{U_i}\:\overline{U_j}\right),\\[0.0em]
&\kappa^p=0,\qquad\kappa^p_i=0,\qquad\kappa_{ij}^p=\kappa_i\frac{\partial\overline{P}}{\partial
x_j},\qquad \kappa_{ij}^\nu=0.
\end{aligned}
~~~\right\}
\end{gather}
When comparing the full-field invariant surface condition
\eqref{160502:1454} with the correspondingly given
Reynolds-decomposed condition [Eq.$\,$(354)] in
\cite{Oberlack15Rev}, one can recognize that in the one-point
limit ($\vr\to\boldsymbol{0}$) both conditions are indeed
equivalent, except on three points:

(i)\label{(i)} Instead of the two independent translation
symmetries $\bar{T}_{\bar{U}_1}$ \eqref{160430:2131} and
$\bar{T}^\prime_c$ \eqref{150424:1633}, the equivalent set of
transformations $\bar{T}_{\bar{U}_1}$ and
$\bar{T}_{\text{tr},1}:=\bar{T}^\prime_c|_{c_1=k_{\text{tr},1}}
\circ\bar{T}_{\bar{U}_1}|_{k_{\bar{U}_1}=-k_{\text{tr},1}}$ has
been used in \cite{Oberlack15Rev}. For more details see also
\cite{Rosteck14} [pp.$\,$228-229].

(ii)\label{(ii)} The invariant (symmetry breaking) constraint
$\bar{U}_2^{*}=\bar{U}_2$ of a constant wall-normal transpiration
velocity $\bar{U}_2=v_0$ (denoted in \cite{Oberlack15Rev} as
$U_T$) has already been directly implemented in both scaling
symmetries $\bar{T}^\prime_s$ \eqref{160424:1418} and
$\bar{T}_{\text{NS}}$ \eqref{160502:1424}, namely by transferring
these full-field symmetries back to their corresponding
Reynolds-decomposed form under the separate conditions $k_s\neq 0$
and $k_{\text{NS}}\neq 0$. Such a procedure, however, is based on
a fallacy, as we will show further below, because consistency
reveals that $k_s$ and $k_{\text{NS}}$ each must be zero, i.e.,
when imposing the constraint $\bar{U}_2^{*}=\bar{U}_2$, the
symmetry breaking cannot be circumvented, no matter which {\it
modus~operandi} is applied.

(iii)\label{(iii)} The in\-finites\-imal generator for $R_{12}$ in
[Eq.$\,$(354)] in \cite{Oberlack15Rev} contains two misprints: The
term ``$-k_{z,2}x_2U_T$" has to be deleted, since a parameter such
as $k_{z,2}$ does not exist, neither in the considered symmetries
nor in the derived invariant solutions [Eq.$\,$(357)] and
[Eq.$\,$(363)]. Similar for the misprinted parameter
``$k_{\text{sc},\text{tr}2}$", which should be replaced by
$k_{\text{NS}}$. Both misprints also appear in \cite{Rosteck14}
[p.$\,$228].

As also already outlined in the previous subsection, we recall
again that before invariant solutions get determined from
\eqref{160502:1454}, we first have to ensure the invariance of two
enclosed system constraints: That of a mean constant wall-normal
velocity $\overline{U_2}=v_0$, or, equivalently
$d\overline{U_2}=0$, and that of a mean constant streamwise
pressure gradient $\partial\overline{P}/\partial x_1=-K$, or,
equivalently $d(\partial_1\overline{P})=0$. Implementing these
into \eqref{160502:1454} will then collectively result into the
following symmetry breaking constraints
\begin{equation}
-k_\text{NS}+k_s=0, \quad c_2=0,\;\;\text{and}\;\;
-3k_\text{NS}+k_s=0,\quad \omega^p_1=0,\label{160603:2239}
\end{equation}
which leads us to the equivalent restrictions
\begin{equation}
k_\text{NS}=0,\quad k_s=0,\quad c_2=0,\quad k_{{\scriptscriptstyle
z}12}=0.\label{160603:2245}
\end{equation}
Consequently, the only invariant structure that can be derived for
the mean velocity profile $\overline{U_1}$ from
\eqref{160502:1454} is that of a linear function
\begin{equation}
\overline{U_1}(x_2)=\alpha\cdot x_2+\beta,\label{160612:1806}
\end{equation}
which, of course, does not constitute a reasonable scaling law
(where $\alpha=(k_{\bar{U}_1}+c_1)/k_{x_2}$ and $\beta$~some
arbitrary integration constant). Hence, in contrast to the
inviscid symmetry analysis performed in the previous subsection
which only fails at a higher-order moment, the current viscous
analysis already fails at the lowest-order moment
$\overline{U_1}$. The reason is that the viscous analysis misses
out one scaling symmetry: Instead of three inviscid scaling
symmetries, $\bar{T}_1$ \eqref{160425:1017}, $\bar{T}_2$
\eqref{160502:1402} and $\bar{T}^\prime_s$ \eqref{160424:1418}, we
only face two possible scaling symmetries for the viscous case,
namely $\bar{T}_{\text{NS}}$ \eqref{160502:1424} and again
$\bar{T}^\prime_s$ \eqref{160424:1418}, which turns out to be
crucial when at least two independent symmetry breaking
constraints are imposed, as can be seen in \eqref{160603:2239}, or
\eqref{160603:2245}, where both scaling symmetries then get
broken.

Although due to the symmetry breaking \eqref{160603:2245} no
logarithmic or algebraic scaling for the mean velocity profile
$\overline{U_1}$ can be derived, the corresponding analysis
carried out in \cite{Oberlack15Rev}, however, nevertheless
succeeded to do so. The mistake lies in the fallacy already
pointed out in \hyperref[(ii)]{(ii)} above. To comprehend the
mistake that has been done in \cite{Oberlack15Rev}, let us repeat
their line of reasoning by first looking at the scaling symmetry
$\bar{T}_{\text{NS}}$ \eqref{160502:1424} in how the invariant
(symmetry breaking) constraint $\bar{U}_2^{*}=\bar{U}_2$ of a
constant wall-normal transpiration velocity $\bar{U}_2=U_T$ has
been implemented under the condition of a non-zero group parameter
$k_{\text{NS}}\neq 0$ associated to that symmetry. The details can
also be found in \cite{Rosteck14} [p.$\,$229]. Although the same
line of reasoning has also been used for the second scaling
symmetry $\bar{T}^\prime_s$ \eqref{160424:1418}, we will discuss
it separately, due to being a special symmetry.

The starting point are the {\it full-field} statistical transport
equations \eqref{160502:1240}-\eqref{150503:1037}, where we
explicitly insert the constraint of a constant wall-normal
transpiration velocity $\overline{U_2}(x_2)=U_T$:
\begin{gather}
\frac{\partial \overline{U_1U_2}}{\partial x_2}+\frac{\partial
\overline{P}}{\partial
x_1}-\nu\frac{\partial^2\overline{U_1}}{\partial x_2^2}=0,\qquad
\frac{\partial \overline{U_2U_2}}{\partial x_2}+\frac{\partial
\overline{P}}{\partial x_2}=0,\qquad
\overline{U_1U_3}=\overline{U_2U_3}=0,
\label{160604:1036}\\[0.5em]
\left.
\begin{aligned}
\frac{\partial\overline{U_1U_2U_2}}{\partial x_2}
+\overline{\frac{\partial P}{\partial
x_1}U_2}+\overline{U_1\frac{\partial P}{\partial x_2}}
-\nu\frac{\partial^2\overline{U_1U_2}}{\partial x_2^2}+2\nu
\overline{\frac{\partial U_1}{\partial x_k}\frac{\partial
U_2}{\partial x_k}}
=0,\hspace{0.7cm}\\[0.5em]
\frac{\partial\overline{U_iU_jU_2}}{\partial x_2}
+\overline{\frac{\partial P}{\partial
x_i}U_j}+\overline{U_i\frac{\partial P}{\partial
x_j}}-\nu\frac{\partial^2\overline{U_iU_j}}{\partial x_2^2}+2\nu
\overline{\frac{\partial U_i}{\partial x_k}\frac{\partial
U_j}{\partial x_k}}=0,\;\text{for $i=j$}.
\end{aligned}
~~~\right\} \label{160604:1037}
\end{gather}
As a result, these equations do not dependent anymore on the mean
wall-normal velocity $\overline{U_2}$, and this observation is
true for all orders in the infinite hierarchy of equations. Hence,
based on the scaling symmetry $\bar{T}_{\text{NS}}$
\eqref{160502:1424} for the initial system
\eqref{160502:1240}-\eqref{150503:1037}, we now may consider a
modified symmetry $\bar{Q}_{\text{NS}}$ that already inherently
respects the required invariant
constraint~$\overline{U_2}^{\,*}=\overline{U_2}=U_T$:
\begin{align}
\bar{Q}_{\text{NS}}: &\quad x_i^*=e^{q_\text{NS}}x_i,\;\;\;\;
\overline{U_1}^{\, *}=e^{-q_\text{NS}\,}\overline{U_1},\;\;\;\;
\overline{U_2}^{\,*}=\overline{U_2},\;\;\;\; \overline{P}^{\,
*}=e^{-2q_\text{NS}\,}\overline{P},\;\;\;\; \overline{U_iU_j}^{\,
*}=e^{-2q_\text{NS}\,}\overline{U_iU_j},\nonumber\\[0.5em]
&\quad \overline{U_iU_jU_k}^{\,
*}=e^{-3q_\text{NS}\,}\overline{U_iU_jU_k},\;\;\;\;
\overline{U_i\frac{\partial P}{\partial x_j}}^{\,
*}=e^{-4q_\text{NS}}\, \overline{U_i\frac{\partial P}{\partial
x_j}},\nonumber\\[0.5em]
&\quad \overline{\frac{\partial U_i}{\partial x_k}\frac{\partial
U_j}{\partial x_k}}^{\, *}
=e^{-4q_\text{NS}}\,\overline{\frac{\partial U_i}{\partial
x_k}\frac{\partial U_j}{\partial x_k}}.\label{160604:1059}
\end{align}
Indeed, transformation \eqref{160604:1059} is admitted as a
symmetry by the equations \eqref{160604:1036}-\eqref{160604:1037}
as can be readily verified. However, important to note here is
that $\bar{Q}_{\text{NS}}$ \eqref{160604:1059} is a different
scaling symmetry than the initially considered
$\bar{T}_{\text{NS}}$ \eqref{160502:1424}, i.e., the latter
symmetry cannot be reduced to the former one. With
$\bar{Q}_{\text{NS}} \eqref{160604:1059}$ we obtained a symmetry
that automatically obeys the invariant constraint
$\overline{U_2}^{\,*}=\overline{U_2}=U_T$ without breaking the
group parameter $q_\text{NS}$ down to zero;\linebreak a result
impossible to achieve with the initial scaling symmetry
$\bar{T}_{\text{NS}}$ \eqref{160502:1424}. Hence, when generating
invariant solutions under the constraint
$\overline{U_2}^{\,*}=\overline{U_2}=U_T$, the Navier-Stokes
scaling symmetry $\bar{T}_{\text{NS}}$ \eqref{160502:1424} has to
be replaced by its appropriate but non-linked modification
$\bar{Q}_{\text{NS}}$~\eqref{160604:1059}.

In its equivalent Reynolds-decomposed form, the symmetry
$\bar{Q}_{\text{NS}}$~\eqref{160604:1059} reads
\citep{Rosteck14,Oberlack15Rev}:
\begin{align}
\bar{Q}_{\text{NS}}: &\quad x_i^*=e^{q_\text{NS}}x_i,\;\;\;\;
\bar{U}_i^{
*}=e^{-q_\text{NS}}\bar{U}_1\delta_{1i}+U_T\delta_{2i},\;\;\;\;
\bar{P}^{*}=e^{-2q_\text{NS}}\bar{P},\nonumber\\[0.5em]
&\quad \tau_{ij}^{*}=
e^{-2q_\text{NS}}\tau_{ij}+e^{-2q_\text{NS}}\bar{U}_i\bar{U}_j -
\bar{U}_i^*\bar{U}_j^*,\nonumber\\
&\quad \phantom{\tau_{ij}^{*}}=e^{-2q_\text{NS}}\tau_{ij}+
\Big(e^{-2q_\text{NS}}-e^{-q_\text{NS}}\Big)\bar{U}_1U_T
\Big(\delta_{1i}\delta_{2j}+\delta_{1j}\delta_{2i}\Big)
+\Big(e^{-2q_\text{NS}}-1\Big)U_T^2\delta_{2i}\delta_{2j}
,\nonumber\\[0.5em]
&\quad \tau_{ijk}^{*}=e^{-3q_\text{NS}}\tau_{ijk}
+e^{-3q_\text{NS}}\Big(\bar{U}_i\bar{U}_j\bar{U}_k+
\bar{U}_i\tau_{jk}+\bar{U}_j\tau_{ik}+ \bar{U}_k\tau_{ij}\Big)
\nonumber\\
&\quad\hspace{5.72cm} -\bar{U}^*_i\bar{U}^*_j\bar{U}^*_k-
\bar{U}^*_i\tau^*_{jk}-\bar{U}^*_j\tau^*_{ik}-\bar{U}^*_k\tau^*_{ij}
,\nonumber\\[0.5em]
&\quad\overline{u_i\frac{\partial p}{\partial x_j}}^{\,
*}=e^{-4q_\text{NS}\,}\overline{u_i\frac{\partial p}{\partial
x_j}}+ e^{-4q_\text{NS}\,}
\bar{U}_i\frac{\partial\bar{P}}{\partial
x_j}-\bar{U}^*_i\frac{\partial\bar{P}^*}{\partial x^*_j}
,\nonumber\\[0.5em]
&\quad \varepsilon_{ij}^{*}
=e^{-4q_\text{NS}}\,\varepsilon_{ij}+e^{-4q_\text{NS}}\,
2\nu\frac{\partial \bar{U}_i}{\partial x_2}\frac{\partial
\bar{U}_j}{\partial x_2}-2\nu\frac{\partial \bar{U}^*_i}{\partial
x^*_2}\frac{\partial \bar{U}^*_j}{\partial x^*_2}
,\label{160604:1325}
\end{align}
which indeed is a symmetry of the corresponding
Reynolds-decomposed transport equations
\eqref{160502:1320}-\eqref{160604:1332} for $\bar{U}_2=U_T$:
\begin{gather}
U_T\frac{\partial\bar{U}_1}{\partial
x_2}+\frac{\partial\bar{P}}{\partial x_1}+\frac{\partial
\tau_{12}}{\partial x_2}-\nu\frac{\partial^2\bar{U}_1}{\partial
x_2^2}=0,\qquad \frac{\partial\bar{P}}{\partial
x_2}+\frac{\partial
\tau_{22}}{\partial x_2}=0,\qquad \tau_{13}=\tau_{23}=0,\label{160604:1333}\\[0.5em]
\left.
\begin{aligned}
U_T\frac{\partial \tau_{12}}{\partial
x_2}+\frac{\partial\tau_{122}}{\partial
x_2}+\tau_{22}\frac{\partial\bar{U}_1}{\partial x_2}+
\overline{\frac{\partial p}{\partial
x_1}u_2}+\overline{u_1\frac{\partial p}{\partial x_2}}
-\nu\frac{\partial^2\tau_{12}}{\partial x_2^2}
+\varepsilon_{12}=0,\hspace{1.15cm}\\[0.5em]
\!\!U_T\frac{\partial \tau_{ij}}{\partial
x_2}+\frac{\partial\tau_{ij2}}{\partial
x_2}+\tau_{i2}\frac{\partial\bar{U}_j}{\partial
x_2}+\tau_{j2}\frac{\partial\bar{U}_i}{\partial x_2}+
\overline{\frac{\partial p}{\partial
x_i}u_j}+\overline{u_i\frac{\partial p}{\partial
x_j}}-\nu\frac{\partial^2\tau_{ij}}{\partial x_2^2}
+\varepsilon_{ij}=0,\;\text{for $i=j$}.
\end{aligned}
~~~\right\}\label{160604:1334}
\end{gather}
Although $\bar{Q}_{\text{NS}}$ \eqref{160604:1325} is
mathematically correctly admitted as a symmetry transformation by
the infinite and unclosed system of statistical equations
\eqref{160604:1333}-\eqref{160604:1334}, it nevertheless has to be
checked whether this symmetry is also consistent with the
underlying deterministic Navier-Stokes equations, in particular
because $\bar{Q}_{\text{NS}}$ \eqref{160604:1325} acts as a purely
statistical symmetry which is not reflected in the original
deterministic equations. Hence, it is necessary to check whether
this symmetry violates the principle of cause and effect. As
explained and discussed in
\cite{Frewer14.1,Frewer15.1,Frewer16.1}, no violation of causality
occurs if at least one (invertible) deterministic transformation
$\mathcal{Q}_{\text{NS}}$ of the Navier-Stokes equations can be
found such that then the symmetry $\bar{Q}_{\text{NS}}$
\eqref{160604:1325} is induced on the statistical level, i.e., $\L
\mathcal{Q}_\text{NS}\R=\bar{Q}_{\text{NS}}$, where $\L\cdot\R$
denotes any statistical averaging operator. Important to note here
is that the deterministic cause $\mathcal{Q}_{\text{NS}}$ itself
need {\it not} to be symmetry in order to induce the statistical
symmetry $\bar{Q}_{\text{NS}}$ as an effect.

The aim is to find at least one (invertible) deterministic
transformation $\mathcal{Q}_{\text{NS}}$ (which itself need not to
be a symmetry) of the Navier-Stokes equations
\begin{align}
\mathcal{Q}_{\text{NS}}: &\quad
t^*=t^*(t,x_i,\bar{U}_i,\bar{P},u_i,p),\quad x_i^*=e^{q_\text{NS}}
x_i,\quad
\bar{U}_i^*=e^{-q_\text{NS}}\bar{U}_1\delta_{1i}+\bar{U}_2\delta_{2i},\quad
\bar{U}_3^*=\bar{U}_3=0,\quad\nonumber\\[0.5em]
&\quad u_i^*=u_i^*(t,x_i,\bar{U}_i,\bar{P},u_i,p),\quad
\bar{P}^*=e^{-2q_\text{NS}}\bar{P},\quad
p^*=p^*(t,x_i,\bar{U}_i,\bar{P},u_i,p),\label{160604:2117}
\end{align}
such that it induces the statistical symmetry
$\bar{Q}_{\text{NS}}$ \eqref{160604:1325}, i.e., such that $\L
\mathcal{Q}_\text{NS}\R=\bar{Q}_{\text{NS}}$.  We will restrict
the analysis only to point transformations, where the
transformations for the fluctuations $u_i^*$ and $p^*$, as well as
for the time $t^*$, are unknown transformations that need to be
determined. We start off with the symmetry transformation of
$\tau^*_{33}$, for which, according to \eqref{160604:1325}, the
transformed fluctuation $u_3^*$ has to be the deterministic cause
for the statistical symmetry-effect
\begin{equation}
\L u_3^{* 2}\R = e^{-2q_\text{NS}} \L u_3^2\R,\label{160605:1141}
\end{equation}
which can only be satisfied if $u_3^*$ transforms as
\begin{equation}
u_3^*=e^{-q_\text{NS}}u_3.\label{160605:0851}
\end{equation}
Then by considering the symmetry transformation of $\tau_{23}^*$
\eqref{160604:1325}
\begin{equation}
\L u_2^*u_3^*\R = e^{-2q_\text{NS}} \L u_2u_3\R,
\end{equation}
this effect, when incorporating the previous result
\eqref{160605:0851}, can only be caused by
\begin{equation}
u_2^*=e^{-q_\text{NS}}u_2,
\end{equation}
but which then is inconsistent to the effect observed by
$\tau_{22}^*$ \eqref{160604:1325}
\begin{equation}
\L u_2^{* 2}\R = e^{-2q_\text{NS}} \L u_2^2\R
+\big(e^{-2q_\text{NS}}-1\big)U_T^2.
\end{equation}
As can be readily seen, a consistent transformation can only be
achieved if
\begin{equation}
\big(e^{-2q_\text{NS}}-1\big)U_T^2=0,\label{160605:1142}
\end{equation}
and since $U_T\neq 0$, we thus yield the result
\begin{equation}
q_\text{NS}=0.
\end{equation}
Hence, for $q_\text{NS}\neq 0$, the statistical symmetry
$\bar{Q}_{\text{NS}}$~\eqref{160604:1325} is violating the
classical principle of cause and effect, since obviously no
deterministic cause $\mathcal{Q}_\text{NS}$~\eqref{160604:2117}
can be found that statistically induces this symmetry-effect
$\bar{Q}_{\text{NS}}$~\eqref{160604:1325}. The statistical
symmetry $\bar{Q}_{\text{NS}}$~\eqref{160604:1325} is thus
inconsistent to its underlying deterministic theory, and can only
be restored if $q_\text{NS}=0$, i.e. if the symmetry gets broken.

The same line of reasoning also applies to the second scaling
symmetry $\bar{T}^\prime_s$ \eqref{160424:1418}. Based on this
symmetry for the initial system
\eqref{160502:1240}-\eqref{150503:1037}, the analysis in
\cite{Oberlack15Rev} considers again a modified symmetry
$\bar{Q}_s$ such that it already inherently respects again the
required invariant
constraint~$\overline{U_2}^{\,*}=\overline{U_2}=U_T$:
\begin{align}
\bar{Q}_{s}: &\quad x_i^*=x_i,\;\;\;\; \overline{U_1}^{\,
*}=e^{q_s}\overline{U_1},\;\;\;\;
\overline{U_2}^{\,*}=\overline{U_2},\;\;\;\; \overline{P}^{\,
*}=e^{q_s}\overline{P},\;\;\;\; \overline{U_iU_j}^{\,
*}=e^{q_s}\overline{U_iU_j},\nonumber\\[0.5em]
&\quad \overline{U_iU_jU_k}^{\,
*}=e^{q_s}\overline{U_iU_jU_k},\;\;\;\;
\overline{U_i\frac{\partial P}{\partial x_j}}^{\, *}=e^{q_s}
\overline{U_i\frac{\partial P}{\partial
x_j}},\nonumber\\[0.5em]
&\quad \overline{\frac{\partial U_i}{\partial x_k}\frac{\partial
U_j}{\partial x_k}}^{\, *} =e^{q_s}\overline{\frac{\partial
U_i}{\partial x_k}\frac{\partial U_j}{\partial
x_k}},\label{160605:0958}
\end{align}
which indeed is a symmetry of the considered full-field system
\eqref{160604:1036}-\eqref{160604:1037}. In its equivalent
Reynolds-decomposed form, this symmetry reads
\citep{Rosteck14,Oberlack15Rev}:
\begin{align}
\bar{Q}_{s}: &\quad x_i^*=x_i,\;\;\;\; \bar{U}_i^{
*}=e^{q_s}\bar{U}_1\delta_{1i}+U_T\delta_{2i},\;\;\;\;
\bar{P}^{*}=e^{q_s}\bar{P},\nonumber\\[0.5em]
&\quad \tau_{ij}^{*}= e^{q_s}\tau_{ij}+e^{q_s}\bar{U}_i\bar{U}_j -
\bar{U}_i^*\bar{U}_j^*,\nonumber\\
&\quad \phantom{\tau_{ij}^{*}}= e^{q_s}\tau_{ij}+
\big(e^{q_s}-e^{2q_s}\big)\bar{U}_1^2 \delta_{1i}\delta_{1j}
+\big(e^{q_s}-1\big)U_T^2\delta_{2i}\delta_{2j}
,\nonumber\\[0.5em]
&\quad \tau_{ijk}^{*}=e^{q_s}\tau_{ijk}
+e^{q_s}\Big(\bar{U}_i\bar{U}_j\bar{U}_k+
\bar{U}_i\tau_{jk}+\bar{U}_j\tau_{ik}+ \bar{U}_k\tau_{ij}\Big)
\nonumber\\
&\quad\hspace{4.56cm} -\bar{U}^*_i\bar{U}^*_j\bar{U}^*_k-
\bar{U}^*_i\tau^*_{jk}-\bar{U}^*_j\tau^*_{ik}-\bar{U}^*_k\tau^*_{ij}
,\nonumber\\[0.5em]
&\quad\overline{u_i\frac{\partial p}{\partial x_j}}^{\,
*}=e^{q_s}\overline{u_i\frac{\partial p}{\partial x_j}}+ e^{q_s}
\bar{U}_i\frac{\partial\bar{P}}{\partial
x_j}-\bar{U}^*_i\frac{\partial\bar{P}^*}{\partial x^*_j}
,\nonumber\\[0.5em]
&\quad \varepsilon_{ij}^{*}
=e^{q_s}\varepsilon_{ij}+e^{q_s}2\nu\frac{\partial
\bar{U}_i}{\partial x_2}\frac{\partial \bar{U}_j}{\partial
x_2}-2\nu\frac{\partial \bar{U}^*_i}{\partial x^*_2}\frac{\partial
\bar{U}^*_j}{\partial x^*_2} ,\label{160605:1035}
\end{align}
which indeed is also a symmetry of the corresponding
Reynolds-decomposed transport equations
\eqref{160604:1333}-\eqref{160604:1334}. However, as in the
previous case for $\bar{Q}_\text{NS}$ \eqref{160604:1325},
although the second scaling
transformation~$\bar{Q}_s$~\eqref{160605:1035} is also
mathematically correctly admitted as a symmetry by its statistical
equations, it nevertheless is inconsistent to its underlying
deterministic description, too, since, also in this case, no
deterministic cause
\begin{align}
\mathcal{Q}_s: &\quad t^*=t^*(t,x_i,\bar{U}_i,\bar{P},u_i,p),\quad
x_i^*=x_i,\quad
\bar{U}_i^*=e^{q_s}\bar{U}_1\delta_{1i}+\bar{U}_2\delta_{2i},\quad
\bar{U}_3^*=\bar{U}_3=0,\qquad\nonumber\\[0.5em]
&\quad u_i^*=u_i^*(t,x_i,\bar{U}_i,\bar{P},u_i,p),\quad
\bar{P}^*=e^{q_s}\bar{P},\quad
p^*=p^*(t,x_i,\bar{U}_i,\bar{P},u_i,p),\label{160605:1112}
\end{align}
can be found such that on its statistical level the symmetry
$\bar{Q}_s$~\eqref{160605:1035} can be observed, that is, such
that $\L \mathcal{Q}_s\R=\bar{Q}_s$, where the deterministic cause
$\mathcal{Q}_s$~\eqref{160605:1112}, of course, need not to be
symmetry of the Navier-Stokes equations itself, in order to induce
a symmetry as a statistical effect. Following the same procedure
as outlined in \eqref{160605:1141}-\eqref{160605:1142} for
$\bar{Q}_\text{NS}$ \eqref{160604:1325}, one again readily sees
that the statistical scaling symmetry
$\bar{Q}_s$~\eqref{160605:1035} only can be made consistent to its
underlying deterministic description if $q_s=0$. Hence, as in the
full-field representation, where we obtained the symmetry-breaking
result \eqref{160603:2245}
\begin{equation}
k_\text{NS}=0,\qquad k_s=0,
\end{equation}
for the two scaling symmetries $\bar{T}_{\text{NS}}$
\eqref{160502:1424} and $\bar{T}^\prime_s$ \eqref{160424:1418}, we
thus also obtain the equivalent result in the Reynolds-decomposed
representation, namely that both correspondingly modified scaling
symmetries $\bar{Q}_\text{NS}$ \eqref{160604:1325} and
$\bar{Q}_s$~\eqref{160605:1035} each must get broken
\begin{equation}
q_\text{NS}=0,\qquad q_s=0,
\end{equation}
when imposing the invariant constraint
$\bar{U}_2^{*}=\bar{U}_2=U_T$ in a consistent manner. Obviously,
this constitutes a plausible result, because the full-field and
the Reynolds-decomposed representation are ultimately equivalent
to each other: Both must give the same mathematical and physical
results with the same conclusions. Worthwhile to note in this
regard is that, in contrast to the classical Navier-Stokes scaling
symmetry $\bar{T}_{\text{NS}}$~\eqref{160502:1424}, which
constitutes a consistent and well-defined symmetry, the new
statistical scaling symmetry $\bar{T}^\prime_s$
\eqref{160424:1418}, as first proposed in \cite{Khujadze04} and
then later generalized in \cite{Oberlack10}, is already
inconsistent and thus unphysical by itself. For more details, we
refer to \cite{Frewer14.1,Frewer15.1}.

For the sake of completeness, let us continue the inconsistent
analysis as performed in \cite{Oberlack15Rev}. This will lead us
to another, independent mistake done therein. When~re\-writing the
full-field invariant surface condition \eqref{160502:1454} into
its Reynolds-decomposed form as proposed in \cite{Oberlack15Rev},
namely by replacing the two full-field scaling symmetries
$\bar{T}^\prime_s$~\eqref{160424:1418} and
$\bar{T}_{\text{NS}}$~\eqref{160502:1424} with their
correspondingly modified Reynolds-decomposed scaling symmetries
$\bar{Q}_s$~\eqref{160605:1035} and $\bar{Q}_\text{NS}$
\eqref{160604:1325}, respectively, we obtain, for $q_s\neq 0$ and
$q_\text{NS}\neq 0$, the following (inconsistent) invariant
surface condition respecting the invariant constraint
$\bar{U}_2^{*}=\bar{U}_2=U_T$:
\begin{align}
\frac{dx_1}{q_\text{NS}x_1+k_{x_1}}=\frac{dx_2}{q_\text{NS}x_2+k_{x_2}}
&=\frac{d\bar{U}_i}{q_\text{NS}\phi_{i}+q_s\psi_i
+\omega_i+\zeta_i+k_{\bar{U}_1}\delta_{1i}}=
\frac{d\bar{P}}{q_\text{NS}\phi^p+q_s\psi^p+\omega^p+\zeta^p}
\nonumber\\[0.5em]
&=\frac{d\big(\partial_i\bar{P}\big)}{q_\text{NS}\phi^p_i+q_s\psi^p_i+\omega^p_i+\zeta^p_i}=
\frac{d\tau_{ij}}{q_\text{NS}\phi_{ij}+q_s\psi_{ij}+\omega_{ij}+\zeta_{ij}}
\quad\;\nonumber\\[0.5em]
&=\frac{d\tau_{ijk}}{q_\text{NS}\phi_{ijk}+q_s\psi_{ijk}+\omega_{ijk}+\zeta_{ijk}}=\frac{d\overline{u_i\partial_j
p}}{q_\text{NS}\phi^p_{ij}+q_s\psi^p_{ij}
+\omega_{ij}^p+\zeta_{ij}^p}\nonumber\\[0.5em]
&=\frac{d\varepsilon_{ij}}{q_\text{NS}\phi^\nu_{ij}+q_s\psi^\nu_{ij}
+\omega_{ij}^\nu+\zeta_{ij}^\nu}, \label{160605:1457}
\end{align}
which is identical to result [Eq.$\,$(354)]\footnote[2]{Up to a
non-essential linear combination in the translation symmetries and
two misprints in the generator $R_{12}$, as mentioned in the
points \hyperref[(i)]{(i)} and \hyperref[(iii)]{(iii)} in the
beginning of this subsection (p.$\,$\pageref{(i)}), respectively.
Further note that the correspondence of the parameters used in
\eqref{160605:1457}-\eqref{160606:1207} to the ones defined in
\cite{Oberlack15Rev}~is: $k_{x_2}=k_{G,2}$,
$q_\text{NS}=k_\text{NS}$, $q_s=k_s$, $c_1=k_{\text{tr,1}}$,
$c_{ij}=k_{ij}$, and $k_{\bar{U}_1}=k_{{\scriptscriptstyle z}1}$.}
given in \cite{Oberlack15Rev}, where the $\phi$-terms result from
the scaling symmetry $\bar{Q}_\text{NS}$ \eqref{160604:1325}
hierarchically given as
\begin{gather}
\left.
\begin{aligned} & \phi_i=-\bar{U}_1\delta_{1i},\qquad
\phi_{ij}=-2\tau_{ij}-2\bar{U}_i\bar{U}_j-\phi_i\bar{U}_j-\phi_j\bar{U}_i,\\[0.5em]
&\phi_{ijk}=-3\tau_{ijk}-3\Big(\bar{U}_i\bar{U}_j\bar{U}_k+\bar{U}_i\tau_{jk}
+\bar{U}_j\tau_{ik}+\bar{U}_k\tau_{ij}
\Big)-\phi_{ij}\bar{U}_k-\phi_{ik}\bar{U}_j-\phi_{jk}U_i\\
&\hspace{4.35cm}-\phi_i\left(\tau_{jk}+\bar{U}_j\bar{U}_k\right)-
\phi_j\left(\tau_{ik}+\bar{U}_i\bar{U}_k\right)-
\phi_k\left(\tau_{ij}+\bar{U}_i\bar{U}_j\right),\\[0.25em]
&\phi^p=-2\bar{P},\qquad
\phi^p_i=-3\partial_i\bar{P},\qquad\phi_{ij}^p=-4\overline{u_i\partial_j
p}-\bar{U}_i\frac{\partial\bar{P}}{\partial
x_j}-\phi_i\frac{\partial\bar{P}}{\partial x_j},\\[0.5em]
&\phi_{ij}^\nu=-4\varepsilon_{ij}-4\nu\frac{\partial\bar{U}_i}{\partial
x_2}\frac{\partial\bar{U}_j}{\partial
x_2}-2\nu\frac{\partial\phi_i}{\partial
x_2}\frac{\partial\bar{U}_j}{\partial
x_2}-2\nu\frac{\partial\phi_j}{\partial
x_2}\frac{\partial\bar{U}_i}{\partial x_2},
\end{aligned}
~~~\right\}
\end{gather}
the $\psi$-terms from the scaling symmetry
$\bar{Q}_s$~\eqref{160605:1035}
\begin{gather}
\left.
\begin{aligned} & \psi_i=\bar{U}_1\delta_{1i},\qquad
\psi_{ij}=\tau_{ij}+\bar{U}_i\bar{U}_j-\psi_i\bar{U}_j-\psi_j\bar{U}_i,\\[0.5em]
&\psi_{ijk}=\tau_{ijk}+\bar{U}_i\bar{U}_j\bar{U}_k+\bar{U}_i\tau_{jk}
+\bar{U}_j\tau_{ik}+\bar{U}_k\tau_{ij}-\psi_{ij}\bar{U}_k-\psi_{ik}\bar{U}_j-\psi_{jk}U_i\\
&\hspace{3.48cm}-\psi_i\left(\tau_{jk}+\bar{U}_j\bar{U}_k\right)-
\psi_j\left(\tau_{ik}+\bar{U}_i\bar{U}_k\right)-
\psi_k\left(\tau_{ij}+\bar{U}_i\bar{U}_j\right),\\[0.25em]
&\psi^p=\bar{P},\qquad
\psi^p_i=\partial_i\bar{P},\qquad\psi_{ij}^p=\overline{u_i\partial_j
p}-\psi_i\frac{\partial\bar{P}}{\partial x_j},\\[0.5em]
&\psi_{ij}^\nu=\varepsilon_{ij}+2\nu\frac{\partial\bar{U}_i}{\partial
x_2}\frac{\partial\bar{U}_j}{\partial
x_2}-2\nu\frac{\partial\psi_i}{\partial
x_2}\frac{\partial\bar{U}_j}{\partial
x_2}-2\nu\frac{\partial\psi_j}{\partial
x_2}\frac{\partial\bar{U}_i}{\partial x_2},
\end{aligned}
~~~~~~~~~\right\}
\end{gather}
the $\omega$-terms from the linear superposition symmetry
$\bar{T}^\prime_{+}$ \eqref{160501:1952} with the specification
\eqref{160601:1953}
\begin{gather}
\left.
\begin{aligned}
&\omega_i=0,\\[0.5em]
&\omega_{ij}=2k_{{\scriptscriptstyle
z}11}x_2\delta_{1i}\delta_{1j}+2k_{{\scriptscriptstyle
z}22}x_2\delta_{2i}\delta_{2j}+2k_{{\scriptscriptstyle
z}33}x_2\delta_{3i}\delta_{3j}+k_{{\scriptscriptstyle
z}12}x_2\big(\delta_{1i}\delta_{2j}+\delta_{1j}\delta_{2i}\big),\\[0.5em]
&\omega_{ijk}=0,\qquad\omega^p=-k_{{\scriptscriptstyle z}12}x_1
-2k_{{\scriptscriptstyle
z}22}x_2,\qquad\omega^p_i=-k_{{\scriptscriptstyle z}12}\delta_{1i}
-2k_{{\scriptscriptstyle
z}22}\delta_{2i},\\[0.5em]
&\omega_{ij}^p=k_{{\scriptscriptstyle z}12}\bar{U}_i\delta_{1j}
+2k_{{\scriptscriptstyle z}22}\bar{U}_i\delta_{2j},\qquad
\omega_{ij}^\nu=0,
\end{aligned}
~~~~~~~~~~~\right\}
\end{gather}
and finally the $\zeta$-terms resulting from the translation
symmetry $\bar{T}^\prime_c$ \eqref{150424:1633}
\begin{gather}
\left.
\begin{aligned} &\zeta_i=c_1\delta_{1i},\qquad
\zeta_{ij}=-\zeta_i\bar{U}_j-\zeta_j\bar{U}_i+c_{ij},\\[0.5em]
&\zeta_{ijk}=-\zeta_{ij}\bar{U}_k
-\zeta_{ik}\bar{U}_j-\zeta_{jk}\bar{U}_i\\
&\hspace{2.26cm}-\zeta_i\left(\tau_{jk}+\bar{U}_j\bar{U}_k\right)-
\zeta_j\left(\tau_{ik}+\bar{U}_i\bar{U}_k\right)-
\zeta_k\left(\tau_{ij}+\bar{U}_i\bar{U}_j\right)+c_{ijk},\\[0.0em]
&\zeta^p=d,\qquad \zeta^p_i=0,\qquad
\zeta_{ij}^p=-\zeta_i\frac{\partial\bar{P}}{\partial x_j},\qquad
\zeta_{ij}^\nu=0.
\end{aligned}
~~~~~~~~~~~~\right\}\label{160606:1207}
\end{gather}
Note that \eqref{160605:1457} coherently extends the condition in
\cite{Oberlack15Rev} up to third order in the velocity
correlations, including the moments for pressure and dissipation.

Anyhow, although the required constraint of a mean constant and
invariant wall-normal velocity $\bar{U}_2^*=\bar{U}_2=U_T$ has
been (inconsistently) implemented into the invariant surface
condition \eqref{160605:1457} without breaking a scaling symmetry,
i.e. for $q_\text{NS}\neq 0$ and $q_s\neq 0$, this has not been
done for the second required system constraint, namely that of a
mean constant and invariant streamwise pressure gradient
$\partial\bar{P}^*/\partial x_1^*=\partial\bar{P}^/\partial
x_1=-K$. Because, when this constraint
$d(\partial_1\bar{P})=0$\linebreak is applied to
\eqref{160605:1457}, it will unavoidably result into the two
symmetry breaking constraints
\begin{equation}
q_\text{NS}\phi^p_1+q_s\psi^p_1=0,\;\;\text{and}\;\; \omega^p_1=0,
\end{equation}
which equivalently turn into the restrictions
\begin{equation}
q_s=3q_\text{NS},\;\;\text{and}\;\; k_{{\scriptscriptstyle
z}12}=0,\label{160606:1340}
\end{equation}
an important result not obtained in \cite{Oberlack15Rev}. The
reason of why this result \eqref{160606:1340} was not obtained, is
that in \cite{Oberlack15Rev} a second, independent mistake was
made: Instead of correctly determining the invariant mean pressure
gradient $\partial_i\bar{P}$ as a function of $x_2$ in the
wall-normal and as a constant in the streamwise direction via its
invariant surface condition~\eqref{160605:1457}, it was
incorrectly determined as a functional residual of the two
momentum~equations \eqref{160604:1333}, namely as
\begin{equation}
\frac{\partial\bar{P}}{\partial
x_1}=-U_T\frac{\partial\bar{U}^\text{inv}_1}{\partial
x_2}-\frac{\partial \tau^\text{inv}_{12}}{\partial
x_2}+\nu\frac{\partial^2\bar{U}^\text{inv}_1}{\partial
x_2^2},\;\;\text{and}\;\;\; \frac{\partial\bar{P}}{\partial
x_2}=-\frac{\partial \tau^\text{inv}_{22}}{\partial
x_2},\label{160606:2042}
\end{equation}
where the parameters for the already determined invariant
solutions of \eqref{160605:1457}, $\bar{U}_1^\text{inv}$,
$\tau^\text{inv}_{12}$ and~$\tau^\text{inv}_{22}$, were then
arranged such that $\partial\bar{P}/\partial x_1$ is a constant
and $\partial\bar{P}/\partial x_2$ only a function of $x_2$. The
reason why the latter procedure is incorrect, is that it is
decisively incomplete: The relations in \eqref{160606:2042} only
give constraint conditions among the parameters of the invariant
solutions $\bar{U}_1^\text{inv}$,
$\tau^\text{inv}_{12}$~and~$\tau^\text{inv}_{22}$, under the {\it
assumption} of a constant pressure gradient
$\partial\bar{P}/\partial x_1=-K$ in the streamwise and a sole
$x_2$-dependence $\partial\bar{P}/\partial x_2=\mathcal{G}(x_2)$
in the wall-normal direction. But, these relations do {\it not}
warrant that the determined pressure $\bar{P}=\bar{P}(x_1,x_2)$
from \eqref{160606:2042}, with gradient
$\partial_i\bar{P}=-K\delta_{1i}+\mathcal{G}(x_2)\delta_{2i}$,
constitutes an invariant function by itself, being compatible to
the invariant pressure solution
$\bar{P}^\text{\,inv}=\bar{P}^\text{\,inv}(x_1,x_2)$ obtained from
the invariant surface condition~\eqref{160605:1457} with gradient
$\partial_i\bar{P}^\text{\,inv}=-K\delta_{1i}+\mathcal{G}^\text{\,inv}(x_2)\delta_{2i}$.
In other words, the pressure solution $\bar{P}$ obtained from
\eqref{160606:2042} is in general not an invariant function under
all symmetries considered, and thus in general not compatible to
the invariant pressure solution $\bar{P}^\text{inv}$ obtained from
\eqref{160605:1457}.\linebreak Generally speaking, the reason for
this is that the residuals \eqref{160606:2042} do not constitute
invariant relations, since the coordinates $x_1$ and $x_2$
themselves do not constitute invariant quantities. Hence, instead
of the incomplete and thus in general incorrect relations
\eqref{160606:2042} as considered in \cite{Oberlack15Rev}, the
following complete and correct relations have to be inquired
\begin{equation}
\frac{\partial\bar{P}^\text{inv}}{\partial
x_1}=-U_T\frac{\partial\bar{U}^\text{inv}_1}{\partial
x_2}-\frac{\partial \tau^\text{inv}_{12}}{\partial
x_2}+\nu\frac{\partial^2\bar{U}^\text{inv}_1}{\partial
x_2^2},\qquad \frac{\partial\bar{P}^\text{inv}}{\partial
x_2}=-\frac{\partial \tau^\text{inv}_{22}}{\partial
x_2},\label{160606:2216}
\end{equation}
which now not only give the correct and consistent constraint
conditions among the parameters of {\it all} invariant solutions
involved, but which will also give, in general, {\it more}
constraint conditions than the (inconsistent) relations
\eqref{160606:2042} may give, simply because the invariant-based
system constraint for \eqref{160606:2216},
$\partial_i\bar{P}^\text{\,inv}=-K\delta_{1i}+\mathcal{G}^\text{\,inv}(x_2)\delta_{2i}$,
is in general more restrictive than the constraint
$\partial_i\bar{P}=-K\delta_{1i}+\mathcal{G}(x_2)\delta_{2i}$ for
\eqref{160606:2042}. For example, for the case presently studied,
\eqref{160606:2042}~will only give one non-zero-constraint,
[Eq.$\,$(361)] {\it or} [Eq.$\,$(365)]\footnote[2]{To note is that
the result for the invariant solution $\tilde{R}_{12}$
[Eq.$\,$(363)] in \cite{Oberlack15Rev} misses the summand $U_T
k_{{\scriptscriptstyle
z}1}k_{G,2}/(k_\text{NS}(k_{G,2}+k_\text{NS}x_2))$, but which
apparently was absorbed into the term
$C_{I,12}/(k_{G,2}+k_\text{NS}x_2)$ of the arbitrary integration
constant $C_{I,12}$, while the result for the invariant solution
$\tilde{R}_{22}$ carries the wrong sign in the $U^2_T$-term.\\
\phantom{x}} in \cite{Oberlack15Rev}, while \eqref{160606:2216}
will not only give more non-zero-constraints, but, additionally,
also two pivotal constraints, namely exactly those two already
obtained before in \eqref{160606:1340}.

The methodological mistake done in \cite{Oberlack15Rev}, namely to
consider \eqref{160606:2042} and not \eqref{160606:2216}, is
critical to their conclusions: (i) Since the correct relation
\eqref{160606:2216} will give the constraint $q_s=3q_\text{NS}$
\eqref{160606:1340}, no logarithmic scaling law for the mean
velocity profile $\bar{U}_1$ can be derived as incorrectly claimed
in \cite{Oberlack15Rev}, because the ansatz $q_s=q_\text{NS}$
would then only lead to $q_s=q_\text{NS}=0$. Hence, only an
algebraic invariant solution for $\bar{U}_1$ can be generated.
(ii) In their ``algebraic solution" [Eq.$\,$(382)] for $k_s\neq
3k_\text{NS}$, the second constraint $k_{{\scriptscriptstyle
z}12}=0$ from \eqref{160606:2216} will give the analytical result
$D_{12}=0$, being different to their DNS-matched value $D_{12}\sim
1$~[Table$\,$10]. Hence, since $D_{12}$ represents the mean
streamwise pressure gradient, the constraint
$k_{{\scriptscriptstyle z}12}=0$ thus can only go along with the
constraint $k_s=3k_\text{NS}$, in order to generate a non-zero
streamwise pressure gradient.

In the following we repeat the (inconsistent) analysis of
\cite{Oberlack15Rev}, in generating several invariant solutions
from \eqref{160605:1457} and matching them to the DNS data of
\cite{Oberlack14}, however, only for the correctly posed
constraints \eqref{160606:1340}. For $q_\text{NS}\neq 0$ and
$q_s\neq 0$, we yield from \eqref{160605:1457} with
\eqref{160606:1340} only a quadratic power-law for the mean
invariant velocity profile~as
\begin{equation}
\bar{U}_1(x_2)=B_1+C_1\left(\frac{x_2}{h}+A\right)^2,\label{160607:2024}
\end{equation}
where we use the parameter notation of \cite{Oberlack15Rev}: The
parameters $A$ and $B_1$ are\pagebreak[4] given by
[Eq.$\,$(379)],\footnote[2]{[Eq.$\,$(379)] in \cite{Oberlack15Rev}
contains two misprints: The parameter $A$ is missing a factor
$1/h$ to be dimensionally correct, and in $C_1$ the non-constant
$k_\text{NS}\, x_2^{k_s/k_\text{NS}-1}$ has to be replaced by
$k_\text{NS}^{k_s/k_\text{NS}-1}$. Moreover, in [Eq.$\,$(380)],
and as well as in [Eq.$\,$(384)], all field variables were
misleadingly denoted in dimensionless ``+"-units, although the
functional expressions themselves are not normalized on $u_\tau$.
Finally note that $B_1$ in [Eq.$\,$(379)] differs by one
translation group parameter to ours defined in
\eqref{160607:2024}. As already mentioned in point
\hyperref[(i)]{(i)} in the beginning of this subsection
(p.$\,$\pageref{(i)}), the reason is that in \cite{Oberlack15Rev}
a different but equivalent linear combination of the two
independent translation symmetries is considered.} as
$A=k_{x_2}/(h\hspace{0.05cm} q_\text{NS})$ and
$B_1=-(k_{\bar{U}_1}+c_1)/(2q_\text{NS})$, while $C_1$ is an
arbitrary integration constant. Note the striking difference that
for the presently considered viscous case ($\nu\neq 0$), the
consistent and correct scaling law for the mean velocity profile
\eqref{160607:2024} carries one (matching) parameter less than the
correspondingly derived (inconsistent) scaling law [Eq.$\,$(380)]
in \cite{Oberlack15Rev}. A consistent analysis shows that
$\gamma=2$, in clear contrast to the non-positive and
non-constantly matched values for $\gamma$ in \cite{Oberlack15Rev}
[Table$\,$10], where the algebraic scaling coefficient $\gamma$ is
declared to be a non-positive and dependent function on the
transpiration rate and the Reynolds number: $\gamma=\gamma(U_T^+,
Re_\tau)<0$.

\begin{figure}[t]
\centering
\begin{minipage}[c]{.48\linewidth}
\FigureXYLabel{\includegraphics[width=.91\textwidth]{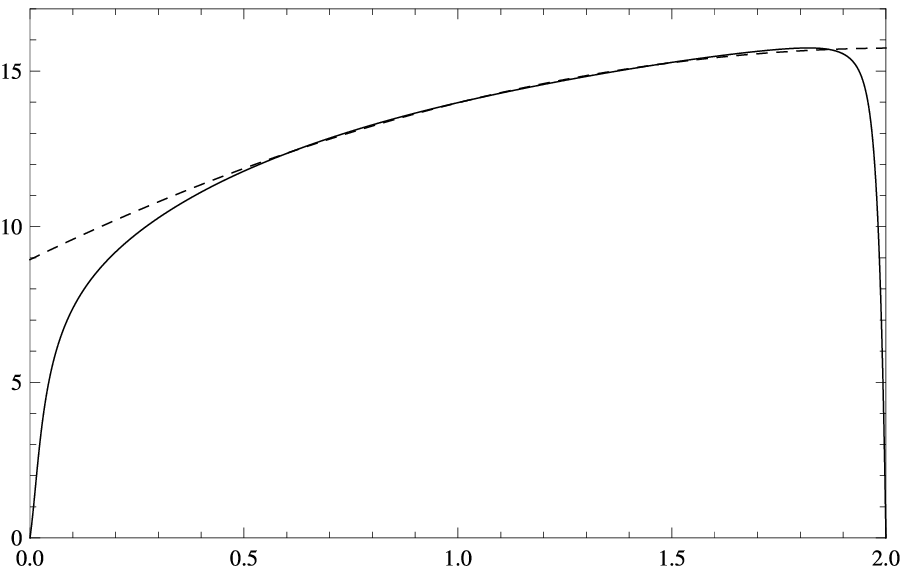}}
{${\scriptstyle\hspace{0.75cm} x_2/h}$}{-1mm}{\begin{rotate}{0}
${\scriptstyle\bar{U}_1^+}$\end{rotate}}{5mm}
\end{minipage}
\hfill
\begin{minipage}[c]{.48\linewidth}
\FigureXYLabel{\includegraphics[width=.91\textwidth]{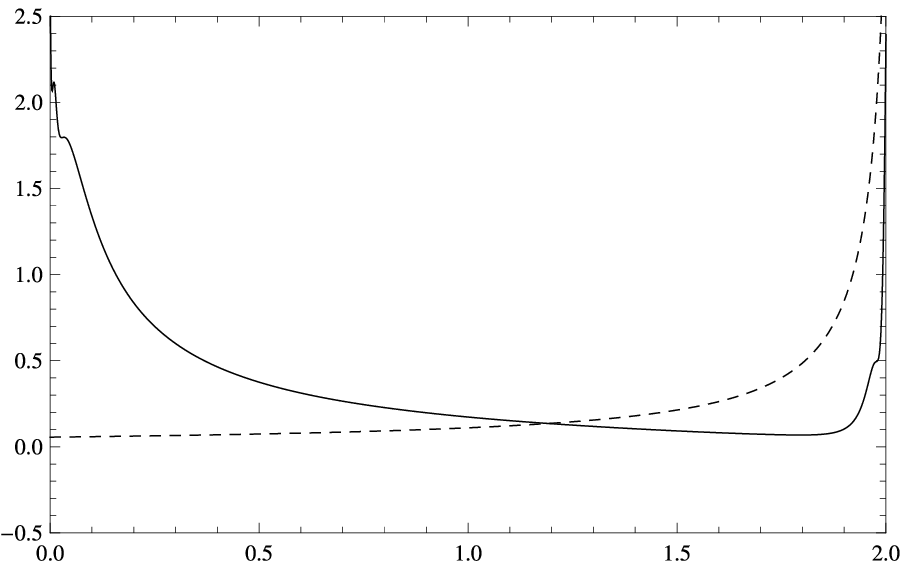}}
{${\scriptstyle\hspace{0.75cm} x_2/h}$}{-1mm}{\begin{rotate}{0}
${\scriptstyle\varepsilon^+}$\end{rotate}}{2mm}
\end{minipage}
\caption{Matching of the two theoretically predicted scaling laws
for the (normalized) mean velocity field $\bar{U}^+_1$
\eqref{160607:2024} and the scalar dissipation $\varepsilon^+$
\eqref{160614:1250} to the DNS data of \cite{Oberlack14} for
$Re_\tau= 480$ and $v^+_0=0.10$. The DNS data is displayed by
solid lines, the corresponding scaling laws by dashed lines. The
matching region $0.5\leq x_2/h\leq 1.6$ has been taken over from
the result determined in \cite{Oberlack15Rev} as listed in
[Table$\,$9]. The resulting best-fitted parameters are given as:
$A=-2.035$, $B_1^+=15.74$, $C_1^+=-1.643$ and $\mu^+=-0.114$.
While the fitting for $\bar{U}_1^+$ is more or less satisfactory,
the fitting of~$\varepsilon^+$~fails.}\vphantom{x}\hrule
\label{fig8}
\end{figure}

With the result \eqref{160607:2024}, all remaining invariant
solutions can be determined from \eqref{160605:1457} with
\eqref{160606:1340} accordingly. For example, the invariant
Reynolds stresses are given as
\begin{gather}
\left.
\begin{aligned}
&\tau_{ij}(x_2)=\beta_{ij}\left(\frac{x_2}{h}+A\right)-\bar{U}_i\bar{U}_j
+\sigma_{ij}\left(\frac{x_2}{h}
+A\right)\ln\left(\frac{x_2}{h}+A\right)+\rho_{i}\bar{U}_j+\rho_{j}\bar{U}_i
+\alpha_{ij},\\[0.5em]
&\tau_{13}(x_2)=0,\qquad \tau_{23}(x_2)=0,
\end{aligned}
~~~ \right\}
\end{gather}
where all $\beta$'s  are arbitrary integration
constants,\footnote[3]{Full arbitrariness in all parameters,
however, is not given, since certain consistency relations have to
be satisfied from the underlying statistical equations
\eqref{160604:1333}-\eqref{160604:1334}. For example, the
parameter $\beta_{12}$ is not arbitrary, but determined as
$\beta_{12}=u_\tau^2+2 C_1 u_\tau/Re_\tau$.} while the remaining
parameters are determined through the group constants as
\begin{gather}
\left.
\begin{aligned}
& \sigma_{ij}=\frac{2h\big(k_{{\scriptscriptstyle
z}11}\delta_{1i}\delta_{1j}+k_{{\scriptscriptstyle
z}22}\delta_{2i}\delta_{2j}+k_{{\scriptscriptstyle
z}33}\delta_{3i}\delta_{3j}\big)}{q_\text{NS}},\qquad
\rho_{i}=\frac{k_{\bar{U}_1}}{q_\text{NS}}\delta_{1i},\\[0.5em]
&\alpha_{ij}=\sigma_{ij}A-\frac{k_{\bar{U}_1}\Big(4B_1\delta_{1i}\delta_{1j}
+2U_T\big(\delta_{1i}\delta_{2j}+\delta_{1j}\delta_{2i}\big)\Big)}
{q_\text{NS}}-\frac{c_{ij}}{q_\text{NS}}.
\end{aligned}
~~~ \right\}
\end{gather}
An interesting measure to verify the predictability of the
invariant functions from \eqref{160605:1457} is the dissipation,
which is given as
\begin{equation}
\varepsilon_{ij}(x_2)=\mu_{ij}\left(\frac{x_2}{h}+A\right)^{-1}
-2\nu\left(\frac{\partial\bar{U}_1}{\partial
x_2}\right)^2\delta_{1i}\delta_{1j},
\end{equation}
where the $\mu$'s are again arbitrary integration constants. For
the statistical DNS data available from \cite{Oberlack14}, we can
only compare to the scalar dissipation defined as
\begin{align}
\varepsilon:=&\frac{1}{2}\sum_{i=1}^3
\varepsilon_{ii}=\frac{1}{2}\sum_{i=1}^3\left[
\mu_{ij}\left(\frac{x_2}{h}+A\right)^{-1}
-2\nu\left(\frac{\partial\bar{U}_1}{\partial
x_2}\right)^2\delta_{1i}\delta_{1j}\right]\nonumber\\[0.5em]
&\equiv \mu\left(\frac{x_2}{h}+A\right)^{-1}
-\nu\left(\frac{\partial\bar{U}_1}{\partial x_2}\right)^2 ,
\end{align}
which, since it was only calculated in the $u_\tau$-normalized
form, has to be transformed accordingly
\begin{align}
\varepsilon^+= &\frac{1}{2}\sum_{i=1}^3 \varepsilon_{ii}^+ =
\frac{1}{2}\sum_{i=1}^3 2\overline{\frac{\partial u_i^+}{\partial
x_k^+}\frac{\partial u_i^+}{\partial x_k^+}}
=\frac{\nu}{u_\tau^4}\cdot\frac{1}{2}\sum_{i=1}^3
2\nu\overline{\frac{\partial u_i}{\partial x_k}\frac{\partial
u_i}{\partial x_k}}=
\frac{\nu}{u_\tau^4}\cdot\frac{1}{2}\sum_{i=1}^3
\varepsilon_{ii}=\frac{\nu}{u_\tau^4}\cdot\varepsilon\nonumber\\[0.5em]
& = \mu^+ \left(\frac{x_2}{h}+A\right)^{-1}-\frac{4C_1^{+
2}}{Re_\tau^2} \left(\frac{x_2}{h}+A\right)^{2}\!
.\label{160614:1250}
\end{align}
To note is that the above scaling law only has one free matching
parameter $\mu^+$, since $A$ and $C_1^+$ are determined by the
scaling law \eqref{160607:2024} of the normalized mean velocity
field $\bar{U}_1^+$. Hence, the scalar dissipation $\varepsilon^+$
will thus be the ultimate litmus test in how far the
Lie-group-based scaling theory, as currently proposed in
\cite{Oberlack15Rev}, is able to consistently predict the scaling
behavior of Navier-Stokes turbulence. As to be expected from the
investigation done in this section, the proposed theory fails: As
shown in Figure \ref{fig8}, the scaling law \eqref{160614:1250}
for the scalar dissipation~$\varepsilon^+$ fails to even roughly
predict the tendency of the DNS data, although for the lowest
order moment, the mean velocity field $\bar{U}_1^+$, the scaling
law \eqref{160607:2024} was matched more or less
satisfactorily.\footnote[2]{That the scaling law
\eqref{160607:2024} for the mean velocity field can be matched
more or less satisfactorily, is not surprising, since this law
involves three independent matching parameters, while the scaling
law \eqref{160614:1250} for the scalar dissipation only involves
one free parameter.}

The reason for this failure is clear: The considered invariant
surface condition \eqref{160605:1457}, as proposed in
\cite{Oberlack15Rev}, involves two unphysical scaling symmetries,
namely $\bar{Q}_\text{NS}$ \eqref{160604:1325} and $\bar{Q}_s$
\eqref{160605:1035}, which both are inconsistent to the underlying
deterministic theory in violating the classical principle of cause
and effect. As a consequence, the theoretically predicted scaling
behavior of the lowest order moment $\bar{U}_1^+$ is incompatible
to the scaling behavior of the higher-order moment
$\varepsilon^+$, as clearly seen in Figure \ref{fig8}, an
incompatibility which also runs through all other higher-order
moments.

\section{Summary and conclusion}

The main motivation of this investigation was to reveal in how far
the study of \cite{Oberlack14} is reproducible. With the data made
available on their institutional
website~\href{http://www.fdy.tu-darmstadt.de/forschung_16/direkte_numerische_simulation/direkte_numerische_simulation.de.jsp}{[fdy]},
we failed to reproduce Fig.$\,$9 ($a$) \& ($c$) in
\cite{Oberlack14}. The critical conclusions made from these
figures can not be confirmed from our analysis: Neither do the
mean velocity profiles (in deficit form) universally collapse onto
a single curve for different transpiration rates at a constant
Reynolds number (Fig.$\,$9 ($a$)), nor does the universally
proposed logarithmic scaling law in the center of the channel
match the DNS data for the presented parameter values
(Fig.$\,$9~($c$)).

No universal scaling behavior in the center of the channel can be
detected as claimed in \cite{Oberlack14}, not even when
considering the case of a constant transpiration rate at different
Reynolds numbers, which led to the incorrect assumption to only
conduct a Reynolds-number independent symmetry analysis. Because,
as we have demonstrated several times, such
\newpage
\newgeometry{left=2.5cm,right=2.5cm,top=2.5cm,bottom=2.2cm,headsep=1em}
\noindent an assumption, of an inviscid ($\nu=0)$ and thus
Reynolds-number independent symmetry analysis, is not justified to
consistently predict the scaling behavior of a channel flow with
uniform wall-normal transpiration for the flow conditions
considered. In particular, we revealed that the associated
$Re_\tau$-{\it independent} scaling group parameter for the mean
velocity field was inconsistently matched in \cite{Oberlack14} to
a $Re_\tau$-{\it dependent} quantity, being proportional
to~$u_\tau$, which, as clearly shown in Figure~\ref{fig6}, or
Table \ref{tab3}, inevitably leads to a strong
$Re_\tau$-dependence in all invariant scaling laws when extending
the scaling theory of \cite{Oberlack14} coherently to higher
orders beyond the mean velocity moment. Hence, a consistent
symmetry analysis to all orders can only be achieved when also
including the viscous terms. This has been attempted in their
subsequent study \cite{Oberlack15Rev}.

But, both the inviscid ($\nu=0)$ as well as the viscous ($\nu\neq
0)$ symmetry analysis, performed in \cite{Oberlack14} and
\cite{Oberlack15Rev}, respectively, is inconsistent {\it per
se}.\footnote[2]{Apart from the additional fact that the symmetry
analysis in \cite{Oberlack15Rev} is also technically flawed, in
that a wrong and not enough constraint relations from the
statistical momentum equations are determined which incorrectly
allow for a logarithmic as well as an algebraic invariant solution
in the mean velocity field. Instead, a correct analysis reveals
that only an algebraic invariant solution of quadratic type can be
obtained. And when excluding even the unphysical symmetries, then
only a featureless linear profile is obtained.} As explained and
discussed in the previous section, this inconsistency is due to
that both their investigations involve several unphysical
symmetries that are inconsistent with the underlying deterministic
description of turbulence, in that they violate the classical
principle of cause and effect: The former inviscid analysis in
\cite{Oberlack14} (when extended to higher-order moments) involves
two unphysical symmetries, namely $\bar{T}_s^\prime$
\eqref{160430:2130} and $\bar{T}^\prime_c$ \eqref{150424:1633},
while the latter symmetry analysis in \cite{Oberlack15Rev}
involves three unphysical symmetries, $\bar{Q}_\text{NS}$
\eqref{160604:1325},\footnote[3]{Recall again that, although the
artificially constructed and unphysical statistical symmetry
$\bar{Q}_\text{NS}$ \eqref{160604:1325} is motivated from the
well-known single physical scaling symmetry of the Navier-Stokes
equations $\bar{T}_\text{NS}$ \eqref{160502:1424}, there is no
connection between them.} $\bar{Q}_s$~\eqref{160605:1035} and
again $\bar{T}^\prime_c$ \eqref{150424:1633}. The consequence: Any
derived set of invariant solutions beyond the lowest-order moment
cannot be consistently matched to the DNS data anymore, as clearly
shown in Figure \ref{fig4} \& \ref{fig8} for the inviscid and
viscous symmetry analysis, respectively. In particular the
matching of the scalar dissipation, being a critical indicator to
judge the prediction quality of any theoretically proposed scaling
laws, failed exceedingly. To gain the mathematical insight into
the reason for this failure, we refer to our foregoing
publications \cite{Frewer14.1,Frewer15.0,Frewer15.0x,Frewer15.1}
and \cite{Frewer16.1}.

\appendix
\section{Friction velocity from both walls as a measure of the
pressure gradient\label{A}}

In a canonical turbulent channel flow of height $2h$ without
wall-normal transpiration, driven by a mean constant streamwise
pressure gradient $-\partial \bar{P}/\partial x_1=K>0$, between
$x_2=0$ (lower plate) and $x_2=2h$ (upper plate), the squared
friction velocity (normalized on the density $\rho$)
\begin{equation}
u_\tau^2=\tau|_{x_2=0}=-\tau|_{x_2=2h}>0,\label{160530:2314}
\end{equation}
where $\tau=\tau(x_2)$ being the total mean shear stress
\begin{equation}
\tau=-\overline{u_1u_2}+\nu\frac{d\bar{U}_1}{dx_2},
\label{160530:2324}
\end{equation}
is simply determined by the pressure gradient and the half-width
of the channel only (see. e.g. \citep{Tennekes72})
\begin{equation}
u_\tau^2=K\cdot h, \label{160530:2201}
\end{equation}
due to the fact that at the center of the channel ($x_2=h$) the
total shear stress is zero, i.e., $\tau|_{x_2=h}=0$, for reasons
of symmetry. In particular, this result \eqref{160530:2201} is
obtained by integrating the mean streamwise momentum equation from
the lower plate upwards
\begin{equation}
0=\int_0^{x_2} \left(K+\frac{d\tau}{dx_2^\prime}\right)dx_2^\prime
=Kx_2 + \tau -\tau|_{x_2=0}=Kx_2+\tau-u_\tau^2,
\end{equation}
\newpage\restoregeometry
\noindent which then reduces to
\eqref{160530:2201} when evaluated at $x_2=h$. However, when
considering a turbulent channel flow with uniform wall-normal
transpiration $v_0$, the total shear stress (including the shear
stress from the transpiration)
\begin{equation}
\mathcal{T}=\tau-v_0\bar{U}_1,
\end{equation}
is obviously  not zero anymore at the center of the channel, i.e.,
$\mathcal{T}|_{x_2=h}\neq 0$, but rather at some different, yet
unknown height position $x_2=x_2^*$ somewhere inside the channel,
i.e., $\mathcal{T}|_{x_2=x_2^*}=0$. Although not knowing this
position $0\leq x_2^*\leq 2h$, one nevertheless can derive the
same relation \eqref{160530:2201} in an averaged sense by
considering the different shear stresses on both walls. Because,
by first integrating the mean streamwise momentum equation once
from the lower plate up to the unknown position
\begin{align}
0=\int_0^{x^*_2} \left(K+\frac{d\mathcal{T}}{dx_2}\right)dx_2
=Kx^*_2 + \mathcal{T}|_{x_2=x_2^*}
-\mathcal{T}|_{x_2=0}&=Kx^*_2-\mathcal{T}|_{x_2=0} \nonumber\\
& =Kx^*_2-\tau|_{x_2=0},
\end{align}
and once from the unknown position up to the upper plate
\begin{align}
0=\int_{x^*_2}^{2h} \left(K+\frac{d\mathcal{T}}{dx_2}\right)dx_2
=2Kh-Kx^*_2 + \mathcal{T}|_{x_2=2h}
-\mathcal{T}|_{x_2=x_2^*}&=2Kh-Kx^*_2+\mathcal{T}|_{x_2=2h}\qquad\nonumber\\
&=2Kh-Kx^*_2+\tau|_{x_2=2h},\;\;
\end{align}
and then by adding both relations, we obtain the
$x_2^*$-independent result
\begin{equation}
0=2Kh+\tau|_{x_2=2h}-\tau|_{x_2=0},
\end{equation}
which, according to the initial definition \eqref{160530:2314},
finally turns into
\begin{equation}
Kh =\frac{\tau|_{x_2=0}-\tau|_{x_2=2h}}{2}=u_\tau^2,
\end{equation}
where we then have, according to \eqref{160530:2324},
\begin{equation}
u_{\tau b}^2 :=\tau_{wb}:=
\tau|_{x_2=0}=\nu\frac{d\bar{U}_1}{dx_2}\bigg|_{x_2=0},\qquad
u_{\tau s}^2 :=\tau_{ws}:=
-\tau|_{x_2=2h}=-\nu\frac{d\bar{U}_1}{dx_2}\bigg|_{x_2=2h},
\end{equation}
the wall shear stresses at the blowing $(b)$ and the suction $(s)$
wall, respectively, and thus overall coinciding with the result
[Eq.$\,$(2.1)] given in \cite{Oberlack14}.

\section[Laminar channel flow with uniform wall transpiration]
{Laminar channel flow with uniform wall
transpiration\footnote[2]{Alternative derivations for laminar
solutions under these flow conditions can also be found, e.g., in
\cite{Chang09} or in \cite{Avsarkisov13}.}\label{B}}

The governing equations are the incompressible Navier-Stokes
equations
\begin{equation}
\left.
\begin{aligned}
\frac{\partial U_k}{\partial x_k}=0,\hspace{1.75cm}\\[0.5em]
\frac{\partial U_i}{\partial t}+ U_j\frac{\partial U_i}{\partial
x_j}=-\frac{\partial P}{\partial x_i}+\nu\Delta U_i,
\end{aligned}
~~~~~\right\}\label{160522:1832}
\end{equation}
which considerably reduces in dimension when considering a
stationary laminar channel flow of width $2h$ driven by a constant
streamwise pressure gradient $K>0$. When additionally considering
permeable walls in which a uniform wall-normal flow $v_0>0$ is
injected at the lower wall (the blowing side $x_2=0$) to be then
also fully uniformly sucked out at the upper wall (the suction
side $x_2=2h$), the overall flow conditions will read:
\begin{equation}
U_1=U_1(x_2), \quad U_2=v_0,\quad U_3=0,\quad -\frac{\partial
P}{\partial x_1}=K,\quad U_1(x_2=0)=U_1(x_2=2h)=0,
\end{equation}
for which the Navier-Stokes equations \eqref{160522:1832} will
reduce to the single equation
\begin{equation}
v_0\frac{dU_1(x_2)}{dx_2}=K+\nu\frac{d^2
U_1(x_2)}{dx_2^2},\;\;\text{with}\;\;
U_1(0)=U_1(2h)=0.\label{160522:1843}
\end{equation}
Two things should be pointed out: (i) If the dependent variable
$U_1$ is not normalized, then equation \eqref{160522:1843}
consists of three parameters which can be varied independently,
the transpiration rate $v_0$, the driving force $K$ and the
viscosity $\nu$. This threefold independent variation turns out to
be necessary when normalizing according to procedure outlined in
\cite{Oberlack14}.\linebreak (ii) The DNS in \cite{Oberlack14} was
performed under the additional constraint of a constant mass
flux.\footnote[2]{To maintain during simulation a constant mass
flux in each time step, the pressure gradient has to adapt
accordingly. However, since we are only interested in the
statistically stationary state, the pressure gradient will still
average out to a constant in the streamwise direction, but in each
case to different values for different transpiration rates and
Reynolds numbers.} This constraint was applied globally
(universally) for all different initially chosen transpiration
rates and Reynolds numbers. Now, since every DNS can also simulate
laminar solutions as a special case, we will construct these in
accord with the simulation performed in \cite{Oberlack14}, i.e.,
we will construct the set of all laminar solutions under the
additional universal constraint of a constant mass flux $Q=Q^*$,
where
\begin{equation}
Q=\rho\cdot \frac{1}{2h}\int_0^{2h} U_1(x_2)dx_2=:\rho\cdot U_B.
\end{equation}
Instead of $Q$ we can also equivalently consider the bulk velocity
$U_B$ (since the density $\rho$ is treated here as constant which
can be absorbed into $Q$, similar to the pressure $P$ in
\eqref{160522:1832} which is also normalized relative to $\rho$).
Note that only the mass flux in the streamwise direction needs to
be considered, since in the wall-normal direction the mass flux is
already constant by construction. Hence, next to equation
\eqref{160522:1843} we thus have to also consider the equation of
a universally fixed bulk velocity $U_B=U_B^*$
\begin{equation}
U_B^*=\frac{1}{2h}\int_0^{2h} U_1(x_2)dx_2,\label{160522:2010}
\end{equation}
that is, equation \eqref{160522:1843} needs to be solved such that
the constraint is always universally satisfied for all different
initially chosen parameters $v_0$, $K$ and $\nu$. The particular
value $U_B^*$ can be chosen arbitrarily from the outset, but once
chosen, it is universally fixed and cannot change anymore during
solution construction.

Before we explicitly solve equation \eqref{160522:1843} under the
constraint \eqref{160522:2010}, it is advantageous to normalize
the expressions appropriately. Two interrelated but different
normalization choices exist: The first one is based on $U^*_B$
along with $h$ (for the independent spatial coordinate). The
system \eqref{160522:1843} and \eqref{160522:2010} then turns into
\begin{gather*}
\frac{v_0}{U_B^*\cdot h}\frac{dU_1(x_2/h\cdot
h)}{d(x_2/h)}=\frac{K}{U_B^*}+\frac{\nu}{U_B^*\cdot h^2}\frac{d^2
U_1(x_2/h\cdot h)}{d(x_2/h)^2},\;\;\text{with}\;\;
U_1(0/h\cdot h)=U_1(2h/h\cdot h)=0,\\[0.5em]
U_B^*=\frac{1}{2h}h\int_{0/h}^{2h/h} U_1(x_2/h\cdot h)d(x_2/h),
\end{gather*}
which, in terms of the dimensionless spatial coordinate
$x_2^\prime=x_2/h$, can be equivalently written~as
\begin{equation}
\left.
\begin{aligned}
v_0^B\frac{d\hat{U}_1(x^\prime_2)}{dx^\prime_2}=w_K^B+\frac{1}{Re_B}\frac{d^2
\hat{U}_1(x^\prime_2)}{dx^{\prime 2}},\;\;\text{with}\;\;
\hat{U}_1(0)=\hat{U}_1(2)=0,\\[0.5em]
U_B^*=\frac{1}{2}\int_{0}^{2} \hat{U}_1(x^\prime_2)dx^\prime_2,
\hspace{3.25cm}
\end{aligned}
~~~~~\right\}\label{160522:2156}
\end{equation}
where $v_0^B=v_0/U_B^*$, $w_K^B=Kh/U_B^*$ and $Re_B=U_B^*h/\nu$
are the (relative to the bulk velocity)\linebreak normalized
transpiration rate, the pressurized forcing rate and the bulk
Reynolds number, respectively. Note that system
\eqref{160522:2156} is yet not fully normalized, since
$\hat{U}_1(x_2^\prime)$ still carries the dimension of velocity.
Obviously, this quantity can be normalized by the remaining
constant velocity scale $w_K^B$, but in this final step we have to
bear in mind that the parameter $w_K^B$ is explicitly needed to
satisfy the constraint equation $U_B=U_B^*$. Hence, only a partial
normalization may be performed in which $w_K^B$ is not completely
absorbed by both equations. This will turn \eqref{160522:2156}
into the equivalent system
\begin{equation}
\left.
\begin{aligned}
v_0^B\frac{d\hat{U}^w_1(x^\prime_2)}{dx^\prime_2}=1+\frac{1}{Re_B}\frac{d^2
\hat{U}^w_1(x^\prime_2)}{dx^{\prime 2}},\;\;\text{with}\;\;
\hat{U}^w_1(0)=\hat{U}^w_1(2)=0,\\[0.5em]
w_K^B=\frac{U_B^*}{\frac{1}{2}\int_{0}^{2}
\hat{U}^w_1(x^\prime_2)dx^\prime_2}, \hspace{3.25cm}
\end{aligned}
~~~~~\right\}\label{160523:0019}
\end{equation}
where $\hat{U}^w_1=\hat{U}_1/w_K^B$ is the normalized
(dimensionless) velocity field relative to the velocity scale
$w_K^B\sim K$ being a measure of the pressure gradient $K$. As
already pointed out in the beginning of this section, three
independent parameters need to be initialized in order to solve
\eqref{160523:0019}: $v_0^B$, $Re_B$ and $U_B^*$, representing
ultimately the transpiration rate $v_0$, the viscosity $\nu$ and
indirectly, via $U_B^*\sim w_K^B$, the pressure gradient $K$,
respectively. Note that the fully normalized
system~\eqref{160523:0019} is uncoupled: The first equation gives
$\hat{U}^w_1$, which then immediately yields the consistent value
for the unknown scale $w_K^B$ by just evaluating the right-hand
side of the second equation.

The second normalization is based on $u_\tau$, as defined through
[Eq.$\,$(2.1)] in \cite{Oberlack14}, and again along with $h$ for
the spatial coordinate. For this choice, system
\eqref{160522:1843} and \eqref{160522:2010} turns into
\begin{gather*}
\frac{v_0}{u_\tau\cdot h}\frac{dU_1(x_2/h\cdot
h)}{d(x_2/h)}=\frac{K}{u_\tau}+\frac{\nu}{u_\tau\cdot
h^2}\frac{d^2 U_1(x_2/h\cdot h)}{d(x_2/h)^2},\;\;\text{with}\;\;
U_1(0/h\cdot h)=U_1(2h/h\cdot h)=0,\\[0.5em]
U_B^*=\frac{1}{2h}h\int_{0/h}^{2h/h} U_1(x_2/h\cdot h)d(x_2/h),
\end{gather*}
which then, again in terms of the dimensionless spatial coordinate
$x_2^\prime=x_2/h$, can be equivalently written~as
\begin{equation}
\left.
\begin{aligned}
v_0^+\frac{d\hat{U}_1(x^\prime_2)}{dx^\prime_2}=u_\tau+\frac{1}{Re_\tau}\frac{d^2
\hat{U}_1(x^\prime_2)}{dx^{\prime 2}},\;\;\text{with}\;\;
\hat{U}_1(0)=\hat{U}_1(2)=0,\\[0.5em]
U_B^*=\frac{1}{2}\int_{0}^{2} \hat{U}_1(x^\prime_2)dx^\prime_2,
\hspace{3.25cm}
\end{aligned}
~~~~~\right\}\label{160524:1008}
\end{equation}
where $u_\tau=\sqrt{Kh}$, $v_0^+=v_0/u_\tau$ and $Re_\tau=u_\tau
h/\nu$ are the friction velocity (measured relative to the
constant streamwise pressure gradient $K>0$), the transpiration
rate based on this scale $u_\tau$ and friction Reynolds number,
respectively. Note again that at this stage system
\eqref{160524:1008} is yet not fully normalized, since
$\hat{U}_1(x_2^\prime)$ still carries the dimension of velocity.
Similarly as discussed before for the first normalization choice,
$\hat{U}_1(x_2^\prime)$ can be obviously normalized by the
constant velocity scale $u_\tau$, but in this step we have to bear
in mind again that the parameter $u_\tau$ is explicitly needed to
satisfy the constraint equation $U_B=U_B^*$. Hence, again, only a
partial normalization may be performed in which $u_\tau$ may not
be completely absorbed by both equations. This will turn
\eqref{160524:1008} into the equivalent system
\begin{equation}
\left.
\begin{aligned}
v_0^+\frac{d\hat{U}^+_1(x^\prime_2)}{dx^\prime_2}=1+\frac{1}{Re_\tau}\frac{d^2
\hat{U}^+_1(x^\prime_2)}{dx^{\prime 2}},\;\;\text{with}\;\;
\hat{U}^+_1(0)=\hat{U}^+_1(2)=0,\\[0.5em]
u_\tau=\frac{U_B^*}{\frac{1}{2}\int_{0}^{2}
\hat{U}^+_1(x^\prime_2)dx^\prime_2}, \hspace{3.25cm}
\end{aligned}
~~~~~\right\}\label{160524:1030}
\end{equation}
where $\hat{U}^+_1=\hat{U}_1/u_\tau$ is the normalized
(dimensionless) velocity field relative to the velocity scale
$u_\tau~\sim\sqrt{K}$ being again a measure of the pressure
gradient $K$. As was also already discussed before, three
independent parameters need to be given again in order to solve
the (uncoupled) system \eqref{160524:1030}: Two, namely $v_0^+$
and $Re_\tau$, in the beginning to solve the first equation and
then one, namely $U_B^*$, in the end to evaluate the second
expression in order to obtain the consistent value for the unknown
scale $u_\tau$.

\begin{table}
\begin{center}
\begin{tabular}{c c | c | c}\hline\\
& & Turbulent flow & Laminar flow\\[0.5em]
$Re_\tau$ & $v_0^+$ & $v_0/U_B^*$ & $v_0^L/U_B^*$\\[0.5em]
250       & 0.05    & 0.0030   & 0.0027\\
250       & 0.10    & 0.0069   & 0.0104\\
250       & 0.16    & 0.0164   & 0.0263\\
250       & 0.26    & 0.0500   & 0.0687\\
250       & $\infty$&          & $\infty$\\[0.5em]
480       & 0.05    & 0.0030   & 0.0026\\
480       & 0.10    & 0.0075   & 0.0102\\
480       & 0.16    & 0.0164   & 0.0259\\
480       & 0.26    & 0.0490   & 0.0681\\
480       & $\infty$&          & $\infty$\\[0.5em]
850       & 0.05    & 0.0026   & 0.0026\\
850       & 0.16    & 0.0160   & 0.0258\\[0.5em]
$\infty$  & $v_0^+\neq 0$ & & $v_0^{+ 2}$\\
$\infty$ & $\infty$ & & $\infty$
\end{tabular}
\caption{Calculated values for $v_0/U_B^*$ for initially given
$Re_\tau$ and $v_0^+$. The values for the turbulent case were
taken from Table$\,$1 [p.$\,$106] in \cite{Oberlack14}, while the
values for the corresponding laminar case were calculated
according to the analytical relation \eqref{160601:0922}, where we
denoted the dimensionalized transpiration rate as $v_0^L$ to
distinguish it from the turbulent flow condition.} \label{tab4}
\end{center}
\vspace{-0.4em}\hrule
\end{table}

The two different normalization choices just discussed above are,
of course, interrelated. That is, system \eqref{160523:0019} can
be bijectively mapped to system \eqref{160524:1030} and vive
versa. The relations are: $v_0^B/v_0^+=u_\tau/U_B^*$ and
$Re_B/Re_\tau=\hat{U}^w_1/\hat{U}_1^+=U_B^*/u_\tau$. Since the
$u_\tau$-normalization is mainly used throughout this study, we
will only show the explicit solution of system
\eqref{160524:1030}, which reads
\begin{equation}
\hat{U}^+_1(x_2^\prime)=\frac{x_2^\prime}{v_0^+}-\frac{2}{v_0^+}
\frac{e^{v_0^+ Re_\tau x_2^\prime}-1}{e^{2v_0^+ Re_\tau}-1},
\;\;\;\text{with}\;\;\;
u_\tau=U_B^*v_0^+\cdot\frac{v_0^+Re_\tau}{v_0^+Re_\tau\cdot\coth(v_0^+
Re_\tau)-1},\label{160524:1523}
\end{equation}
or, in the non-normalized (dimensionalized) form, as:
\begin{equation}
\frac{U^L_1(x_2/h)}{u^L_\tau}=\frac{x_2/h}{v_0^+}-\frac{2}{v_0^+}
\frac{e^{v_0^+ Re_\tau x_2/h}-1}{e^{2v_0^+ Re_\tau}-1}, \qquad
u^L_\tau=U_B^*v_0^+\cdot\frac{v_0^+Re_\tau}{v_0^+Re_\tau\cdot\coth(v_0^+
Re_\tau)-1},\label{160524:1524}
\end{equation}
where we used the notation $U_1=U_1^L$ and $u_\tau=u_\tau^L$ from
Section \ref{S3} \& \ref{S4} to distinguish these quantities from
the corresponding turbulent flow behavior. The initial
(dimensionalized) system parameters $v_0$, $K$ and $\nu$ as given
\eqref{160522:1843} are then related to the three independently
chosen ones $v_0^+$, $Re_\tau$ and $U_B^*$ as follows:
\begin{equation}
\left.
\begin{aligned}
\!\!\!\!\!\! v_0=U_B^*v_0^{+
2}\frac{v_0^+Re_\tau}{v_0^+Re_\tau\cdot\coth(v_0^+
Re_\tau)-1},\;\;\; K=\frac{U_B^{* 2} v_0^{+ 2}}{h}
\left(\frac{v_0^+Re_\tau}{v_0^+Re_\tau\cdot\coth(v_0^+
Re_\tau)-1}\right)^2,\\[0.5em]
\nu=\frac{U_B^* v_0^+
h}{Re_\tau}\frac{v_0^+Re_\tau}{v_0^+Re_\tau\cdot\coth(v_0^+
Re_\tau)-1}.\hspace{4cm}
\end{aligned}
~~\right\}
\end{equation}
Hence, note that when initializing in the $u_\tau$-normalization
the two independent system parameters $v_0^+$ and $Re_\tau$, then
the transpiration parameter normalized on the universal bulk
velocity, i.e. $v_0/U_B^*$, is determined as
\begin{equation}
\frac{v_0}{U_B^*}=v_0^{+
2}\frac{v_0^+Re_\tau}{v_0^+Re_\tau\cdot\coth(v_0^+ Re_\tau)-1},
\label{160601:0922}
\end{equation}
which converges to $v_0^{+2}$ in the limit $Re_\tau\to\infty$ at a
fixed transpiration rate $v_0^+$. In other words, although the
transpiration rate $v_0^+$ inside the $u_\tau$-normalization can
be chosen independently from $Re_\tau$, it is not so for the
bulk-velocity-normalized transpiration rate $v_0/U_B^*$, which is
even bounded when the Reynolds number goes to infinity:
$\lim_{Re_\tau\to\infty}v_0/U_B^*=v_0^{+2}$, a property also to be
expected in the turbulent case, but where the value of course is
unknown; see Table \ref{tab4}.

\bibliographystyle{jfm}
\bibliography{BibDaten}

\end{document}